\documentclass[final,3p,times,twocolumn]{elsarticle}

\usepackage{amssymb}
\usepackage{geometry}
\usepackage[colorlinks=true, 
            linkcolor=blue, 
            citecolor=blue, 
            urlcolor=blue]{hyperref} 
\usepackage{amsmath,amsfonts}
\usepackage{algorithmic}
\usepackage{algorithm}
\usepackage{array}
\usepackage[caption=false,font=normalsize,labelfont=sf,textfont=sf]{subfig}
\usepackage{textcomp}
\usepackage{stfloats}
\usepackage{pifont} 
\usepackage{arydshln}
\usepackage{makecell}
\usepackage{url}
\usepackage{verbatim}
\usepackage{graphicx}
\usepackage{natbib}
\usepackage{titlesec}
\usepackage{booktabs}
\usepackage{multirow}
\usepackage{enumitem} 
\usepackage{cellspace} 
\newenvironment{compactitemize}{%
  \vspace{0.7\baselineskip}
  \begin{itemize}[
    noitemsep,          
    topsep=-2pt,          
    partopsep=0pt,       
    parsep=0pt,          
    leftmargin=*         
  ]
}{%
  \end{itemize}%
  \vspace{-0.3\baselineskip}
}

\begin{document}

\begin{frontmatter}



\title{Advancing Image Super-resolution Techniques in Remote Sensing: \\ A Comprehensive Survey}

\author[inst1]{Yunliang Qi\fnref{label1}}
\author[inst2]{Meng Lou\fnref{label1}}

\author[inst1]{Yimin Liu\corref{cor1}}

\author[inst1]{Lu Li\corref{cor1}}
\ead{lilu@zhejianglab.org}

\author[inst3]{Zhen Yang}
\author[inst1]{Wen Nie}

\cortext[cor1]{Corresponding author: Lu Li and Yimin Liu.}
\fntext[label1]{Authors contributed equally to this paper.}

\affiliation[inst1]{organization={Research Center for Space Computing System, Zhejiang Lab},
            city={Hangzhou},
            postcode={311100},
             state={Zhejiang},
             country={China}}

\affiliation[inst2]{organization={School of Computing and Data Science, The University of Hong Kong},
            city={Hong Kong SAR},
            country={China}}

\affiliation[inst3]{organization={School of Information Science and Engineering, Lanzhou University},
           city={Lanzhou},
           postcode={730000},
            state={Gansu},
            country={China}}

\begin{abstract}
Remote sensing image super-resolution (RSISR) is a crucial task in remote sensing image processing, aiming to reconstruct high-resolution (HR) images from their low-resolution (LR) counterparts. Despite the growing number of RSISR methods proposed in recent years, a systematic and comprehensive review of these methods is still lacking. This paper presents a thorough review of RSISR algorithms, covering methodologies, datasets, and evaluation metrics. We provide an in-depth analysis of RSISR methods, categorizing them into supervised, unsupervised, and quality evaluation approaches, to help researchers understand current trends and challenges. Our review also discusses the strengths, limitations, and inherent challenges of these techniques. Notably, our analysis reveals significant limitations in existing methods, particularly in preserving fine-grained textures and geometric structures under large-scale degradation. Based on these findings, we outline future research directions, highlighting the need for domain-specific architectures and robust evaluation protocols to bridge the gap between synthetic and real-world RSISR scenarios.
\end{abstract}




\end{frontmatter}


\section{Introduction}
\label{sec:sample1}
\begin{figure}[t]
  \centering
  \includegraphics[width=0.498\textwidth]{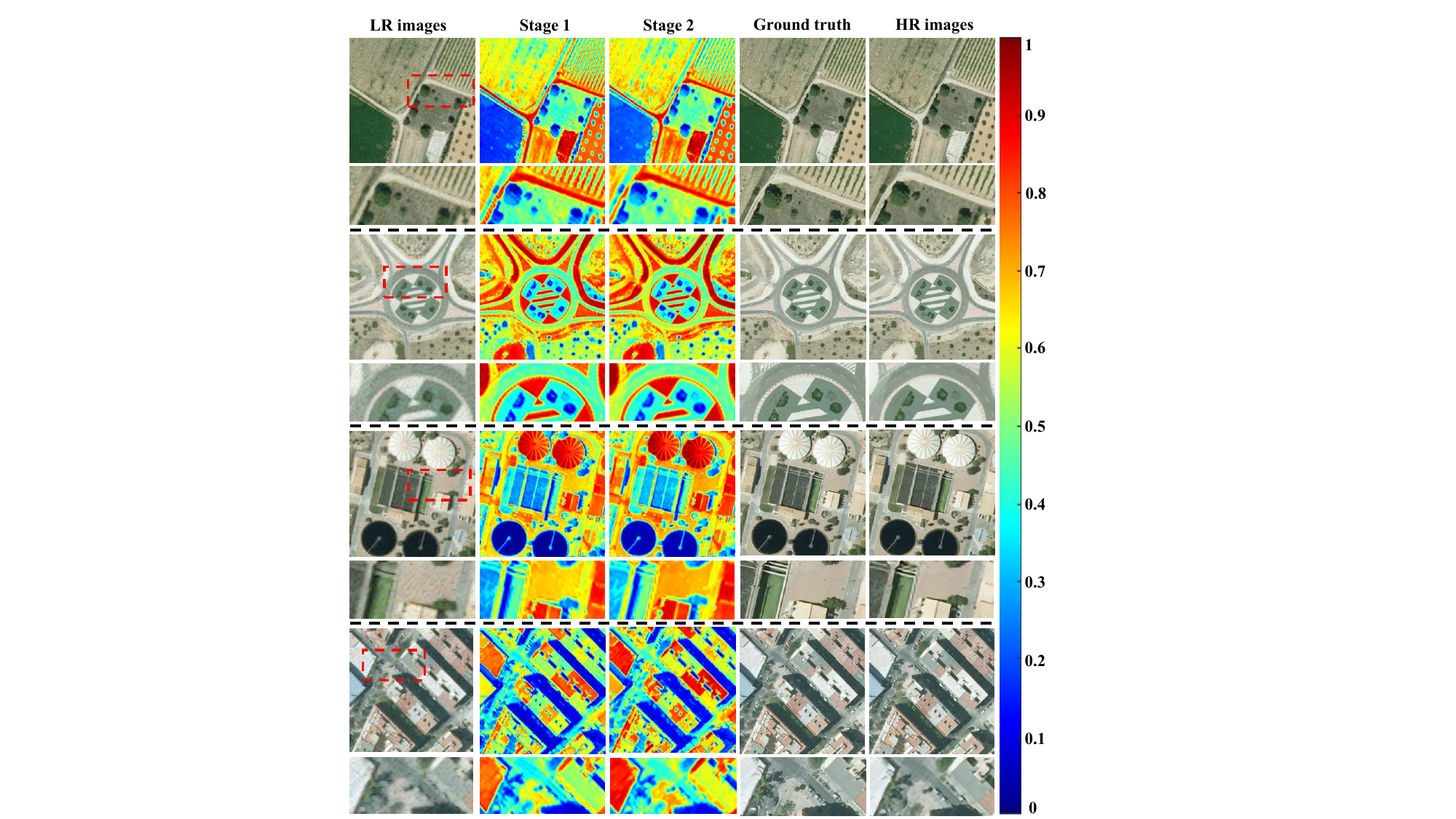}
  \caption{Several examples of RSISR. The first column shows the LR images, while the fifth column shows the results of 4$\times$ SR using the SRCNN method \cite{7115171,10.1007/978}. The second and third columns show heatmap visualizations for the strongest-response channel of the feature maps generated by the first and second convolutional layers, respectively. The fourth column displays the HR ground truth. As can be seen, SR images contain richer details and texture information. The relevant dataset and SR implementation codes are from \cite{Fernandez2017single}.}\label{fig1}
\end{figure}

Image super-resolution (SR) involves reconstructing high-resolution (HR) images from low-resolution (LR) images \cite{farsiu2004fast}. As a fundamental computer vision task, SR has far-reaching applications in various domains, including remote sensing \cite{wang2022comprehensive}, medical imaging \cite{shi2013cardiac}, and security investigation \cite{zou2011very}. However, remote sensing images (RSIs) often suffer from limited spatial resolution due to sensor constraints, such as resolution, cost, orbit altitude, and transmission bandwidth. Despite these limitations, high-quality HR images are essential for real-world remote sensing applications. To address this challenge, remote sensing image super-resolution (RSISR) aims to enhance the detailed information of RSI by generating HR images from LR images, as shown in Fig. \ref{fig1}. RSISR has been extensively applied in land cover mapping \cite{xie2021super}, agricultural monitoring \cite{jia2024enhancing}, disaster assessment \cite{fu2022toward}, and urban monitoring \cite{li2016improved}.
\par
Typically, RSISR approaches are categorized as single-frame or multi-frame based on input count. Single-frame methods utilize a single LR image to reconstruct an HR output, whereas multi-frame techniques exploit multiple LR observations from diverse perspectives or spectral bands to enhance reconstruction fidelity. Multi-frame SR benefits from complementary information across temporal, multi-view, or multi-sensor data, enabling improved recovery of fine details and high-frequency components. However, practical deployment is constrained by the limited availability of multi-temporal or multi-sensor imagery \cite{Fernandez2017single} and the computational complexity of accurate image registration. Satellite-based applications, for instance, face challenges in acquiring multi-temporal data due to long revisit intervals. Consequently, several studies \cite{7284770,chavez2014super,demirel2011discrete,wang2021channel} have focused on single-frame RSISR to circumvent these limitations. Motivated by these considerations, this survey primarily examines single-frame RSISR methodologies.
\par
RSISR has been an active research area for over three decades, with a plethora of algorithms proposed to date. As illustrated in Fig. \ref{fig1}, the number of publications on RSISR methods has experienced a rapid surge in recent years. Early RSISR approaches can be broadly categorized into three paradigms: interpolation-based methods \cite{yang2015remote,ling2013interpolation}, transform domain-based methods \cite{ma2019achieving,wang2023multi}, and reconstruction-based methods \cite{shao2019remote}. Interpolation-based methods offer the advantages of low computational complexity and real-time implementation, but their performance is often limited by the quality of the LR images, making it challenging to recover high-frequency details and real structures. In contrast, reconstruction-based methods can better preserve the perceptual properties of LR images \cite{Fernandez2017single}, but they often suffer from high computational complexity and dependence on the detail preservation of LR images. Transform domain-based methods, on the other hand, transform the image into another domain (e.g., wavelet domain or sparse representation domain) and perform HR reconstruction by enhancing high-frequency components. Various transform domain-based methods have been proposed, including wavelet domain-based methods \cite{ma2019achieving,yang2019multi}, sparse representation methods \cite{he2012learning,shao2019remote,dong2016hyperspectral}, and hybrid methods \cite{wu2016new}. 
\par
In recent years, the rapid development of artificial intelligence (AI) has led to the widespread adoption of deep learning techniques in RSISR. A plethora of deep learning-based RSISR methods have emerged, leveraging various models such as convolutional neural networks (CNNs) \cite{tuna2018single,gargiulo2019advances}, generative adversarial networks (GANs) \cite{wang2023review,jia2022multiattention,tu2024rgtgan}, Transformers \cite{kang2024efficient,xiao2024ttst,xiao2024remote}, and Mamba \cite{zhi2024mambaformersr} to perform RSISR tasks.
\par
Despite the numerous papers on RSISR methods published annually, there is a scarcity of comprehensive survey papers that thoroughly summarize these algorithms. As shown in Table \ref{tab:survey_comparison},  we analyzed and compared existing RSISR reviews across key dimensions such as methodological scope, categorization strategy, and coverage of emerging deep learning architectures. Although these survey works \cite{Fernandez2017single,yang2015remote,liu2021research,wang2022review,karwowska2022using,wang2022comprehensive,lepcha2023image,chen2023review,wang2023review,al2024single} are excellent, they still face significant challenges and limitations as follows:
\begin{itemize}
    \item \textbf{Limited Scope:} Early review papers \cite{yang2015remote,Fernandez2017single} primarily focused on traditional RSISR algorithms, which are insufficient to meet current needs due to the lack of investigation of deep learning methods. Additionally, some surveys only cover specific types of methods, such as deep learning-based methods \cite{wang2022review,wang2022comprehensive,chen2023review} and GAN-based methods \cite{wang2023review}.

    \item \textbf{Outdated Coverage:} Liu et al. \cite{liu2021research} conducted an excellent survey on RSISR, but it was published in 2021 and does not cover algorithms published from 2022 to 2025.
    
    \item \textbf{Inconsistent Categorization:} Existing surveys lack a unified and inclusive framework for classifying the diverse range of RSISR algorithms.

    \item \textbf{Limited Emerging Architectures Coverage:} None comprehensively covers the latest and impactful deep learning paradigms like Transformers, Diffusion model, and Mamba.
\end{itemize}

\begin{figure}
  \centering
  \includegraphics[width=0.495\textwidth]{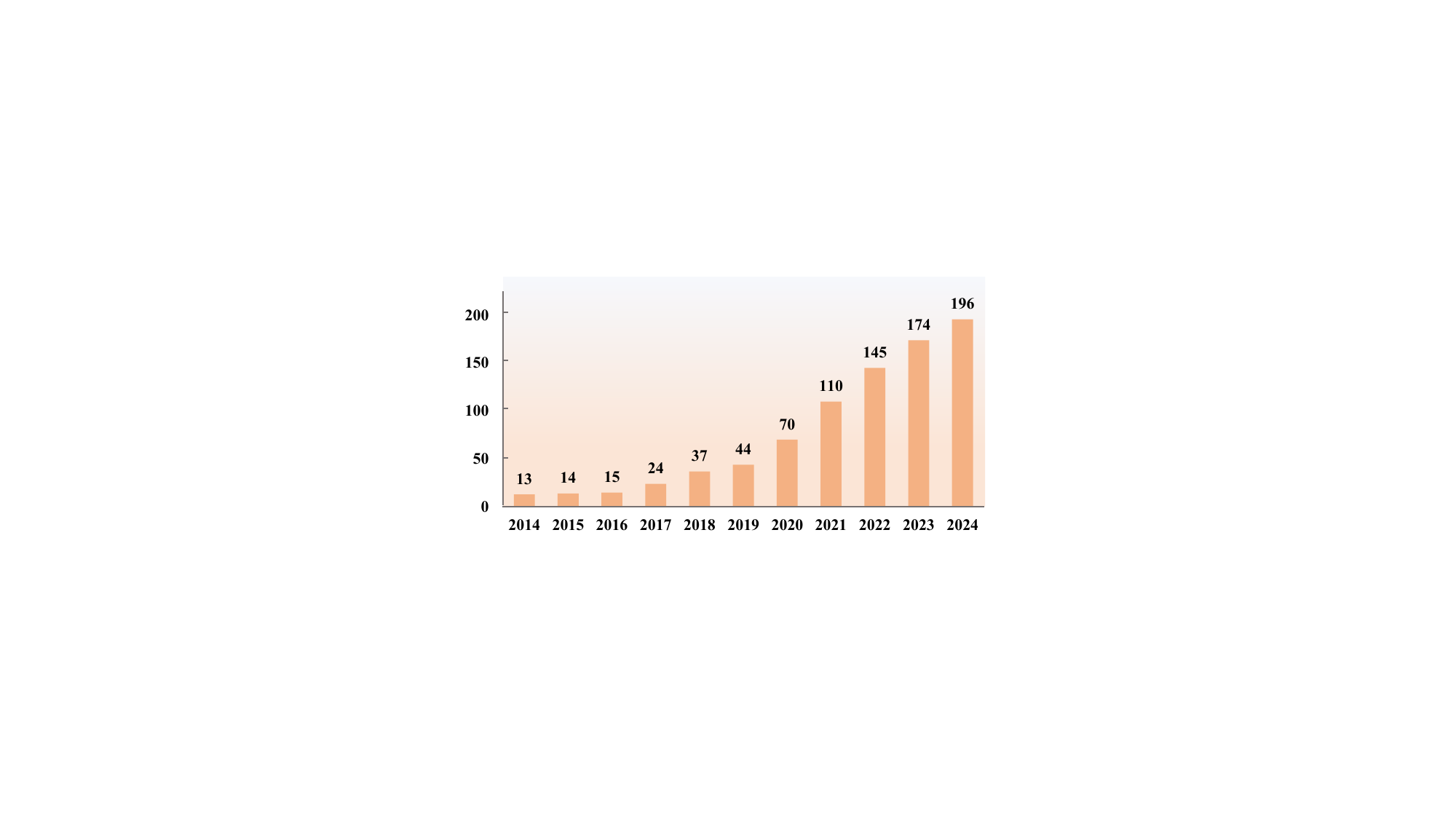}
  \caption{The number of papers about SR algorithms for RSI since 2014 (according to Web of Science).}\label{fig2}
\end{figure}

\begin{figure*}
  \centering
  \includegraphics[width=0.98\textwidth]{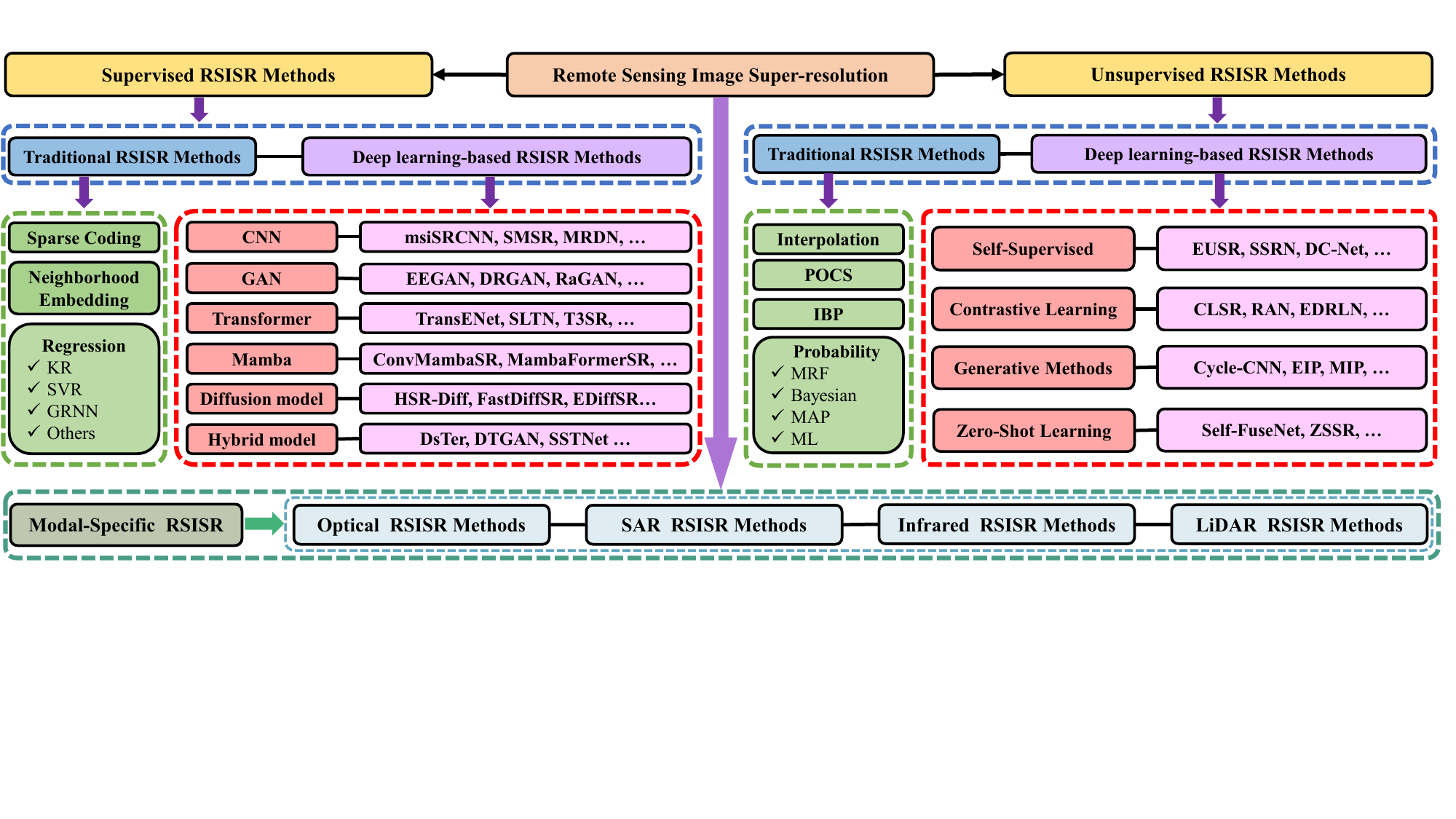}
  \caption{Taxonomy of the existing RSISR methods.}\label{fig3}
\end{figure*}

\begin{table*}[h]
  \centering
  \caption{Comparison of existing RSISR Surveys. Only peer-reviewed surveys are summarized in this table.}
   \resizebox{0.98\textwidth}{!}{
  \begin{tabular}{!{\vrule width 0.8pt}l!{\vrule width 0.8pt}c!{\vrule width 0.8pt}c!{\vrule width 0.8pt}c!{\vrule width 0.8pt}c!{\vrule width 0.8pt}c !{\vrule width 0.8pt} c !{\vrule width 0.8pt} c!{\vrule width 0.8pt}c!{\vrule width 0.8pt}}
  \Xhline{1pt}
  \multirow{2}{*}{Surveys/Reference} &  \multirow{2}{*}{Year} & \multirow{2}{*}{RSI} & \multirow{2}{*}{Methodological Scope}  & \multirow{2}{*}{Categorization Strategy}  & \multicolumn{3}{c!{\vrule width 1.1pt}}{Deep Learning} \\
  \cline{6-8}
      &  & & &  & Transformer & Diffusion Model & Mamba  \\
    \Xhline{1pt}
    Yang et al. \cite{yang2015remote} & 2015 & \checkmark &  Traditional  & Mathematical Modeling Paradigm & \ding{55} & \ding{55} & \ding{55} \\
    \Xhline{1pt}
    Fernandez et al. \cite{Fernandez2017single} & 2017 & \checkmark & Traditional & Task-driven Functional Taxonomy	 & \ding{55} & \ding{55} & \ding{55} \\
    \Xhline{1pt}
    Liu et al. \cite{liu2021research} & 2021 & \checkmark & Traditional + Deep learning & Methodological Chronology Perspective & \ding{55} & \ding{55} & \ding{55} \\
    \Xhline{1pt}
    Wang et al. \cite{wang2022review} & 2022 & \checkmark & Deep learning & Architectural Design-centric Taxonomy & \checkmark & \ding{55} & \ding{55}\\
    \Xhline{1pt}
    Karwowska and Wierzbicki \cite{karwowska2022using} & 2022 & \checkmark & Traditional + Deep learning & Platform-aware Operational Framework	& \ding{55} & \ding{55} & \ding{55} \\
    \Xhline{1pt}
    Wang et al. \cite{wang2022comprehensive} & 2022 & \checkmark & Deep learning & Model Development Workflow & \ding{55} & \ding{55} & \ding{55} \\
    \Xhline{1pt}
    Lepcha et al. \cite{lepcha2023image} & 2023 & \ding{55} & Traditional + Deep learning & Algorithmic Hierarchy and Taxonomy	 & \checkmark & \ding{55} & \ding{55} \\
    \Xhline{1pt}
    Chen et al. \cite{chen2023review} & 2023 & \checkmark & Deep learning & Hyperspectral-Specific Representation & \checkmark & \ding{55} & \ding{55} \\
    \Xhline{1pt}
    Wang et al. \cite{wang2023review} & 2023 & \checkmark & Deep learning & Generative Adversarial Modeling Paradigm & \checkmark & \ding{55} & \ding{55} \\
    \Xhline{1pt}
    Mekhlafi and Liu \cite{al2024single} & 2024 & \ding{55} & Traditional + Deep learning & Distillation-based Optimization Taxonomy & \ding{55} & \ding{55} & \ding{55}  \\
    \Xhline{1pt}
    Ours & 2025 & \checkmark & Traditional + Deep learning & Learning Supervision Paradigm Taxonomy & \checkmark & \checkmark & \checkmark \\
    \Xhline{1pt}
  \end{tabular}}
  \label{tab:survey_comparison}
\end{table*}

In this paper, we address these critical gaps by presenting a detailed, comprehensive, and systematic review of RSISR methods developed over the past two decades, covering more than 400 references. Furthermore, we provide detailed summaries of the background, taxonomy, recent development characteristics, datasets, and evaluation metrics fundamental to RSISR. To the best of our knowledge, this work is the most comprehensive, up-to-date, and well-structured review on RSISR to date, effectively filling the identified gaps in the existing literature. The main contributions of this paper are summarized as follows:

\begin{itemize}
\item \textit{A novel \& inclusive categorization:} We introduce the learning supervised classification paradigm to systematically categorize existing RSISR algorithms, providing a more unified and comprehensive framework than previous taxonomies.

\item \textit{An unprecedented temporal coverage:} We encompass the latest research progress up to 2025, including very recent publications missed by prior surveys.

\item \textit{A pioneering coverage of emerging architectures:} This is the \textbf{first RSISR survey} to thoroughly integrate and discuss the impact of the three most significant emerging deep learning architectures: Transformers, Diffusion models, and Mamba.

\item \textit{A systematic summary of the development characteristics of RSISR methods:} Recent development features are summarized based on an in-depth survey of more than 400 RSISR papers. Further, development trends and challenges can be extrapolated. 

\item \textit{A detailed discussion of future prospects:} Based on a comprehensive investigation and practical experience, future prospects of RSISR methods are discussed. These insights are expected to encourage further exploration of fundamental challenges and inspire broader engagement in this rapidly evolving field. 
\end{itemize}

The rest of this paper is organized as follows. Section \ref{sec2} introduces the background and taxonomy of RSISR methods. Section \ref{sec3} and \ref{sec4}  elaborate on the supervised and unsupervised RSISR methods, respectively. Section \ref{sec:modal} surveys RSISR methods from a modality-specific perspective, followed by a summary of datasets and evaluation metrics in Section \ref{sec6}. Section \ref{sec7} discusses the performance evaluation methods for RSISR. Next, Section \ref{sec5} summarizes the recent development characteristics of the RSISR and Section \ref{sec8} discusses future prospects. Finally, Section \ref{sec9} concludes this paper. 

\section{Background and Taxonomy}
\label{sec2}
This section first elaborates on the image modeling background of the RSISR solution to better understand the mechanism of RSISR. Then, the category structure of the RSISR method in this paper is introduced.
\subsection{Background of RSISR}
RSISR has evolved over two decades, encompassing diverse technical branches such as multispectral image super-resolution (MSISR) \cite{qian2022selfs2}, hyperspectral image super-resolution (HSISR) \cite{liu2024spectral}, and pan-sharpening \cite{he2025pan}. RSISR has emerged as a critical research field due to the inherent resolution limitations of satellite and aerial imaging systems, which directly affect the performance of downstream applications such as land-use monitoring \cite{xie2021super}, disaster assessment \cite{fu2022toward}, and urban monitoring \cite{li2016improved}.

Generally speaking, the process of RSI acquisition can be understood as a sequence of degradation operations involving blurring, downsampling, and noise. Specifically, RSI acquisition can be modeled as follows:
\begin{equation}\label{eq1}
{I_{LR}} = {\mathcal{D}}\left( {{\mathcal{B}}\left( {{I_{HR}}} \right)} \right) + {\mathcal{N}}
\end{equation}
where $\mathcal{B}$ is the blurring operator, which represents the blurring factor introduced by the optical system, such as low-pass filtering, $\mathcal{D}$ represents the downsampling operator of the corresponding sensor resolution, and $\mathcal{N}$ represents the noise disturbance in the imaging process. $I_{HR}$ and $I_{LR}$ are the corresponding HR image and LR image of the same target scene, respectively. It can be seen from Eq.(\ref{eq1}) that the RSISR task is a process of inversely solving RSI-degraded imaging, which is an ill-posed problem because different HR images may obtain the same LR image after passing through a specific imaging model.

\subsection{Taxonomy of RSISR Algorithms}
RSISR algorithms can be classified in different ways, such as model architecture-based, input data modality-based, application-oriented, processing scope-based, loss function-based, etc. Given its fundamental impact on model design, training requirements, and real-world applicability, this review adopts the supervised vs. unsupervised dichotomy as the main axis of classification. Other taxonomies, including those based on architecture, data modality, application focus, processing scope, and loss strategies, are discussed within each primary category to provide a multi-faceted understanding of RSISR methods. Fig. \ref{fig3} shows a taxonomy of existing RSISR methods in this paper. Specifically, unsupervised methods and supervised methods can be grouped into traditional methods and deep learning based methods, respectively. It should be mentioned that most existing traditional RSISR methods are unsupervised, which mainly focus on the intrinsic structural characteristics and prior knowledge of the image. In addition, most deep learning-based RSISR methods are supervised, which mainly rely on large-scale labeled data to learn a more complex mapping. Furthermore, supervised traditional methods can be divided into sparse coding-based, neighborhood embedding-based, and regression-based methods, while unsupervised traditional methods can be divided into interpolation-based, Projection onto Convex Sets (POCS)-based, Iterative Back Projection (IBP)-base,d and probability-based methods. According to the type of adopted model, supervised deep learning methods can be divided into CNN-based, GAN-based, Transformer-based, Diffusion-based, Mamba-based, and hybrid model-based methods. In contrast, unsupervised deep learning methods can be divided into self-supervised learning, contrastive learning, generative methods, and zero-shot learning methods. An overview of representative
RSISR methods are given in Table \ref{tab1}.

\begin{table*}
  \centering
  \caption{An overview of representative RSISR methods. Only peer-reviewed methods are summarized in this table.}
   \resizebox{0.98\textwidth}{!}{
  \begin{tabular}{!{\vrule width 1pt}l!{\vrule width 1pt}c!{\vrule width 1pt}c!{\vrule width 1pt}c!{\vrule width 1pt}c!{\vrule width 1pt}c c!{\vrule width 1pt}c c!{\vrule width 1pt}c!{\vrule width 1pt}}
  \Xhline{1pt}
    \multirow{2}{*}{Name/Reference} & \multirow{2}{*}{Year} & \multirow{2}{*}{Category}  & \multirow{2}{*}{Image type}  & \multirow{2}{*}{\centering Innovations} & \multicolumn{2}{c!{\vrule width 1.1pt}}{Supervised} & \multicolumn{2}{c!{\vrule width 1.1pt}}{Deep Learning}    & \multirow{2}{*}{Code} \\
    \cline{6-9}
    & & & &   & Yes  & No  &  Yes  & No   &                   \\
    \Xhline{1pt}
NeedFS \cite{chan2009neighbor} & 2009 & Neighbor embedding & Natural & Edge-aware feature-selected neighbor embedding & \checkmark & & & \checkmark & \\
\Xhline{1pt}
Zhang et al. \cite{zhang2011super} & 2011 & POCS & Optical & Global weighted projection convex set optimization & & \checkmark  & & \checkmark & \\
\Xhline{1pt}
He et al. \cite{he2012learning} & 2012 & Sparse coding & SAR & Coherence-optimized coupled multi-dictionary & \checkmark &  &  & \checkmark &  \\
Zhou et al. \cite{zhou2012interpolation} & 2012 & Interpolation & Natural & Multisurface maximum a posteriori fusion & & \checkmark  & & \checkmark & \\
\Xhline{1pt}
ANR \cite{timofte2013anchored} & 2013 & Neighbor embedding & Natural & Anchored collaborative coding super-resolution
 & \checkmark &  & & \checkmark & \checkmark (Matlab)\\
\Xhline{1pt}
DGNE \cite{6805627} & 2014 & Neighbor embedding & Natural & Dual-geometric neighbor embedding & \checkmark & & & \checkmark & \\
Ma et al. \cite{ma2014robust} & 2014 & Kernel regression & Optical & Robust locally weighted kernel regression with Taylor approximation &   & \checkmark  & & \checkmark & \\
SVR\_SRM \cite{zhang2014example} & 2014 & SVR &  Hyperspectral &  Example-based support vector regression & \checkmark & & & \checkmark & \\
\Xhline{1pt}
JOR \cite{dai2015jointly} & 2015 & Regression & Natural & Adaptive multi-regressor ensemble selection & \checkmark & &  & \checkmark & \checkmark (Matlab)\\
SRCNN \cite{7115171} & 2015 & \textbf{CNN} & Natural & End-to-end convolutional super-resolution & \checkmark & & \checkmark & & \checkmark (Matlab) \\
\Xhline{1pt}
SSME \cite{xinlei2016super} & 2016 & Sparse coding   & Optical &  Geometric regularities embedding &   & \checkmark  & & \checkmark & \\
GRNN-SMUF \cite{rs8080625} & 2016 & GRNN & Optical & First application in urban flood super-resolution mapping & \checkmark & & & \checkmark & \\
Liu et al. \cite{liu2016improved} & 2016 & POCS & Infrared & Variable thresholds and contrast constraints via visual mechanism & & \checkmark  & & \checkmark & \\
\Xhline{1pt}
LGCNet \cite{lei2017super} & 2017 & CNN & Optical & Multifork-structured local-global CNN prior integration & \checkmark &  & \checkmark & &   \\
\Xhline{1pt}
Haut et al. \cite{haut2018new} & 2018 & Self-Supervised & Optical & Unsupervised convolutional generative super-resolution & & \checkmark  &  \checkmark & & \\
Solanki \cite{solanki2018efficient} & 2018 & Interpolation & Optical & Wavelet-enhanced edge-directed artifact-free super-resolution & & \checkmark  & & \checkmark & \\
Irmak \cite{irmak2018map} & 2018 & Probability theory & Hyperspectral & Quadratic optimization in abundance map domain & & \checkmark  & & \checkmark & \\
EEIBP \cite{nayak2018enhanced} & 2018 & IBP & Natural & Hybrid spline-adaptive evolutionary ibp optimization & & \checkmark & & \checkmark & \\
\Xhline{1pt}
CSAE \cite{shao2019remote} & 2019 & Sparse coding  & Optical & Coupled sparse autoencoder mapping & \checkmark & & & \checkmark & \\
EEGAN \cite{jiang2019edge} & 2019 & GAN & Optical & Edge-enhanced GAN with noise purification and mask processing & \checkmark &  & \checkmark & & \\
FSVMGRNN \cite{li2019enhanced} & 2019 & GRNN & Optical & Combination of SVM and GRNN & \checkmark & & & \checkmark & \\
\hdashline
DRSEN \cite{gu2019deep} & 2019 & CNN & Optical & Residual squeeze-excitation channel attention & \checkmark &  & \checkmark & & \\
Haut et al. \cite{haut2019remote} & 2019 & CNN & Optical & Visual attention mechanism & \checkmark &  & \checkmark & & \\
MPSR \cite{dong2019transferred} & 2019 & CNN & Optical & Multi-perception attention with adaptive hierarchical fusion & \checkmark &  & \checkmark & & \\
\hdashline
RSMAP \cite{zhang2019sea} & 2019 & Probability theory & SAR & Rayleigh-based sparse Bayesian deconvolution & & \checkmark  & & \checkmark & \\
\Xhline{1pt}
CA-FRN \cite{li2020fused} & 2020 & CNN & Optical & Parameter-efficient channel-attentive recursive fusion & \checkmark &  & \checkmark & & \checkmark (Pytorch)  \\
MSAN \cite{zhang2020scene} & 2020 & CNN & Optical & Multiscale activation fusion with scene-adaptive & \checkmark &  & \checkmark & & \\
NLASR \cite{wang2020non} & 2020 & CNN & Optical & Non-local global context with multi-scale channel-spatial attention & \checkmark &  & \checkmark & & \\
\hdashline
CDGANs \cite{lei2019coupled} & 2020 & GAN & Optical & Coupled-discriminator GAN with dual-path adversarial learning & \checkmark &  & \checkmark & & \\
EUSR \cite{sheikholeslami2020efficient} & 2020 & Self-Supervised & Optical & Dense skip-connection efficient unsupervised super-resolution & & \checkmark  &  \checkmark & & \\
\Xhline{1pt}
DSSR \cite{dong2020remote} & 2021 & CNN & Optical & Dense multi-prior attention with chain optimization & \checkmark &  & \checkmark & & \\
MHAN \cite{haut2019remote} & 2021 & CNN & Optical & High-order attention with frequency-aware refinement & \checkmark &  & \checkmark & & \checkmark (Pytorch)\\
\hdashline
CGAN \cite{guo2021remote} & 2021 & GAN & Optical & Cascade generative adversarial networks with scene-content constraints & \checkmark &  & \checkmark & & \\
\hdashline
MIP \cite{wang2021unsupervised} & 2021 & Generative method & Optical & Unsupervised migration prior generative super-resolution & & \checkmark  &  \checkmark & & \checkmark (Pytorch) \\
Cycle-CNN \cite{zhang2020nonpairwise} & 2021 & Generative method & Optical & Unsupervised cycle convolutional super-resolution & & \checkmark  &  \checkmark & & \\
EIP \cite{wang2021enhanced} & 2021 & Generative method & Optical & Enhanced latent prior generative super-resolution & & \checkmark  &  \checkmark & & \checkmark (Pytorch) \\
\hdashline
SSRN \cite{chen2021hyperspectral} & 2021 & Self-Supervised & Hyperspectral & Self-supervised spectral-spatial residual fusion & & \checkmark  &  \checkmark & & \\
Wang et al. \cite{wang2021improved} & 2021 & POCS & Hyperspectral & Gradient interpolation and adaptive iteration termination & & \checkmark  & & \checkmark & \\
\Xhline{1pt}
FeNet \cite{wang2022fenet} & 2022 & CNN & Optical & Attention-guided lightweight lattice hierarchical fusion & \checkmark &  & \checkmark & & \checkmark (Pytorch)\\
Zhang et al. \cite{zhang2022single} & 2022 & CNN & Optical & Blur-noise-degradation-modeled RBAN with adversarial UNet & \checkmark &  & \checkmark & & \checkmark (Pytorch)\\
\hdashline
Fusformer \cite{hu2022fusformer} & 2022 & Transformer & Hyperspectral & Transformer-based spatial residual estimation & \checkmark &  & \checkmark & & \checkmark (Pytorch) \\
TransENet \cite{lei2021transformer} & 2022 & Transformer & Optical & Transformer-based multistage enhancement fusion & \checkmark &  & \checkmark & & \\
\hdashline
MA-GAN \cite{jia2022multiattention} & 2022 & GAN & Optical & Multi-attention GAN with pyramid-convolutional mechanisms & \checkmark &  & \checkmark & & \checkmark (Pytorch) \\
Mutai et al. \cite{mutai2022cubic} & 2022 & IBP & Optical & Wavelet-guided b-spline edge-preserving fusion & & \checkmark & & \checkmark & \\
Ye et al. \cite{ye2022bayesian} & 2022 & Probability theory & Hyperspectral & Bayesian fusion for high-resolution hyperspectral imaging & & \checkmark  & & \checkmark & \\
\Xhline{1pt}
JASISR \cite{deka2023joint} & 2023 & Sparse coding  & Optical & Adaptive joint local-nonlocal strategy & & \checkmark & & \checkmark & \\
HAUNet \cite{wang2023hybrid} & 2023 & CNN & Optical & Hybrid-attention U-Net with cross-scale lightweight interactionh) & \checkmark &  & \checkmark & & \checkmark (Pytorch) \\
HSR-Diff \cite{wu2023hsr} & 2023 & Diffusion model & Hyperspectral & Conditional Diffusion HSI-MSI Progressive Denoising & \checkmark &  & \checkmark & &   \\
ESSAformer \cite{zhang2023essaformer} & 2023 & Transformer & Hyperspectral & Efficient spectral-correlation kernel attention with linear complexity & \checkmark &  & \checkmark & & \checkmark (Pytorch) \\
FSSBP \cite{tao2023fssbp} & 2023 & IBP & Multispectral & Spatial-spectral consistent back-projection fusion & \checkmark &  &  & \checkmark & \checkmark (Matlab) \\
\hdashline
RAN \cite{liu2023ran} & 2023 & Contrastive learning & Optical & Contrastive region-aware graph super-resolution &  \checkmark & & \checkmark &  & \\
CLSR \cite{mishra2023clsr} & 2023 & Contrastive learning & Optical & Contrastive semi-supervised multi-modal super-resolution &  & \checkmark  & \checkmark  & & \\
\hdashline
DRSR \cite{xiao2023degrade} & 2023 & Self-Supervised & Hyperspectral & Self-supervised degradation-guided adaptive super-resolution & & \checkmark  &  \checkmark & & \checkmark (Pytorch)\\
SelfS2 \cite{qian2022selfs2} & 2023 & Self-Supervised & Multispectral & Self-supervised spectral-spatial deep prior & & \checkmark  &  \checkmark & & \\
\hdashline
Self-FuseNet \cite{mishra2023self} & 2023 & Zero-Shot learning & Optical & Forward self-fusion single-image super-resolution & & \checkmark  &  \checkmark & & \\
Cha et al. \cite{cha2023meta} & 2023 & Zero-Shot learning & Optical & Meta-learning multi-task zero-shot super-resolution & & \checkmark  &  \checkmark & & \\
\Xhline{1pt}
TSFNet \cite{wang2024two} & 2024 & CNN & Optical & Two-stage spatial-frequency adaptive degradation disentanglement & \checkmark &  & \checkmark & & \checkmark (Pytorch) \\
\hdashline
SCDM \cite{chen2024spectral} & 2024 & Diffusion model & Hyperspectral & Spectral-cascaded Diffusion Model with Image Condition Guidance & \checkmark &  & \checkmark & & \checkmark (Pytorch) \\
FastDiffSR \cite{meng2024conditional} & 2024 & Diffusion model & Optical & Fast Conditional Diffusion with Residual Attention Efficiency & \checkmark &  & \checkmark & & \checkmark (Pytorch) \\
LWTDM \cite{an2023efficient} & 2024 &  Diffusion model & Optical & Lightweight Cross-Attention Diffusion  & \checkmark &  & \checkmark & & \checkmark (Pytorch) \\
SDP \cite{liu2024spectral} & 2024 & Diffusion model & Hyperspectral & Unsupervised Spectral Diffusion Prior MAP Integration    & & \checkmark & \checkmark & & \checkmark (Pytorch) \\
SSDiff \cite{zhong2024ssdiff} & 2024 & Diffusion model & Optical & Spatial-spectral diffusion with frequency-modulated alternating fusion & \checkmark &  & \checkmark & & \checkmark (Pytorch) \\
EVADM \cite{lu2024effective} & 2024 & Diffusion model & Optical & Variance-aware spatial attention enhanced efficient diffusion & \checkmark &  & \checkmark & & \checkmark (Pytorch) \\
EDiffSR \cite{10353979} & 2024 & Diffusion model & Optical & Efficient diffusion with simplified attention and prior enhancement & \checkmark &  & \checkmark & & \checkmark (Pytorch) \\
\hdashline
SRDSRAN \cite{10375518} & 2024 & GAN & Optical & Stratified dense sampling residual attention with chain training & \checkmark &  & \checkmark & & \checkmark (Pytorch) \\
RGTGAN \cite{tu2024rgtgan} &  2024 & GAN & Optical & Gradient-assisted GAN with deformable convolution & \checkmark &  & \checkmark & & \checkmark (Pytorch) \\
\hdashline
ESTNet \cite{10746331} & 2024 & Transformer & Optical & Efficient swin transformer with channel-group attention & \checkmark &  & \checkmark & & \checkmark (Pytorch) \\
TTST \cite{xiao2024ttst} & 2024 & Transformer & Optical & Top-k token selective transformer with multi-scale context & \checkmark &  & \checkmark & & \checkmark (Pytorch) \\
SymSwin \cite{jiao2024symswin} & 2024 & Transformer & Optical & Symmetric multi-scale Swin transformer with wavelet loss & \checkmark &  & \checkmark & & \checkmark (Pytorch) \\
TTSR \cite{wang2024ttsr} & 2024 & Transformer & Optical & Local-global deformable transformer with terrain-optimized loss & \checkmark &  & \checkmark & & \\
LSwinSR \cite{10683775} & 2024 & Transformer & UAVs & Swin transformer with semantic segmentation validation & \checkmark &  & \checkmark & & \checkmark (Pytorch) \\
\hdashline
ConvMambaSR \cite{zhu2024convmambasr} & 2024 & Mamba & Optical & State-space and CNN integrated global-local fusion & \checkmark &  & \checkmark & & \\
FreMamba \cite{10817590} & 2024 & Mamba & Optical & Frequency-Mamba Fusion Framework & \checkmark &  & \checkmark & & \checkmark (Pytorch) \\
MambaFormerSR \cite{zhi2024mambaformersr} & 2024 & Mamba & Optical & Light Mamba-Transformer SR Fusion & \checkmark &  & \checkmark & & \\
\hdashline
EDRLN \cite{wang2024efficient} & 2024 & Contrastive learning & Optical & Contrastive degradation-aware efficient super-resolution &  \checkmark & & \checkmark &  & \checkmark (Pytorch) \\
TransCycleGAN \cite{zhai2024transcyclegan} & 2024 & Generative method & Optical & Transformer-integrated unpaired CycleGAN super-resolution & & \checkmark  &  \checkmark & & \\
TM-GAN \cite{10509697}& 2024 & Hybird & RGB-D & Transformer-based multi-modal generative adversarial network & \checkmark & & \checkmark &  & \\
\Xhline{1pt}
Sharifi et al. \cite{10829708} & 2025 & Transformer & Optical & Multihead integrated spatial-channel attention & \checkmark &  & \checkmark & & \\
\hdashline
IRR-DiffSR \cite{10947187} & 2025 & Diffusion model & Optical & Image reconstruction representation-guided efficient diffusion & \checkmark &  & \checkmark & & \checkmark (Pytorch) \\
SGDM \cite{wang2025semantic}  & 2025 & Diffusion model & Optical & Semantic-guided diffusion with imaging characteristic decoupling & \checkmark &  & \checkmark & & \checkmark (Pytorch) \\
\hdashline
DRGAN \cite{10781453} & 2025 &  GAN & Optical & Dynamic-convolution self-attention dense GAN & \checkmark &  & \checkmark & & \\
Guo  et al. \cite{guo2025structured} & 2025 & Probability theory & SAR & Structured Bayesian modeling for moving radar imaging & & \checkmark  & & \checkmark & \\
ViT-ISRGAN \cite{10836746} & 2025 & Hybird & Multispectral & Vision transformer with spatial-spectral attention & \checkmark & & \checkmark &  & \\
    \Xhline{1pt}
  \end{tabular}}
  \label{tab1}
\end{table*}

\section{Supervised RSISR Methods}
\label{sec3}
Recent advances in machine learning have enabled supervised learning-based methods to surpass traditional approaches \cite{wang2020deep}. Generally, the supervised SR algorithms require training with HR-LR image pairs to learn the mapping relationship from LR images to HR images. In this section, we provide a comprehensive survey of supervised RSISR algorithms, categorizing them into two primary paradigms: traditional methods and deep learning-based approaches.

\subsection{Traditional RSISR Methods}
\subsubsection{Sparse Coding-based Methods}
\begin{figure}[t]
  \centering
  \includegraphics[width=0.46\textwidth]{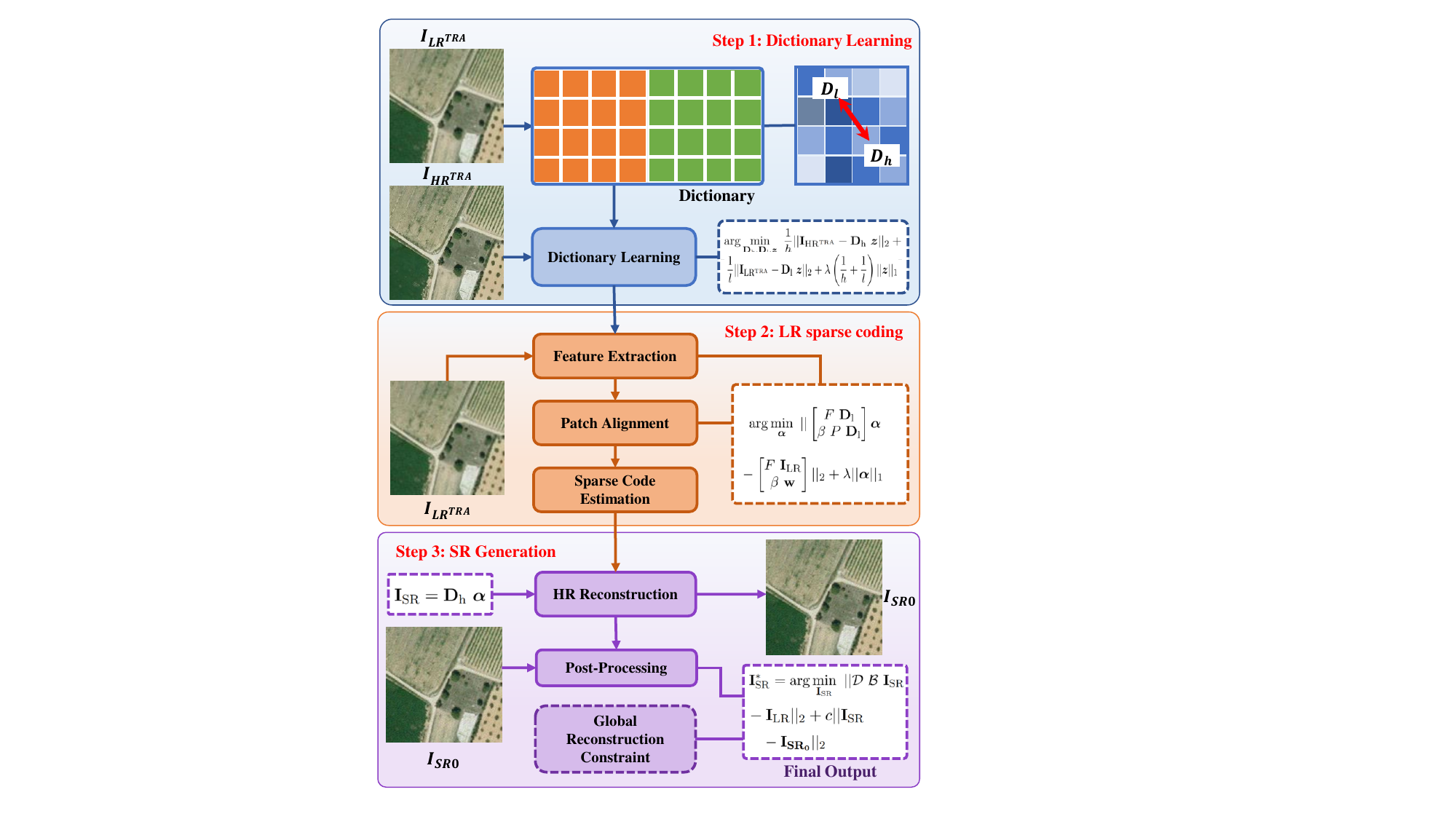}
  \caption{RSISR framework based on sparse representation.}\label{fig4}
\end{figure}
Sparse coding is a learning-based method that represents a signal via an overcomplete dictionary with a small number of non-zero coefficients. To the best of our knowledge, Yang et al. \cite{yang2008image,yang2010image} proposed the first sparse coding-based SR method, which mainly consists of three algorithmic steps: dictionary learning, sparse coding, and HR reconstruction, as shown in Fig. \ref{fig4}. Specifically, an over-complete dictionary is first learned from the training patches, then the test patches are represented by sparse coefficients, and finally, the weighted coefficients are used to reconstruct the HR image.

Some studies based on sparse representation have been conducted to reconstruct the HR images in remote sensing. For example, Zheng et al. \cite{zhihui2011single} proposed an improved sparse representation method for joint RSISR and denoising. Specifically, they employed Batch Orthogonal Matching Pursuit (Batch-OMP) for LR encoding and achieving HR patch reconstruction via jointly-trained LR-HR dictionaries with sparse representation consistency. In addition, Zhang et al. \cite{zhang2013remote} divided the over-complete dictionary into the primitive dictionary pair and the residual dictionary pair. The former is used to generate the initial HR image from the LR image, and the latter is learned to reconstruct the residual loss information. Similar work in Gou et al. \cite{gou2014remote} was performed to improve performance, which presented a non-local paired dictionary learning model involving an estimated dictionary and a residual dictionary. Considering that the sparse coefficients of the observed images of LR and HR images can hardly be kept consistent, Shao et al. \cite{shao2019remote} proposed a novel coupled sparse autoencoder (CSAE) to learn the mapping relationship between LR and HR images. Besides, Deka et al. \cite{deka2023joint} proposed an adaptive joint sparse representation-based method for RSISR. The main novelty is that its dictionary learning and sparse reconstruction only depend on the input LR images.  A sparse structural manifold embedding (SSME) method proposed by Wang and Liu \cite{xinlei2016super} incorporated the geometric properties of images into the neighbors calculation to recover the structure and edge detail information of HR images.

Although sparse representation-based methods are effective in RSISR tasks, they still face many challenges, such as low computational efficiency, high dependence on dictionaries, and low robustness. In addition, with the massive increase in remote sensing data, the real-time requirements of RSISR are becoming increasingly higher, which greatly limits the development of sparse representation methods.

\subsubsection{Neighborhood Embedding-based Methods}
\begin{figure}[t]
  \centering
  \includegraphics[width=0.48\textwidth]{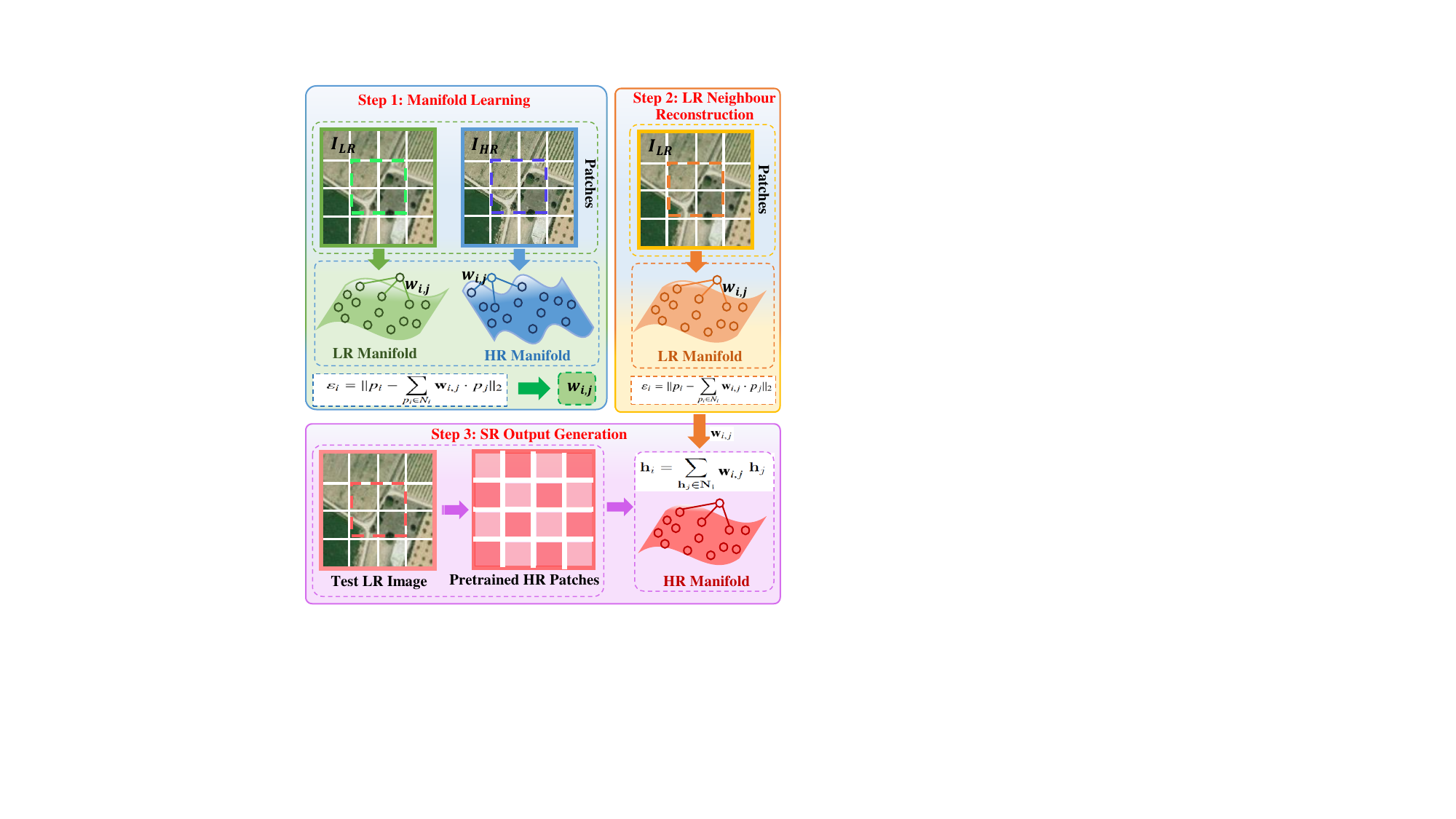}
  \caption{NE-based RSISR method}\label{fig5}
\end{figure}
Neighborhood embedding (NE) based methods consider that small patches of LR and HR images form low-dimensional nonlinear manifolds with uniform local geometry \cite{roweis2000nonlinear}. That is, as long as the number of samples is sufficient, the corresponding image in the HR domain can be reconstructed by weighted averaging the neighbors in the LR domain using the same weights. Generally, the NE-based RSISR method consists of the following three steps: manifold learning, LR neighborhood reconstruction, and super-resolution generation, as shown in Fig. \ref{fig5}. Specifically, manifold learning calculates the reconstruction error by neighborhood weighting for HR and LR manifolds from a specific training set and performs a minimization optimization. Given a patch $p_i$ from the neighborhood $N_i$, the reconstruction error $\varepsilon$ of HR or LR manifolds can be expressed as:
\begin{equation}\label{eq2}
{\varepsilon}_i  = ||{p_i} - \sum\limits_{{p_i} \in {N_i}} {{{\bf{w}}_{i,j}} \cdot {p_j}} |{|_2}
\end{equation}

where ${p_j}$ is an element from the neighborhood $N_i$ and ${{\bf{w}}_{i,j}}$ is the reconstruction weight matrix. For Eq. (\ref{eq2}), the following constraints hold: $\sum {{w_{i,j}}}  = 1$ and $w_{i,j}=0$ for any ${p_j} \notin {N_i}$ apply to $w_{i,j}$. Furthermore, LR neighborhood reconstruction step only computes the neighborhood reconstruction weights for the LR image, representing the LR patch as a neighborhood weighted sum based on Eq. (\ref{eq2}). Finally, super-resolution generation is performed using the LR reconstruction weights to generate high-resolution patches in the HR manifold, ensuring that the local geometric structure is preserved.

To the best of our knowledge, Chang et al. \cite{chang2004super} first proposed the NE-based single image super-resolution method, which employed local linear embedding (LLE) \cite{roweis2000nonlinear} and achieved a good super-resolution performance. Specifically, they perform local compatibility and smoothness constraints between patches in the target HR image via overlap, demonstrating the flexibility and effectiveness of the proposed method. In some studies, feature extraction strategies are utilized to improve performance. For example,  Chan et al. \cite{chan2009neighbor} proposed a NE-based super-resolution method by combining edge detection and feature selection, named NeedFS, which performs a specific selection of training patches based on joint features and edge detection. In addition, Xu et al. \cite{XU2021108033} proposed a two-direction self-learning SR method based on a random oscillation+horizontal propagation+vertical propagation strategy. Bevilacqua et al. \cite{bevilacqua2012low} proposed a low-complexity SR method based on nonnegative NE, which utilized a compact and efficient feature representation strategy to reduce the algorithmic complexity. A work by Timofte et al. \cite{timofte2013anchored} combines sparse dictionary learning and anchored neighborhood regression for fast super-resolution tasks. In particular, they used global collaborative coding to reduce the super-resolution map to a pre-computed projection matrix, which is extremely effective in improving the algorithm speed. Yang et al. \cite{6805627} proposed a dual-geometric neighbor embedding (DGNE) method for the SR task, which employed the multiview features and local spatial neighbors to find a manifold embedding for the image.

NE-based RSISR methods take into account local neighborhood information and can effectively restore high-frequency information of the image, especially in details such as textures and edges. In addition, it retains the structural information of the image and can better avoid over-smoothing. Compared with deep learning methods, the NE-based methods have lower computational complexity and are suitable for resource-constrained environments, especially for large-scale remote sensing data processing. Although the NE-based methods are effective in RSISR, it has obvious disadvantages. For example, NE-based methods focus on the local neighborhoods, and their performance may degrade when processing global information in a larger range. In addition, the choice of neighborhood size has a great impact on the super-resolution results. Specifically, an inappropriate neighborhood size may lead to information loss or overfitting.

\subsubsection{Regression-based Methods}
\begin{figure}[t]
  \centering
  \includegraphics[width=0.475\textwidth]{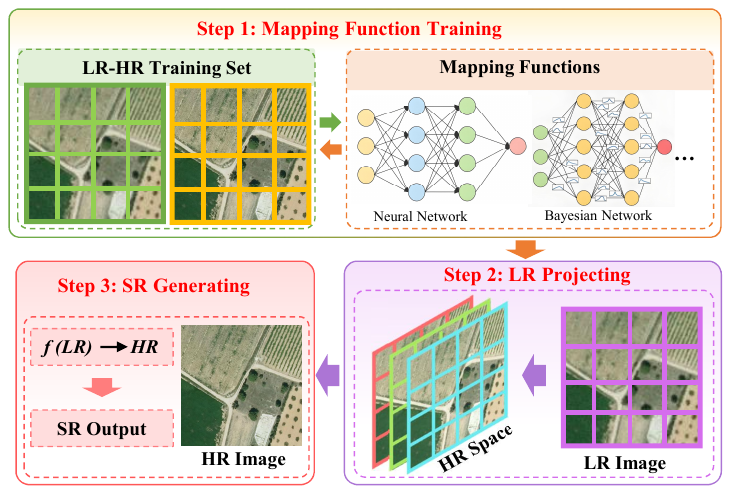}
  \caption{The schematic diagram of the regression-based RSISR method.}\label{fig6}
\end{figure}

The regression-based RSISR methods learn the function mappings between LR and HR images, which regard super-resolution as a regression problem. More specifically, these methods usually extract features from LR images based on paired datasets to train regression models, and then use the trained models to predict and reconstruct HR images. Typically, regression-based methods consist of three main steps: regression model training, projecting of LR images to HR image space, and final super-resolution result generation, as shown in Fig. \ref{fig6}.
In the past two decades, various regression-based super-resolution methods have been proposed, including kernel regression-based methods \cite{zhang2011scale,zhang2014example,zhou2014single,kanakaraj2020adaptive}, Bayesian networks \cite{babacan2010variational,akhtar2015bayesian}, and deep neural network methods \cite{liebel2016single,dong2014learning}. Considering the taxonomy rules of RSISR methods in this paper, the survey of deep learning-based RSISR methods is missing in this section, which can be found in the corresponding sections later. Specifically, the detailed investigations on these types of RSISR methods are as follows:

\emph{\textbf{Kernel Regression.}} In 2010, Kim and Kwon \cite{kim2010single} representatively proposed a super-resolution method based on kernel ridge regression (KRR), which uses the ${\ell ^2}$-norm to train and optimize the mapping regression function. KRR is one of the simplest mapping functions for SR tasks. Furthermore, more complex super-resolution schemes have been proposed \cite{he2011single}. In addition, Zhou et al. \cite{zhou2014single} proposed a novel single-image super-resolution method that uses kernel hill regression to connect LR image patches based on HR coding coefficients. A non-local steering kernel regression (NLSKR) method was used by Zhang et al. \cite{zhang2016single} to perform single image super-resolution tasks, which designed an effective regularization term based on the complementary properties of local structural regularity and non-local self-similarity of images, aiming to preserve sharp edges and produce fine details in the generated images. Similar works can also be found in their previous studies \cite{zhang2012single}. In the RSISR task, a recent work proposed a new synthetic aperture radars (SARs) image SR method by Kanakaraj et al. \cite{9242259} which extended kernel regression to the adaptive importance sampling unscented Kalman filter (AISUKF) framework to deal with speckle noise. Besides, a robust locally weighted regression method proposed by Ma et al. \cite{ma2014robust} also uses the kernel regression to perform RSISR task.

\emph{\textbf{Support Vector Regression (SVR).}} As a kernel regression method, the SVR-based RSISR method has also been widely studied. More concretely, SVR methods involve using a high-dimensional feature space to find nonlinear mappings, which is widely used for RSISR tasks and achieves good performance. For example, Zhang et al. \cite{zhang2011scale} proposed an SVR-based RSISR method that learns prior knowledge between HR and LR images. In particular, bilinear interpolation is employed in their method to upsample to the same resolution and preserve high-frequency information. Besides, Zhang et al. \cite{zhang2014example} proposed an SVR-based super-resolution mapping method (SVR\_SRM) for the land cover map. Specifically, SVR\_SRM learns the nonlinear relationship between coarse-resolution pixels and corresponding labeled sub-pixels to generate fine-resolution land cover maps from the selected best-matching training data. However, SVR\_SRM method is sensitive to fraction errors, resulting in many linear artifacts and speckles in the super-resolution results. To this end, Zhang et al. \cite{zhang2015improvement} again proposed an improved SVR\_SRM method, which mainly used the back-projection operation and the local smoothness prior model to improve the performance of RSISR.

\emph{\textbf{General Regression Neural Network (GRNN)}.} General Regression Neural Networks (GRNN) is a memory-based neural network model \cite{specht1991general}, which has been used in some studies to improve the performance of RSISR. For example, Li et al. \cite{rs8080625} proposed a new GRNN-based method for super-resolution mapping of urban ﬂooding (SMUF) in RSIs. In the GRNN-SMUF method \cite{rs8080625}, a local SMUF model is constructed based on GRNN to describe the relationship between the sub-pixel distribution within a mixed pixel and the scores of the eight neighboring pixels of the mixed pixel. In 2019, Li et al. \cite{li2019enhanced} once again proposed a new SMUF method based on a support vector machine and a general regression neural network, and named it FSVMGRNN-SMUF. Different from GRNN-SMUF, the FSVMGRNN-SMUF method uses a combination of several SVMs to build a local SMUF model and achieves better performance.

\emph{\textbf{Other Regression Methods.}} In addition to the above-mentioned regression-based methods, some other regression-based RSISR approaches have also been proposed. For example, Yang et al. \cite{yang2013fast} trained several simple mapping functions to perform super-resolution tasks. In addition, Dai et al. \cite{dai2015jointly} proposed an adaptive optimal regressor selection method for the super-resolution task, which selects the most appropriate regressor for each input patch, jointly producing the minimum super-resolution error for all training data. Yang et al. \cite{yang2016fast} proposed a soft-assignment based multi-regression method for RSISR, which reconstructs a high-resolution (HR) patch through a dictionary corresponding to its K-nearest training patch clusters. Gao et al. \cite{gao2018self} proposed a self-dictionary regression-based method for hyperspectral image super-resolution.

In summary, traditional supervised SR research work mainly focused on the period before 2016, which was attributed to the fact that deep learning methods had not been fully explored before that time. Generally speaking, the calculation of conventional supervised SR methods is relatively simple and suitable for application scenarios with limited hardware resources. Nevertheless, they rely heavily on the relationship between local pixels, which cannot fully extract deep features and high-level semantic information of images, resulting in suboptimal performance. In addition, in high super-resolution (such as $\times$4 or $\times$8) reconstruction tasks, these conventional methods often fail to recover sufficient details, resulting in blurry or distorted results. For example, regression-based methods are easily affected by noise, resulting in a decrease in reconstruction quality.
\subsection{Deep learning-based RSISR Methods}
As mentioned above, conventional supervised RSISR methods have many shortcomings, such as the inability to extract deep features and high-level semantic information of images. In addition, these methods often rely on manually designed feature extractors, which cannot automatically learn complex patterns and structures in images. With the rapid development of artificial intelligence (AI) technology, deep learning has attracted widespread attention, and a large number of deep learning-based RSISR methods have been studied, which can perform end-to-end training on a large amount of data and automatically learn the details and high-level semantics of images to overcome the shortcomings of traditional methods.

In this section, supervised deep learning-based RSISR methods are surveyed and discussed in detail from the perspectives of CNN-based, GAN-based, Transformer-based, Mamba-based, and hybrid methods. More specifically, in order to provide a comprehensive and in-depth discussion of the methods in this category, we conducted a detailed discussion of the literature up to 2025.

\subsubsection{CNN-based RSISR}
The prototype of a convolutional neural network (CNN) originated from LeNet-5 proposed by LeCun et al. \cite{726791}, which was the first CNN model to be successfully applied to visual recognition. Since 2012, benefiting from the rapid development of computing resources and hardware technologies such as GPU, CNN has been widely studied and many CNN models have been proposed, such as AlexNet \cite{2012ImageNet}, VGG \cite{Simonyan15}, ResNet \cite{He_2016_CVPR}, and OverLoCK \cite{lou2025overlock}, etc. These CNN architectures are widely used in computer vision tasks such as image segmentation \cite{9356353}, object detection \cite{7485869}, and image enhancement \cite{Qi2022ImageEnhancement}, etc.

\begin{figure}[t]
  \centering
  \includegraphics[width=0.48\textwidth]{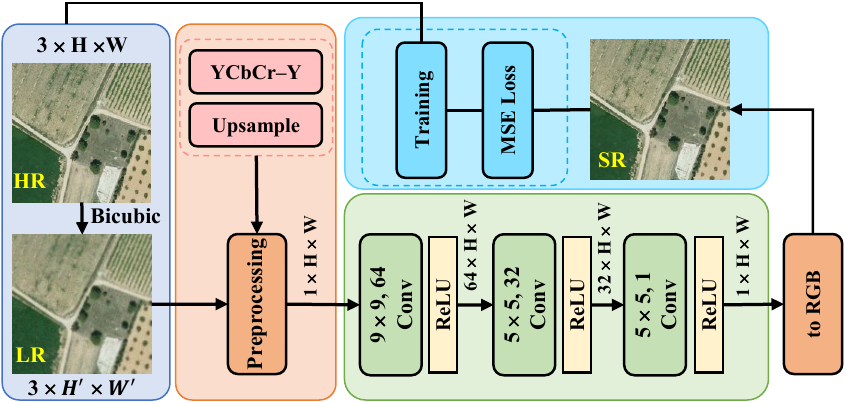}
  \caption{The schematic diagram of the SRCNN-based RSISR method \cite{Liebel2016Single-image}.}\label{fig7}
\end{figure}

In recent years, CNN-based super-resolution methods have been widely studied. Dong et al. \cite{7115171,10.1007/978} first applied CNN to the super-resolution task and achieved good performance. In addition, benefiting from the great success of deep learning, many CNN-based RSISR methods have also been proposed. To the best of our knowledge, Leibel and Korner \cite{Liebel2016Single-image} proposed the ﬁrst CNN-based RSISR method, namely msiSRCNN, which utilizes remote sensing data (Sentinel-2 images) to train and fine-tune SRCNN \cite{7115171,10.1007/978}, as shown in Fig. \ref{fig7}. Specifically, they fine-tuned SRCNN based on the luminance component of the YCbCr color space of RSIs and performed a performance evaluation.

\textit{\textbf{Multi-Scale Features.}} Many studies utilized multi-scale features to improve the performance of RSISR. For example, Fu et al. \cite{8518584} proposed a multi-scale CNN framework (RSCNN) for the RSISR task. Specifically, RSCNN employed convolution kernels of different sizes to extract multi-scale features of images and improved the reconstruction performance of HR images. The idea of extracting multi-scale features based on convolution kernels of different sizes is also involved in these works \cite{keshk2021obtaining,ran2020remote}. A multi-scale residual neural network (MRNN) proposed by Lu et al. \cite{lu2019satellite} also uses complementary features in the patches of different sizes from LR satellite images. Dong et al. \cite{9194276} proposed a second-order multi-scale super-resolution network (SMSR) for RSISR. In addition, Wang et al. \cite{wang2020remote} proposed an adaptive multi-scale fusion network (AMFFN) for RSISI. Specifically, several adaptive multi-scale feature extraction (AMFE) modules and the squeeze-and-excition adaptive gating modules are utilized for feature extraction and fusion. In the work of Huan et al. \cite{Huan2022Remote}, a symmetric multi-scale super-resolution network (AMSSRN) was designed to fully extract and utilize image features. Specifically, they proposed a residual multi-scale block (RMSB) and a residual multi-scale erosion block (RMSDB) to extract shallow and deep features of the image. Multi-scale feature representation is also involved in the RSISR work of Jiang et al. \cite{jiang2018deep},  who proposed a deep distillation recurrent network (DDRN) that performs multi-path feature sharing through ultra-dense residual blocks (UDBs) and a multi-scale purification unit (MSPU). Besides, DeepNESRM method \cite{yin2023super} uses a multi-level feature fusion CNN to eliminate score errors while combining multi-scale information to optimize the mixed pixel problem. On the other hand, residual connections play an important role in solving the gradient vanishing problem of deep networks and are widely used in RSISI tasks. For example, Huan et al. \cite{huan2021end} proposed a new pyramidal multi-scale residual network (PMSRN) for RSISR, which adopts hierarchical residual-like connections to improve the performance. Deeba et. al \cite{deeba2021multi} proposed a transferred wide residual deep neural network model for RSISR, which reduces the memory cost of the network by increasing the width of the residual network and reducing its depth. In addition, Ye et al. \cite{10480122} proposed a lightweight RSISR framework, which involves a novel multi-scale residual attention information distillation group to extract richer regional features. More specifically, they adopted a high- and low-frequency separation reconstruction strategy to enhance the high-frequency details of the image. Considering that most CNN methods ignore low-weight background features, Wu et al. \cite{10720033} proposed a background-based multi-scale feature enhancement network (BMFENet) for RSISR, which constructs a large kernel feature supplement block (LFSB) to generate background feature weights for enhancing the attention of ignored information. Similar work includes the multiscale residual dense network (MRDN) proposed by Kong et al. \cite{10453218} and the dual-path feature reuse multi-scale network (DFMNet) proposed by Xiao et al. \cite{xiao2025dual}.       

\textit{\textbf{Attention mechanism.}} With the remarkable achievements of deep learning in computer vision tasks, the attention mechanism has been widely studied, which dynamically weights input features to enhance the network's ability to focus on important information while suppressing irrelevant or redundant information. In recent years, many researchers have applied the attention mechanism to the RSISR task and achieved remarkable success. Gu et al. \cite{gu2019deep} proposed a Deep Squeeze and Excitation Network (DRSEN) for the RSISR task, which involves a Residual Squeeze and Excitation Block (RSEB) to fuse the input of the current block and its internal features, and model the interdependencies and relationships between channels. Besides, Dong et al. \cite{dong2019transferred} proposed a multi-perception attention network (MPSR) for RSISR, which combines an enhanced residual block (ERB) and a residual channel attention group (RCAG) to more effectively perform weighted fusion of multi-level information. A channel-attention-based fused recurrent network (CA-FRN) proposed by Li et al. \cite{li2020fused} performs sufficient attention on high-frequency information.

Similar RSISR methods include Haut et al. \cite{haut2019remote}, which focuses on valuable high-frequency information and ignores uninformative low-frequency features based on the attention mechanism. Considering the necessity of describing RSIs from multiple scenes, Zhang et al. \cite{zhang2020scene} proposed a multi-scale attention network (MSAN) for RSISR, which employs a scene-adaptive super-resolution strategy to more accurately describe the structural characteristics of different scenes. In addition, Chen et al. \cite{chen2023remote} proposed a novel residual split attention network (RSAN) for RSISR, which can simultaneously maintain the overall structure and local details of the image, as shown in Fig. \ref{fig4} (a). Recently, research on digital elevation model super-resolution (DEM-SR) has also been carried out, which is of great significance for accurately providing remote sensing geographic information. For example, Chen et al. \cite{10585320} proposed an integrating attention mechanism (IAM) network method for DEM-SR, which can well capture long-range geographic feature dependencies and local context information, as shown in Fig. \ref{fig4} (b). In order to solve the problem of the limited global receptive field in RSISR, some studies have been performed. For instance, Wang et al. \cite{wang2020non} proposed a non-local up-down convolutional attention network (NLASR), which mainly designed a feature enhancement module (NLEB) to model spatial long-range dependencies. Wang et al. \cite{wang2023hybrid} proposed a hybrid attention U-type network (HAUNet) for the RSISR task to fuse multi-scale global semantic information. In addition, some research works are devoted to solving the problem of high- and low-frequency information aliasing in CNN-based RSISR methods. For example, Peng et al. \cite{peng2021pre} designed a gated convolution unit to separate the frequency domain information flow, and further proposed a pre-training of a gated convolution neural network (PGCNN) for RSISR. Besides, Zhao et al. \cite{zhao2024structure} used a structure-texture dual preservation strategy to constrain edge generation. Zhang et al. \cite{zhang2022single} further proposed a residual balanced attention network (RBAN) and combined adversarial training to improve the authenticity of high-frequency textures. Aiming at the large-scale super-resolution reconstruction of RSIs, Gao et al. \cite{dong2020remote} proposed a dense sampling super-resolution network (DSSR) RSISR method, which introduced a dense sampling mechanism, a wide feature attention module, and a chain training strategy to improve the network performance. For multi-level feature fusion, Wang et al. \cite{10543045} proposed a multi-scale spatial-spectral network (M3SN) method, which couples 3D convolution with spectral attention, aiming to deeply mine shallow and deep spatial-spectral features in hyperspectral image super-resolution tasks, thereby improving reconstruction performance. With similar motivations, Chen et al. \cite{chen2021remote} proposed a residual aggregation-segmentation attention fusion network (RASAF) to achieve cross-channel interaction for RSISR. 

On the other hand, some attention-based strategies have been studied to improve the local detail recovery capability of RSISR methods. For example, Ma et al. \cite{ma2021remote} proposed a dense channel attention mechanism to enhance high-frequency feature reuse. Patnaik et al. \cite{10453952} combined multi-scale residual attention to capture fine-grained texture for RSISR, and Zhang et al. \cite{zhang2020remote} introduced hybrid high-order attention (HOA) to mine deep statistical associations. In addition, Huang et al. \cite{huang2021deep} proposed a deep residual dual attention network (DRDAN) method, which introduced a residual dual attention module (RDAB) and a local multi-level fusion module to fuse global and local information. Wang et al. \cite{wang2022fenet} proposed a lightweight feature enhancement network (FeNet) for RSISR, which constructed a lightweight lattice block (LLB) and used channel separation and weight sharing to reduce the computation amount. In the field of lightweight research, Wang et al. \cite{wang2021lightweight}proposed a feedback ghost residual dense network (FGRDN), which introduces a spatial-channel attention module (SCM) to extract key features and reduces redundant parameters through a feedback mechanism and a Ghost module. The enhanced residual convolutional neural network (ERCNN) proposed by Ren et al. \cite{ren2021remote} also uses a feature attention module, which enhances the discriminative learning ability across feature maps and combines a dual-brightness scheme to optimize high-frequency recovery.

\textit{\textbf{Transform domain.}} In recent years, some studies have combined CNN with a transform domain to perform RSISR, which utilizes the frequency band decomposition capability of domain transform (such as wavelet transform and Fourier transform) and the feature modeling advantages of deep learning to improve the reconstruction performance. For example, Wang et al. \cite{wang2018aerial} proposed an RSISR method combining wavelet multi-scale decomposition with a multi-branch CNN. Specifically, they trained the network independently and predicted sub-images in different directions and frequency bands, respectively, and then synthesized high-resolution aerial images through inverse wavelet transform, which verified the applicability of frequency domain modeling for complex scenes. Two studies in 2019 further deepened the frequency domain framework in RSISR. Ma et al. \cite{ma2019achieving} proposed a recursive residual network (Recursive Res-Net) to predict high-frequency components, innovatively replaced low-frequency subbands with low-resolution images to preserve details, and optimized network efficiency by removing batch normalization layers. In addition, Yang et al. \cite{yang2019hyperspectral} proposed a multi-scale wavelet 3D convolutional neural network (MW-3D-CNN) for hyperspectral image super-resolution, which uses three-dimensional convolution to extract spatial spectral features and uses L1 loss to optimize multi-subband wavelet coefficient prediction, realizing joint spatial-spectral reconstruction of hyperspectral data. Furthermore, the work of Aburaed et al. \cite{aburaed2020super} also verified the effectiveness of wavelet analysis for reconstructing high-frequency details. In 2021, Deeba et al. \cite{deeba2021plexus} proposed a fast wavelet super-resolution framework (FWSR), which accelerates image restoration through approximate sub-band input and sub-pixel layer, and enhances the ability to retain high-frequency features. Sandana et al. \cite{thool2023combined} proposed a CNN combining wavelet and multi-wavelet transform (CWM-CNN), which integrates discrete wavelet and multi-wavelet transform to extract shallow features, and combines a compact three-layer CNN to achieve parameter-efficient high-resolution reconstruction. Recently, Wang et al. \cite{wang2024two} proposed a two-stage joint space-frequency network (TSFNet) to address the problem of large-scale super-resolution. Specifically, they designed an amplitude-guided phase adaptive filtering module (AGPF) to decouple frequency domain degradation features and achieved progressive detail optimization through cross-stage feature fusion, achieving good performance in maintaining structural integrity.

\textbf{\textit{Spatial-Spectral Joint Modeling.}} In recent years, some studies have focused on joint spatial-spectral modeling to achieve higher quality reconstruction by fusing the spatial context and spectral correlation of images. For example, early work \cite{lanaras2018super} employed CNN to directly learn the mapping from low-resolution bands to high-resolution bands end-to-end based on the inherent resolution differences of Sentinel-2 multi-bands (10m/20m/60m), while implicitly utilizing the spatial features of high-resolution bands to guide the reconstruction of low-resolution bands. Yin et al. \cite{yin2021cascaded} proposed a DeepRivSRM method for RSISR. Specifically, they extracted the river spectral features in mixed pixels by spectral unmixing and then combined the spatial context for super-resolution mapping. To further enhance physical consistency, some studies have injected explicit physical model constraints. For example, Arun et al. \cite{arun2020cnn} proposed a collaborative unmixing framework based on sparse coding for RSISR, which jointly optimized the sparse dictionary and spectral reconstruction, and fused the spatial-spectral prior through the encoding-decoding architecture to ensure spectral fidelity. In addition, Wagner et al. \cite{wagner2019deep} employed a deep residual network (ResNet) to fuse shallow spectral features with deep spatial features via jump connections to achieve end-to-end spectral-spatial joint optimization. Vasilescu et al. \cite{vasilescu2023cnn} innovatively embedded the modulation transfer function (MTF) of the sensor into the multi-objective loss function to simulate the sensor degradation process and force the network to learn the physical consistency of the inverse mapping. In addition, they combined high-frequency detail alignment loss (such as 10m band gradient matching) to improve the accuracy of spatial detail reconstruction. In the recent work of Chen et al. \cite{10499252}, a spatial-spectral attention module (DSSA) combined with a dilated convolution method was proposed to extract global spatial features, and the 3D Inception module was used to mine multi-scale spectral information to solve the problem of spectral-spatial information imbalance in hyperspectral images. In addition, Mei et al. \cite{mei2020spatial} proposed two end-to-end joint optimization frameworks for RSISR, namely simultaneous spatial–spectral joint SR (SimSSJSR) and separated spatial–spectral joint SR (SepSSJSR). Specifically, spectral unmixing, regularized deconvolution, and endmember similarity constraints are employed to ensure spectral consistency. Müller et al. \cite{muller2020super} performed panchromatic band (PAN) and multispectral band fusion, which used CNN training to improve the spatial resolution of multispectral satellite images.

\textbf{\textit{Local-Global Feature Fusion.}} This type of method usually captures the local detail features and global prior information of RSIs, and uses a multi-scale feature fusion mechanism to improve the quality of super-resolution reconstruction. For example, the local–global combined networks (LGCNet) proposed by Lei et al. \cite{lei2017super} adopts a multi-branch structure to achieve multi-level representation learning. Besides, Xu et al. \cite{xu2018high} proposed a deep memory connected network (DMCN) for RSISR, which achieves feature integration by building local-global memory connections. These studies all emphasize the integration of scene-level semantic information while maintaining local texture details. They generally introduce spatial downsampling units to optimize computational efficiency and balance model complexity and performance by compressing the size of feature maps.

\textbf{\textit{Others CNN-based Methods.}} In addition to the above-mentioned CNN-based methods, some other CNN-based RSISR approaches have also been proposed. For example, Chang and Luo \cite{chang2019bidirectional} proposed a bidirectional ConvLSTM network, which reuses local low-level features through a recursive inference block and combines a bidirectional ConvLSTM layer to adaptively select complementary information. Zhu et al. \cite{zhu2020super} proposed a degradation model based on satellite imaging and ground post-processing to improve the SR performance of real satellite images. Deeba et al. \cite{deeba2020single} proposed the Wide Remote Sensing Residual Network (WRSR), which optimizes the efficiency of single-image super-resolution by increasing the width of the residual network, reducing the depth, and using weight normalization to reduce memory costs. Wei and Liu \cite{wei2021construction} introduced dilated convolution into residual dense blocks while keeping parameters and receptive fields unchanged, and combined them with cloud services to optimize RSI reconstruction. In addition, Zhao et al. \cite{10613839} utilized super-Laplacian gradient prior to constrain SR reconstruction and enhanced geometric structure recovery through second-order supervision. Some recent studies have used multimodal fusion to improve performance. For example, the DeepOSWSRM method proposed by Yin et al. \cite{yin2024super} fuses Sentinel-1 SAR and Sentinel-2 optical images to solve the cloud occlusion problem and further generate high-resolution water distribution maps. In addition, Dai et al. \cite{10843849} utilized the polarization information and spatial features of polarimetric SAR to extract and fuse polarization context features through a dual-branch architecture to improve the super-resolution performance of PolSAR images.
\subsubsection{GAN-based RSISR}
\begin{figure}[t]
  \centering
  \includegraphics[width=0.48\textwidth]{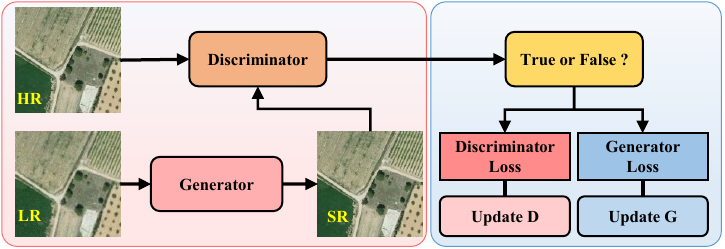}
  \caption{The schematic diagram of the GAN-based RSISR method.}\label{fig8}
\end{figure}
In 2017, Ledig et al. \cite{ledig2017photo} first applied generative adversarial networks (GANs) to image super-resolution. Since then, many GAN-based RSISR methods have been proposed, and GAN-based methods have become one of the most important types of RSISR methods. In general, the GAN-based RSISR methods consist of generators and discriminators. The generator reconstructs the LR image into an HR image, and the discriminator determines whether the image is a true HR (GT HR), as shown in Fig. \ref{fig8}. Specifically, a two-stage training strategy is adopted in this type of method. First, the generator is frozen to train the discriminator, thereby enhancing its discrimination ability. Then the discriminator is frozen to train the generator, and the generation result is continuously optimized through adversarial feedback. At the same time, reconstruction loss and perceptual loss are used to improve image structure restoration and visual quality.

\textbf{\textit{Network architecture optimization.}}
Some studies improve the generator and discriminator structures to enhance feature extraction and reconstruction capabilities. For example, Ma et al. \cite{ma2018super} reduced the computational burden by removing the batch normalization layer and used transfer learning to alleviate the problem of insufficient data. Subsequently, in their follow-up research, the DRGAN method \cite{ma2019super} designed a dense residual generator and combined it with Wasserstein GAN to improve training stability. Guo et al. \cite{guo2021remote} proposed a cascaded GAN method for RSISR, which introduced scene constraints and edge enhancement modules to improve performance. In addition, Wang et al. \cite{wang2020ultra} innovatively constructed a two-dimensional matrix topology structure to significantly increase the connection density of the network for RSISR. The EEGAN method \cite{jiang2019edge} proposed a dual-branch network structure to suppress noise through an edge enhancement subnet. Furthermore, the TDRRDB-EEGAN method \cite{chen2022super} improved the edge purification module of EEGAN and combined the intermediate image with the enhanced edge to generate clear results. Pang et al. \cite{pang2023use} replaced the generator with ResNet-50 and inserted a fully connected layer in the discriminator, significantly improving the quality of super-resolution images while maintaining the reconstruction speed. Liu et al. \cite{liu2020super} proposed a cascaded conditional Wasserstein GAN combined with residual dense blocks to optimize gradient propagation. Guo et al.\cite{guo2022ndsrgan} designed a multi-level dense network and matrix mean discriminator to optimize the reconstruction of real aerial images. Similarly, Sustika et al. \cite{sustika2020generative} combined the residual dense network (RDN) with GAN to balance objective indicators and perceptual quality. In 2023, the MSRBGAN method proposed by Zhao et al. \cite{zhao2023forest} introduced multi-scale residual blocks and GAM attention to improve the performance of RSISR. The DRGAN method in recent work \cite{10781453} proposed a dynamic convolution and a self-attention mechanism to further optimize feature extraction.

\textit{\textbf{Attention Mechanisms.}}
Attention mechanisms and multi-scale modeling are used in some studies. A typical example is that SCSE-GAN \cite{moustafa2021satellite} first introduced spatial-channel attention into the generator to enhance feature representation. In addition, Li et al. \cite{li2021single} improved performance by capturing long-range dependencies based on local and global attention. Zhang et al.\cite{zhang2023super} proposed a coordinated attention mechanism and residual stacking strategy to optimize landslide scene reconstruction. Jia et al. \cite{jia2022multiattention} integrate pyramid convolution residuals, pixel attention, and branch attention to improve the performance of RSISR. Furthermore, Wang et al. \cite{wang2023msagan} optimizes RSISR algorithm by combining multi-scale structure with channel-spatial attention. The CEEGAN proposed by Ren et al. \cite{ren2023context} designs an edge feature enhancement module (EFEM) to fuse multi-scale contextual information and optimize edge recovery for RSISR. In 2024, MFFAGAN proposed by Tang et al. \cite{10461030} developed a multi-level feature fusion module to enhance key information extraction through hybrid attention. The TDEGAN method \cite{guo2024tdegan} introduces Shuffle attention and artifact loss functions to effectively restore high-frequency textures. Similarly, SRDSRAN \cite{10375518} and SRGAN-MSAM-DRC \cite{hu2024super} use layered dense sampling and multi-scale attention mechanisms to significantly improve the performance of large-scale RSISR. 

\textbf{\textit{Loss Functions and Training Strategies.
}}
Some studies have focused on optimizing loss functions and training strategies to further improve the performance of RSISR. In 2019, Lei et al. \cite{lei2019coupled} proposed a coupled discriminator to alleviate the low-frequency region discrimination ambiguity through a dual-path network and a random gate mechanism. In addition, Fan et al. \cite{fan2024rmsrgan} used a U-Net discriminator combined with a CBAM module to enhance detail discrimination for RSISR. Similarly, a RGTGAN method proposed by Tu et al. \cite{tu2024rgtgan} designed a gradient-assisted texture enhancement module (GTEM) and a dense-intern deformable convolution (DIDConv) to improve cross-view alignment for RSISR. On the other hand, the RWDM method \cite{zhang2022single} constructs a real degradation training dataset and combines it with the residual balanced attention network (RBAN) to improve the reconstruction effect of the actual scene. In addition, the BLG-GAN method \cite{chung2023enhancing} uses a two-stage training strategy to migrate real LR images to the bicubic degradation domain to alleviate the simulation data bias. MCWESRGAN method \cite{karwowska2023mcwesrgan} uses Wasserstein loss and a multi-column discriminator to accelerate training for RSISR. Zhu et al. \cite{zhu2023improved} introduced the relative average GAN (RaGAN) reconstruction loss to optimize edge detail learning, and Guo et al. \cite{guo2023improved} improved texture fidelity based on Charbonnier loss and semantic perception loss.

\textbf{\textit{Cross-modal and Multi-task fusion.
}}
Recently, cross-modal methods have been studied. For example, Kong et al. \cite{kong2023super} combined Planet Fusion with Landsat data to generate high spatiotemporal resolution NDVI series. In addition, Sun et al. \cite{10528671} fused high-resolution optical images with digital surface models (DSM) to preserve terrain features through slope loss. The VHR-GAN method \cite{pineda2020generative} uses PeruSat1 satellite data to achieve super-resolution of Sentinel-2 images. In addition, Huang Hejing \cite{huang2020super} combines the wavelet transform to optimize the frequency domain feature expression. The MRSISR method \cite{wang2023multik} proposes a multi-frame super-resolution framework that uses temporal information to optimize video satellite image reconstruction. Some other examples are the Enlighten-GAN method \cite{gong2021enlighten} and recent work \cite{wang2024super}, which both integrate spatial image modality and frequency domain modality information, and optimize the RSISR algorithm by combining perceptual loss with pixel reconstruction loss.

\subsubsection{Transformer-based RSISR}
In recent years, Transformer has been widely applied in the field of natural language processing due to its unique self-attention mechanism and global modeling capabilities. Different from CNN, Transformer efficiently captures long-range dependencies and achieves global information interaction by dynamically calculating the correlation between all elements in the input sequence \cite{vaswani2017attention}. Furthermore, Vision Transformer (ViT) has proven that it can achieve competitive or even better performance than CNN in computer vision tasks such as image classification and object detection, especially in the ability to model complex textures and global structures \cite{dosovitskiy2020image}.

In the field of RSISR, the conventional CNN-based methods still face challenges due to the inherent properties of RSIs, such as the coexistence of multi-scale targets, complex spectral-spatial coupling characteristics, and low processing efficiency of large-size images. In addition, the conventional CNN-based approaches often struggle to capture global contextual information due to their limited local receptive fields, frequently resulting in edge blurring or undesirable artifacts. In contrast, Transformer architecture's global attention mechanism offers a promising solution to address these limitations. To the best of our knowledge, Ye et al. \cite{ye2021super} proposed the first Transformer-based RSISR method, which combines CNN and Transformer modules to achieve efficient super-resolution reconstruction of LR RSIs. Since then, various Transformer-based RSISR methods have been proposed, including transform-domain-based, multi-scale features, multi-modal fusion, special modality input, improved self-attention mechanisms, and the detailed survey in this section is as follows: 

\textit{\textbf{Multi-Scale Features.}} Many studies considered the complexity of RSI scenes and utilized multi-scale feature fusion to improve performance. Specifically, multi-scale feature fusion methods effectively balance the global structure coherence and local detail authenticity through dynamic weight allocation. Early studies such as TransENet proposed by Lei et al. \cite{lei2021transformer} built a multi-stage enhanced structure based on Transformer, which embeds and fuses multi-scale high/low-dimensional features through an encoder-decoder architecture, breaking through the limitation of traditional models that rely on upsampling layers and ignore high-dimensional spatial features. In addition, the MSFNet proposed by Shi et al. \cite{shi2023multisource} achieved hierarchical feature propagation and adaptive cross-scale fusion across iteration stages based on multi-scale implicit constraints and row-column decoupling Transformer modules. Shang et al. \cite{shang2023hybrid} proposed a hybrid-scale hierarchical Transformer network (HSTNet) for RSISR, which explicitly modeled single-scale and cross-scale long-range dependencies via hybrid-scale recursive feature mining and enhanced the discriminative ability of high-dimensional features. In 2024, some studies have made breakthroughs in balancing computational efficiency and multi-scale perception. For example, SymSwin proposed by Jiao et al. \cite{jiao2024symswin} involved a symmetric multi-scale window mechanism (SyMW) and cross-receptive field adaptive attention (CRAA), which dynamically fuses contextual features of different sizes and enhances supervision through frequency domain wavelet transform loss (UWT). In addition, Lu et al. \cite{lu2024enhanced} proposed a Multi-Scale and Global Representation Enhancement-based Transformer (MSGFormer), which designed the dual-window self-attention (DWSA) and multi-scale deep convolutional attention (MDCA), and tracking back structure (TBS) to capture local-global multi-scale features at a lower computational cost. For geographic data, Wang et al. \cite{wang2024ttsr} introduced local-global deformable blocks (LGDB) to fuse the multi-scale heterogeneity of terrain, supplemented by spatial-channel coupled attention, which significantly alleviated the local smoothing problem in DEM reconstruction.

\textbf{\textit{Transform Domain.}} The transform domain-based methods mainly separate the high- and low-frequency information of the image via frequency domain decomposition (such as wavelet transform and Fourier transform), and utilize the global attention mechanism of Transformer to strengthen the feature correlation. For example, the CFMB-T method proposed by Cao et al. \cite{cao2023cfmb} in 2023 employs SwinIR as the architecture, which separates the frequency domain features of infrared RSIs through wavelet decomposition, and adds degradation priors and recursive modules to improve performance. The dual SR framework \cite{sun2024transformer} in 2024 further combined self-supervised learning with wavelet fusion, and modeled the long-range spatiotemporal dependency of RSIs through the Transformer attention mechanism, effectively solving the problem of periodic resolution degradation caused by the rotating synthetic aperture system. Recently, Xie et al. \cite{10742930} proposed a RSISR framework based on Swin Transformer, named CFFormer, which enhances feature expression through frequency domain channel Fourier blocks (CFB) and global attention modules (GAB), and optimizes gradients by combining a jump joint fusion mechanism. In addition, Liu and Yang \cite{liu2025wtt} proposed a frequency domain feature hierarchical architecture based on discrete wavelet transform (DWT), which learns the correlation between full-band and high-frequency features through a self-attention mechanism. In summary, Transform-domain-based methods usually outperform pure spatial domain models, especially showing stronger detail reconstruction capabilities in complex degradation scenarios.

\textbf{\textit{Multi-modal Fusion.}} Some studies have used the complementary information fusion method of hyperspectral (HSI) and multispectral (MSI) modalities to improve the performance of RSISR. For example, Hu et al. \cite{hu2022fusformer} first proposed the Fusformer framework for the hyperspectral super-resolution (HISR) task. Specifically, the low-resolution hyperspectral image (LR-HSI) is used as the core to preserve the spectral integrity, and the spatial residual is estimated on the LR-HSI, thereby simplifying the complex mapping into small spatial residual learning. Li et al. \cite{li2023spectral} proposed a spectral learning based Transformer network (SLTN). Specifically, they designed a spectral response function (SRF) guided multi-level feature extraction module (MFEM) and a cascaded nonlinear mapping learning module (NMLM) to address the spectral complexity and degradation problems of remote sensing scenes.  

\textbf{\textit{Special Modality-guided.}} Recent studies have employed special modality input to improve the performance of Transformer in RSISR tasks. For example, Cai and Zhang \cite{cai2022t} proposed a texture transfer transformer-based RSISR (T3SR) method, which first introduced the texture transfer paradigm into the remote sensing field, and solved the single image texture loss problem through a two-stage framework of shallow texture transfer and U-Transformer feature fusion, while greatly reducing the model's dependence on reference images. In addition, the hybrid attention model proposed by Alireza and Mahdi \cite{10829708} combines multi-head attention and spatial-channel attention mechanisms, which is optimized for the multispectral characteristics of Sentinel-2, and achieves SOTA performance. GeoSR framework for drone images proposed by Zhao and Li \cite{10683775} pioneered a task-driven evaluation paradigm, which uses semantic segmentation accuracy as a supplementary modality to traditional PSNR/SSIM, and establishes a positive correlation between super-resolution quality and 8.6\% accuracy of object classification. These studies further illustrated the potential of special modality-guided reconstruction, such as texture/gradient/semantic maps in improving RSISR performance.

\textbf{\textit{Improved Self-attention Mechanisms.}} In order to solve the problems of high computational complexity, insufficient local receptive field, and inefficient cross-stage feature fusion in the traditional global self-attention mechanism, some research based on improving the self-attention mechanism has been carried out to improve performance. Specifically, most of these studies are mainly based on enhancing context modeling capabilities, optimizing feature fusion efficiency, and reducing computational complexity. For example, Lu et al. \cite{lu2023cross} proposed a cross-spatial pixel integration and cross-stage feature fusion-based transformer network (SPIFFNet) for RSISR. Specifically, the spatial pixel integration attention (CSPIA) module is designed to introduce contextual information into the local window, and the cross-stage feature fusion attention (CSFFA) module is designed to achieve adaptive feature transfer. The ESSAformer method proposed by Zhang et al. \cite{zhang2023essaformer} developed a kernelized self-attention (ESSA) based on spectral correlation coefficient, which reduces the computational complexity to linear and strengthens the interaction of spectral features. In addition, Mao et al. \cite{mao2024desat} proposed a distance-enhanced strip
attention transformer (DESAT), which enhances spatial correlation by fusing strip window attention with a distance prior, and designed an attention-enhanced upsampling module. Kang et al. \cite{10767430} recently proposed an Aggregation Connection Transformer (ACT-SR) for RSISR, which constructs an aggregate connection attention block and uses a series-parallel hybrid connection to fuse spatial-channel features, supplemented by a gated feedforward network to enhance nonlinear expression capabilities. In addition, the scale-aware backprojection Transformer (SPT) proposed by Hao et al. \cite{10753509} innovatively combined backprojection learning with Transformer to construct a scale-aware self-attention layer, thereby improving feature learning efficiency. Kang et al. \cite{kang2024efficient} proposed an efficient Swin Transformer (ESTNet) via channel attention, which reduces the number of parameters by 83\% by synergizing group attention with efficient channel attention. In recent work, Xiao et al. \cite{xiao2024ttst,xiao2024remote} proposed a Top-k Token Selective Transformer (TTST) for RSISR, which designed a three-level self-attention optimization framework involving Residual Token Selective Group (RTSG), multi-scale feature fusion (MFL) and Global Context Attention (GCA), further solving the redundant calculation and single-scale modeling limitations of self-attention in specific tasks.

\subsubsection{Mamba-based RSISR}
Recently, empowered by a state space model (SSM) involving near-linear complexity modeling long-distance dependencies, Mamba \cite{gu2023mamba} has demonstrated superior performance compared to Transformer in natural language processing (NLP).
Specifically, Mamba introduces a selective mechanism into the traditional SSM to improve computational efficiency and significantly enhance the ability to model long sequences. Building on success in sequence modeling, Mamba has recently demonstrated remarkable potential across diverse vision domains, achieving state-of-the-art performance in fundamental tasks, including image classification \cite{liu2024vmamba}, object detection \cite{dong2024fusion}, and semantic segmentation \cite{ma2024u}. However, the research on Mamba for low-level vision tasks is still in its infancy. In particular, a small number of studies have begun to use Mamba to perform RSISR tasks.

In recent work, Zhu et al. \cite{zhu2024convmambasr} proposed a ConvMambaSRF method that integrates SSM and CNN for RSISR. Specifically, they used SSM and CNN to model global dependencies and extract local details, respectively, and realized multi-level feature fusion through the global-detail reconstruction module (GDRM). Besides, Zhi et al. \cite{10663411} proposed a MambaFormerSR method combining Mamba and Transformer for RSISR. They designed a state-space-attention fusion module (SAFM) and combined it with a convolutional Fourier feedforward network (CTFFN) to strengthen frequency domain attention. The spatial-frequency dual-path framework proposed by Xiao et al. \cite{10817590} for RSISR uses a frequency selection module (FSM) and a state space module (VSSM) to extract high-frequency details and capture long-range spatial correlations, respectively, and further designs a hybrid gating module (HGM) to dynamically fuse multi-level features, achieving outstanding performance. To improve computational efficiency, Zhang and Wang \cite{zhao2024mamba} combined SSM with residual connections to implement an equivalent attention mechanism with linear complexity, and experimentally demonstrated that it significantly enhances cross-region feature associations while preserving local textures.

Mamba-based RSISR method successfully solves the local receptive field limitations of traditional CNN and the high computational bottleneck of Transformer by leveraging the linear complexity global modeling capability of SSM, demonstrating a new paradigm for large-scale RSI reconstruction. Although some studies, such as hybrid architecture design strategies \cite{zhu2024convmambasr,10663411} and multi-modal feature collaboration \cite{zhao2024mamba}, have significantly improved reconstruction performance, existing methods still face challenges such as blurred high-frequency details at extremely low-resolution and insufficient temporal modeling of multi-phase images. In the future, the research on RSISR based on Mamba may focus on the following aspects: (a) lightweight 3D Mamba architecture design for time-series video RSISR and dynamic scene reconstruction, (b) physical degradation-aware SSM optimization, combined with imaging models to enhance the robustness of unknown degradation scenes, (c) cross-modal state space modeling, integrating multi-spectral, SAR, and other multi-source remote sensing data to build a unified representation framework for heterogeneous features. With further breakthroughs in hardware perception algorithms and adaptive selectivity mechanisms, Mamba has the potential to become a universal model for RSISR tasks.
\subsubsection{Diffusion Model-based RSISR}
The diffusion model-based RSISR method iteratively reconstructs HR images from LR inputs through a conditional reverse diffusion process, leveraging probabilistic refinement to resolve ill-posedness of SR and geospatial ambiguities under sensor-specific degradation. In recent years, numerous diffusion model-based RSISR methods have been proposed. 

To the best of our knowledge, Liu et al. \cite{liu2022diffusion} proposed the first diffusion model-based RSISR method, which uses LR images as conditional information to generate HR images. Wu et al. \cite{wu2023hsr} proposed an HSR-Diff method for RSISR, which uses a conditional Transformer to fuse HR images and LR images, and then optimizes the reconstruction through hierarchical feature iteration. The efficient hybrid conditional diffusion model (EHC-DMSR) for RSISR proposed by Han et al. \cite{han2023enhancing} combines Transformer-CNN to extract conditional features and introduces Fourier high-frequency constraints to accelerate reasoning. In addition, Meng et al. \cite{meng2024conditional} proposed the FastDiffSR method to improve the sampling strategy and reduce the number of diffusion steps. An et al.\cite{an2023efficient} proposed a lightweight diffusion model LWTDM, which uses a cross-attention encoder-decoder to simplify the denoising network and integrates denoising diffusion implicit models (DDIMs) to accelerate sampling. The DiffALS method \cite{sui2024denoising} introduces a noise discriminator (ND) through an adversarial learning strategy to enhance the diversity of detail generation. Furthermore, the TCDM method proposed by Zhang et al. \cite{zhang2024tcdm} designs a conditional truncated noise generator (CTNG) for large-scale super-resolution and embeds pixel-level constraints in combination with the texture consistency diffusion process to accelerate reasoning. Xiao et al. \cite{10353979} proposed an EDiffSR method for RSISR, which improves efficiency by simplifying channel attention. Some methods focus on semantics and global optimization. For example, Wang et al. \cite{wang2025semantic} combined vector map semantic guidance and proposed a Semantic Guided Diffusion Model (SGDM). Zhu et al. \cite{zhu2025taming} introduced low-rank adaptation (LoRA) to alleviate the distribution difference between natural images and RSI. Weng et al. \cite{10947187} proposed an image reconstruction representation-diffusion
model for RSISR, which uses a pre-trained encoder to guide high-frequency reconstruction.

\subsubsection{Hybrid RSISR Methods}
In recent years, some studies have adopted hybrid architectures to improve the performance of RSISR. For example, He et al. \cite{he2022dster} proposed a hybrid model (DsTer) that combines densely connected Transformer and ResNet, which enhances spectral super-resolution capabilities through multi-level Transformer feature fusion and demonstrates its advantages in long-range dependency modeling on natural and remote sensing datasets. In addition, Tu et al. \cite{tu2022swcgan} proposed a GAN method (SWCGAN) combining Swin Transformer and CNN for RSISR, which improves the multi-scale feature extraction capability through residual dense blocks and outperforms the traditional CNN method on the UCMerced dataset. However, the adversarial training of the SWCGAN method is prone to introducing artifacts. Wang et al. \cite{wang2023landsat} proposed a dilated Transformer generative adversarial network (DTGAN), which introduced dilated convolution into the self-attention module of the Transformer and dynamically focused on local and global features of different scales, significantly improving the performance of land classification tasks. 

In terms of lightweight design, Peng et al. \cite{peng2023context} proposed a context-aware lightweight super-resolution network (CALSRN), which aggregated the global features of Swin Transformer and the local features of CNN by dynamic weights, reducing the number of parameters by 30\% while ensuring the quality of reconstruction. However, the CALSRN method has limited modeling capabilities for complex ground structures. In addition, Lin et al. \cite{10549773} proposed a distillation Transform-CNN Network (DTCNet), which employs a Transformer teacher network to guide a lightweight CNN student network and avoids zero-filling information loss through adaptive upsampling. For cross-sensor and frequency domain optimization, Hou et al. \cite{10562345} proposed a CNN-Swin Transformer RSISR method (CSwT-SR) based on amplitude-phase learning, which jointly enhances spatial and frequency domain feature representation and outperforms prevailing degradation model methods in blind super-resolution tasks. Besides, Wang et al. \cite{wang2024lightweight} proposed a hybrid CNN-Transformer network named RepCHAT, which compresses the number of parameters through structural re-parameterization and introduces a frequency domain multi-scale feature extraction module to improve performance. In recent work, an improved SRGAN method named ViT-ISRGAN proposed by Yang et al. \cite{10836746} employs Vision Transformer and spatial-spectral residual attention to improve performance for four typical ground objects (urban, water, farmland, and forest) in Sentinel-2. For infrared image super-resolution, Zhang et al. \cite{huang2024infrared} designed the residual Swin Transformer block (RSTAB) and proposed a SwinAIR-GAN method based on U-Net, which fuses multi-frequency features through average pooling and introduces artifact discrimination loss, greatly improving performance.

Recently, some studies have focused on exploring the generation ability and stability of hybrid architectures. For example, Guo et al. \cite{guo2024activated} proposed a sparse-activated sub-pixel transformer network (SSTNet), which enhances edge sharpness and texture details through sparse activation in sub-pixel space, but a sparse constraint term requires additional design. Besides, Liu et al. \cite{liu2025remote} proposed an improved RSISR method, which designs a multi-stage hybrid Transformer generator and combines Charbonnier loss with TV loss to improve training stability. The Fusing Transformers and CNN method (ConvFormerSR) for RSISR proposed by Li et al. \cite{10345595} designed a cross-sensor feature fusion module (FFM) and spectral loss function, and achieved high-spectral consistency reconstruction of Landsat-8 and Sentinel-2 data by enhancing the high-order spatial interaction of Transformer. MSWAGAN method \cite{10494508} designs a multi-scale sliding window attention (MSWA) to capture local multi-scale features, and combines the Transformer to model long-range pixel associations without increasing the number of parameters. Zhu et al. \cite{10509697} proposed a Transformer-based Multi-modal Generative Adversarial Network (TMGAN) for multimodal data, which designed a global Transformer generator and fused depth and spectral information through self-attention to improve performance. The Enhanced Swin Transformer with U-Net GAN (ESTUGAN) proposed by Yu et al. \cite{yu2023estugan} employs a Swin Transformer-based generator and a U-Net discriminator, which suppresses artifacts through region-aware adversarial learning. Specifically, ESTUGAN uses Best-buddy loss and Back-projection loss to enhance reconstruction fidelity. In the Trans-CNN GAN method proposed by Lin et al. \cite{lin2024trans}, the global attention of the Transformer is used to compensate for the local defects of the generator. Huo et al. \cite{huo2024stgan} proposed a STGAN model based on GANs and self-attention mechanism, which achieved better high-frequency detail recovery ability in RSISR through a multi-CNN-Swin Transformer module (MCST) and an improved association attention module (RAM-V).

\section{Unsupervised RSISR Methods}
\label{sec4}
Supervised RSISR methods usually need massive amounts of labeled training data, which face many challenges in practical applications, such as high cost of acquiring real HR data, poor annotation transferability across sensor scenarios, and difficulty in eliminating deviations between simulated data and real physical degradation models. In contrast, unsupervised methods alleviate the strict reliance on labeled data, which leverage the intrinsic characteristics of images, such as self-similarity, multi-source information complementarity, and domain-specific physical priors. Therefore, it is still necessary to conduct research on unsupervised RSISR methods. In this section, we conduct a detailed investigation of the unsupervised RSISR methods from the perspectives of traditional methods and deep learning.

\subsection{Traditional RSISR Methods}
Although deep learning has emerged as the dominant approach for RSISR, traditional unsupervised methods still hold significant research value due to their demonstrated advantages, such as model robustness, physical interpretability, and cross-scenario generalization capabilities. In addition, these conventional techniques remain particularly promising in resource-constrained applications, particularly in scenarios involving small-sample learning, low-resource consumption, and specialized tasks like zero-shot super-resolution. Furthermore, their well-established theoretical frameworks offer valuable insights for advancing mainstream algorithms through improved architecture design and interpretability enhancement. Generally, traditional unsupervised RSISR mainly focuses on reconstruction-based methods, which involve the following aspects: interpolation-based, frequency domain-based, iterative back projection (IBP)-based, and regularization constraint-based.
\subsubsection{Interpolation-based Methods}
As one of the most important basic technologies in the field of image reconstruction, interpolation-based methods predict the spatial distribution of LR pixels, and typical representatives are nearest neighbor interpolation, bilinear interpolation, and bicubic interpolation. Although these methods have limitations in reconstructing complex textures, it is still necessary to reveal their prior-guided spatial adaptive mechanisms for RSISR. In this section, we systematically review the classical RSISR algorithms based on interpolation.

In early work \cite{tao2003superresolution}, wavelet transform and spatial interpolation were combined to improve reconstruction performance, which used a spectrum separation processing strategy to improve resolution while suppressing interpolation artifacts. In addition, Zhou et al. \cite{zhou2012interpolation} proposed an interpolation super-resolution method based on multi-surface fitting, which focuses on spatial structure-driven surface modeling and maximum a posteriori probability fusion to achieve high-frequency detail preservation. To address the problems of blur and block effects caused by the loss of high-frequency information in conventional interpolation, Han et al. \cite{han2015wavelet}  proposed a RSISR framework that combines frequency domain segmentation enhancement with spatial domain interpolation optimization based on bicubic interpolation and discrete wavelet transform (DWT), which significantly improves the reconstruction performance and preserves the details of the ground objects. A similar work is the multi-band collaborative super-resolution framework based on dual-tree complex wavelet transform (DT-CWT) and improved edge-directed interpolation (INEDI) proposed by Solanki et al. \cite{solanki2018efficient}. In addition, Mareboyana and Moigne \cite{mareboyana2018super} proposed an edge-directed radial basis function (EDRBF) driven super-resolution method, which injects an edge-constrained radial basis adaptive interpolation mechanism and co-optimizes sub-pixel registration, showing strong robustness to registration errors.

\subsubsection{POCS-based Methods}
Projection onto Convex Sets (POCS) is an iterative optimization method based on convex set constraints, which approximates the feasible solution of RSISR reconstruction by alternately projecting onto a priori knowledge constraint sets of the degradation model. Earlier works \cite{patti1997superresolution,patti1997robust} have demonstrated that POCS has significant potential in improving super-resolution performance. A more concrete work is the two-stage regularization framework based on POCS proposed by Aguena and Mascarenhas \cite{aguena2006multispectral} for multispectral-panchromatic fusion, which achieves spectral-spatial feature collaborative enhancement through sequential and parallel projection or least squares synthesis. In addition, Xie et al. \cite{xie2009blind} proposed a robust POCS iterative framework based on improved sub-pixel shift estimation for blind image super-resolution in 2009, but it has the disadvantage of slow convergence speed. In order to solve the problem of imbalanced image data of the Chang'e-1 three-line array, Zhang et al. \cite{zhang2011super} proposed an RSISR method based on global weighted POCS, which significantly improved the performance of lunar surface topography reconstruction. In 2016, Liu et al. \cite{liu2016improved} proposed an improved POCS algorithm based on visual mechanism for RSISR, which achieves infrared image target edge enhancement and background noise suppression through variable correction threshold consistency constraint and human visual contrast constraint. In order to solve the problem of inaccurate PSF estimation in traditional POCS, Fan et al. \cite{fan2017projections} proposed a slant-edge estimation strategy based on the linear relationship of PSF of LR images, which greatly improved the performance of image reconstruction. In addition, Dai et al. \cite{dai2017study} proposed an improved POCS algorithm for RSISR, which employs iterative curvature-based interpolation (ICBI) to generate high-resolution initial estimates, effectively alleviating the edge blur and detail loss problems in POCS reconstruction. A similar work is that Wang et al. \cite{wang2021improved} proposed to use gradient interpolation instead of nearest neighbor interpolation to solve the problem of edge blur and iterative subjectivity in hyperspectral super-resolution, and designed an adaptive stopping criterion based on the mean square error of adjacent iterations to improve the quality of full-band reconstruction.

\subsubsection{IBP-based Methods}
Iterative Back Projection (IBP)-based super-resolution methods mainly improve the resolution of RSIs through error feedback iterative optimization. To the best of our knowledge, IBP was first introduced as a multi-frame super-resolution algorithm in the work of Irani and Peleg et al. \cite{irani1991improving}. Early work in \cite{lu2002pyramid} proposed a sub-pixel migration multi-observation model, which established super-resolution equations and employed the IBP algorithm to obtain HR images. Besides, Li et al. \cite{li2006improved} proposed an improved IBP algorithm for RSISR by integrating inverse-forward elastic registration with block-wise processing to address local affine transformations, which enhanced geometric adaptability and high-frequency preservation. Their subsequent work \cite{li2007efficient} further demonstrated the effectiveness in stereo image fusion of ALOS satellite's PRISM sensors. A similar work is that Yan and Lu \cite{yan2009super} combined IBP with the Papoulis-Gerchberg extrapolation method to solve the anisotropic resolution problem of MRI.

In order to deal with the inherent defects of IBP, many optimization strategy studies have been conducted. For example, Patel et al. \cite{patel2011hybrid} combined infinite symmetric exponential filter (ISEF) to suppress the checkerboard effect. Bareja and Modi \cite{bareja2012effective} introduced Canny edge detection and error difference to enhance high-frequency reconstruction. In addition, Nayak et al. \cite{nayak2013spatial} innovatively combined the cuckoo search algorithm with IBP to achieve global optimization and improve super-resolution performance.
Some studies utilize regularization strategies to improve performance. For example, Wang et al. \cite{cong2013effective} proposed a super-resolution method based on wavelet edge detection and principal component analysis (PCA), which improved the accuracy of face super-resolution recognition by 12\% by extracting high-frequency information. Nayak et al. \cite{nayak2014morphology} designed a mathematical morphological edge regularization technique for the reconstruction task, which suppressed the ringing artifacts caused by IBP through adaptive weights and significantly improved the visual quality of strong edge areas.

In addition to the above, some related works focused on the research of algorithm efficiency and multimodal fusion. For example, Nayak and Patra \cite{nayak2018enhanced} proposed an evolutionary edge-preserving IBP method (EEIBP) for super-resolution tasks. Specifically, EEIBP optimizes the initial estimate through non-uniform B-spline interpolation and combines an adaptive back-projection kernel driven by covariance, which greatly reduces the reconstruction time and effectively improves edge sharpness. Mutai et al. \cite{mutai2022cubic} incorporated a two-stage process of cubic B-spline approximation and discrete wavelet transform, which greatly improved the super-resolution performance by reducing blur through pre-filtering. In addition, Tao et al. \cite{tao2023fssbp} proposed a spatial-spectral joint back projection (SSBP) method to improve the quality of panchromatic sharpening while maintaining spectral consistency.

\subsubsection{Probability-based Methods}
Probability-based methods mainly utilize statistical modeling and probabilistic inference to perform image reconstruction. Specifically, these methods inherently incorporate prior knowledge of the imaging scene, enabling unsupervised reconstruction without relying on paired high-low-resolution training datasets.
Generally, probabilistic RSISR methods can be systematically categorized into four paradigms: Bayesian inference, maximum a posteriori (MAP) estimation, maximum likelihood (ML) estimation, and Markov random field (MRF) modeling.

\textit{\textbf{Bayesian-based Methods.}} Generally speaking, Bayesian-based methods transform the ill-posed problem of super-resolution reconstruction into a well-posed inversion model by combining image prior probability density with physical constraints. In the past two decades, Bayesian RSISR has been widely studied, and many Bayesian RSISR methods have been proposed. To the best of our knowledge, Tipping and Bishop \cite{tipping2002bayesian} first proposed a Bayesian super-resolution method based on marginalization in 2002. Specifically, they modeled high-resolution images by injecting Gaussian process priors to achieve joint estimation of point spread function and registration parameters. Besides, Molina et al. \cite{molina2005new} integrated the sensor characteristics into the Bayesian framework and combined the observation process of multispectral and panchromatic images to effectively improve the spatial resolution of Landsat ETM+ while maintaining the spectral fidelity. Subsequently, Molina et al. \cite{molina2008variational,molina2006parameter,molina2006hierarchical} further constructed a hierarchical Bayesian model with parameter priors and used variational inference methods to achieve simultaneous estimation of HR multispectral images and model parameters. In addition, Babacan et al. \cite{babacan2010variational} proposed a joint super-resolution reconstruction method based on variational Bayesian, which simultaneously estimates high-resolution images and motion parameters through probabilistic modeling to achieve adaptive robust reconstruction. For pansharpening, Wang et al. \cite{wang2018high} proposed a Bayesian fusion model based on the geometric-spectral-spatial consistency hypothesis, which uses the alternating direction multiplier method (ADMM) to solve and verify the generalization ability of hyperspectral images. In order to solve the problem of parameter sensitivity, Armannsson et al. \cite{armannsson2021comparison} innovatively introduced Bayesian optimization into Sentinel-2 super-resolution, and significantly improved the robustness of methods such as ATPRK and S2Sharp through a small-scale parameter optimization mechanism. Li et al. \cite{li2021forward} proposed a multi-prior Bayesian model based on the fusion of TV and Laplace priors, which significantly improved the accuracy of target contour reconstruction in sea and land scenes. Furthermore, Tan et al. \cite{tan2021novel} established a two-dimensional spatial structure prior based on Markov random field (MRF), and characterized the spatial correlation of the scene through an n-order neighborhood system, which significantly improved the suppression of artifacts compared with previous methods. Besides, Ye et al. \cite{ye2022bayesian} proposed a Bayesian hyperspectral super-resolution method based on texture decomposition, which effectively suppressed spectral variation through Gaussian process spectral prior. In recent work, Shen et al. \cite{shen2024super} and Guo et al. \cite{guo2025structured} achieved breakthroughs in RSISR of Bayesian forward-looking radar imaging in low signal-to-noise ratio and maneuvering scenarios by using a gamma-normal conjugate model and structured Student's t prior, respectively.

\textit{\textbf{MAP-based Methods.}}
RSISR methods based on maximum a posteriori probability (MAP) have also been widely studied. For example, the early work of Chantas et al. \cite{chantas2007super} proposed a MAP framework with local adaptive edge-preserving priors, which achieved joint restoration and registration under degraded observations through a two-step reconstruction strategy. Subsequently, some studies began to optimize from the perspective of prior models. Wang et al. \cite{wang2010spectral} proposed a MAP method based on high-frequency fusion of panchromatic images for RSISR, which introduced spectral fidelity prior constraints to iteratively optimize the framework. In addition, Belekos et al. \cite{belekos2010maximum} designed a multi-channel non-stationary Gaussian prior model to enhance the collaborative reconstruction capability of multispectral data. Li et al. \cite{li2009super} first introduced the Hidden Markov tree (HMT) into the MAP framework and used the multi-scale dependence of wavelet coefficients to improve the reconstruction performance. Improved methods for specific application scenarios have also emerged. For example, Guan et al. \cite{guan2014maximum} proposed a MAP radar angle super-resolution method based on antenna pattern and target scattering convolution prior modeling. Zhang et al. \cite{zhang2019sea} proposed a Rayleigh sparse MAP (RSMAP) algorithm for sea surface target imaging, which combines the distribution characteristics of sea clutter with the sparse prior of the target to improve the angular resolution. In terms of noise robustness, Vrigkas et al. designed a global M-estimation framework to achieve parameter adaptive adjustment. Besides, the regional adaptive total variation (RSATV) model proposed by Yuan et al. \cite{yuan2014remote} for RSISR utilized spatial information classification to suppress artifacts. On the other hand, Li et al. \cite{li2017super} first achieved RSISR on short-time sequence images of the GF-4 geosynchronous satellite, verifying the feasibility of synchronous orbit observation data. In addition, some studies have constructed the MAP-MRF framework to perform hyperspectral super-resolution technology. Irmak et al. \cite{irmak2016super} achieved multi-frame reconstruction optimization through abundance map decomposition and endmember extraction. Further, the extended study \cite{irmak2018map} combined virtual dimension endmember determination with texture-constrained secondary optimization, and its performance exceeded the mainstream method while maintaining spectral consistency, forming a complete technical chain for multi-source to single-source reconstruction.

\textit{\textbf{ML-based Methods.}}
Some studies use maximum likelihood (ML) methods to transform the ill-posed problem of RSISR into a well-posed problem. Different from the MAP method, the ML method omits explicit prior constraints and is more robust. Tan et al. \cite{tan2018penalized} proposed a penalized ML method for I/Q dual-channel joint noise modeling for RSISR, which constructed a joint regularization term that integrated the square-Laplacian characteristic to suppress noise amplification. Furthermore, the extended study \cite{tan2018q} proposed an I/Q channel joint probability model containing phase information, which achieved high-precision target estimation by maximizing the likelihood of complex signals. In addition, Wu et al. \cite{wu2021super} proposed a fast ML algorithm with dynamic adjustment of adaptive acceleration factors, which significantly reduced the computational complexity while maintaining super-resolution performance.

\textit{\textbf{MRF-based Methods.}}
The Markov Random Field (MRF) based method mainly models the spatial dependencies between pixels and their neighborhoods, which performs effective integration of global spatial context information through the energy function of the local neighborhood. MRF-based RSISR has been studied for nearly two decades, and many MRF-based RSISR methods have been proposed. For example, Li et al. \cite{li2013super} proposed a spatial-temporal MRF super-resolution model (STMRF\_SRM), which integrates historical medium-resolution images with current coarse-resolution data to effectively enhance the spatiotemporal consistency of forest mapping. Aghighi et al. \cite{aghighi2015fully} further proposed a full-space adaptive MRF-SRM, which dynamically balances spectral and spatial weights based on endmember analysis and local energy matrix to achieve automatic parameter calibration. In the latest progress, Welikanna et al. \cite{welikanna2024fuzzy} fused the fuzzy C-means and spectral angle quantity to optimize the MRF posterior energy function, which reduces the influence of the point spread effect.
\subsection{Deep learning-based RSISR Method}
The majority of deep learning-based RSISR methods are supervised, which require a large amount of labeled training data. However, the ground truth is normally not available in
RSISR, thus, almost all supervised methods utilize synthetic degraded simulation datasets that are actually not realistic. Therefore, it is desirable to have unsupervised RSISR methods. In this section, we investigate the unsupervised deep learning RSISR methods, mainly involving self-supervised learning, contrastive learning, generative-based, and zero-shot learning.

\subsubsection{Self-supervised Learning}
Self-supervised RSISR methods are usually inspired by internal degradation modeling or cross-modal data association to generate pseudo-supervisory signals, thereby avoiding the reliance on external HR training data. In recent years, self-supervised RSISR methods have emerged. In 2018, Haut et al. \cite{haut2018new} proposed the first unsupervised convolutional generation model, which learned the LR-HR feature relationship through stacked convolution and downsampling layers, and employed LR reconstruction loss to constrain the generation quality. In addition, Choi and Kim \cite{choi2020no} proposed a new degradation model to generate LR samples, and combined it with morphological transformation to enhance HR images, thereby achieving super-resolution of KOMPSAT-3 images. Sheikholeslami et al. \cite{sheikholeslami2020efficient} proposed an efficient unsupervised super-resolution (EUSR) model for RSISR, which uses dense skip connections and bottleneck compression modules to reduce the amount of computation while maintaining reconstruction performance. For cross-modal data, Chen et al. \cite{chen2021hyperspectral} proposed a self-supervised spectral-spatial residual network (SSRN), which generates HR hyperspectral images through the mapping relationship between LR multispectral and hyperspectral images without HR ground truth supervision. In terms of multi-frame super-resolution, Nguyen et al. \cite{nguyen2021self} proposed a frame-to-frame self-supervised framework that fuses satellite image sequence features and improves the resolution of SkySat satellite images. In the latest progress, Hong et al. \cite{hong2023decoupled} proposed a decoupled-and-coupled network (DC-Net) for RSISR, which effectively alleviated the distribution difference of hyperspectral and multispectral data through the decoupled-coupled structure and self-supervised constraints. Qian et al. \cite{qian2022selfs2} proposed an RSISR method based ona  deep image prior. Specifically, they used convolutional neural networks and 3D separable convolutions to perform super-resolution restoration on the coarse resolution band of Sentinel-2 satellite images without the need for additional training data, significantly improving the super-resolution performance. An interesting work is a self-supervised degradation guidance method proposed by Xiao et al. \cite{xiao2023degrade}, which effectively adapts to various unknown degradations through contrastive learning and a bidirectional feature modulation network, significantly improving the RSISR performance.

\subsubsection{Contrastive Learning}
Traditional methods rely on bicubic downsampling to synthesize LR images for training, which cannot reflect the complex degradation patterns of real scenes. Contrastive learning, as a self-supervised learning method, unsupervisedly compares the degradation features and HR detail differences between samples to mine potential degradation representations, significantly improving the model's adaptability to unknown degradation and the physical consistency of super-resolution reconstruction. In recent years, some RSISR methods based on contrastive learning have been proposed. Mishra and Hadar \cite{mishra2023clsr} proposed a semi-supervised contrastive learning framework termed CLSR, which significantly reduced the reliance on large-scale diverse training samples by comparing the potential differences between real degradation and ideal bicubic degradation. Further, the extended research \cite{mishra2023accelerating} proposed a two-stage contrast framework for RSISR. In the first stage, "artificial style images" with HR textures are generated through contrast training, and then neural style transfer is combined to achieve unsupervised super-resolution. An interesting contrastive learning-based RSISR method is the region-aware network (RAN) proposed by Liu et al. \cite{liu2023ran}. In RAN, they integrated the degradation prior extracted by contrastive learning with the regional structure analysis module, and captured cross-block self-similarity through graph neural networks and attention mechanisms, effectively improving the fidelity of texture reconstruction. In the latest progress, the EDRLN network proposed by Wang et al. \cite{wang2024efficient} uses mutual affine convolution to replace the traditional convolution layer and integrates a lightweight pixel attention mechanism in the contrastive learning framework, which reduces the number of model parameters while maintaining excellent performance in multiple degradation scenarios.

\subsubsection{Generative-based Methods}
Unsupervised generative methods establish data distribution mappings rather than relying on paired samples to perform RSISR. In recent years, some studies have used GAN or cyclic structures to learn the joint distribution of degradation and reconstruction, and achieve resolution enhancement through implicit image priors.
For example, Wang et al. \cite{wang2019unsupervised} proposed a Cycle-CNN based on a cyclic structure. They used two generative CNNs to model the downsampling and super-resolution processes, respectively, and verified the effectiveness of unpaired training on GaoFen-2 satellite data. In 2021, the extended research \cite{zhang2020nonpairwise} proposed a new Cycle-CNN that integrates the bidirectional mapping of the downsampling network and the SR network, achieving improved robustness on the UC Merced dataset, and further demonstrated the ability of the cyclic generative structure to suppress noise and blur. In addition, Wang et al. \cite{wang2021enhanced} proposed an enhanced image prior (EIP) method for RSISR, which encodes the reference image into the latent space based on GAN, and combines the cyclic update strategy to transfer texture features, achieving advanced performance. Similarly, the MIP method \cite{wang2021unsupervised} constructs a transfer image prior and uses an implicit noise update mechanism to achieve feature transfer for RSISR. Zhang et al. \cite{zhang2020unsupervised} proposed an unsupervised RSISR method based on GAN, which achieved competitive performance by reconstructing SR images through a generator and then downsampling to train the discriminator. A recent work is TransCycleGAN proposed by Zhai et al. \cite{zhai2024transcyclegan}, which integrates the Transformer module into the CycleGAN framework and uses transposed self-attention to capture global context, effectively removing degenerate features in a pseudo-supervisory manner.

\subsubsection{Zero-shot Learning}
Zero-shot learning-based RSISR method only uses the intrinsic features of the image to build a training set for super-resolution reconstruction. Yang and Wu \cite{yang2022enhanced} proposed an enhanced zero-shot super-resolution method, which generates enhanced images and builds training sets through a content adaptive resampler (CAR) network. In addition, they introduced a convolutional block attention module (CBAM) and a residual module, which significantly improved the quality of image detail reconstruction. In \cite{bose2021zero}, the authors proposed a zero-shot RSISR framework based on self-tessellations and a cascaded attention sharing mechanism, which achieves high-quality reconstruction by exploiting the structural continuity of RSIs without pairing HR images. Self-FuseNet proposed by Mishra and Hadar \cite{mishra2023self} adopts a forward generation paradigm and accelerates image reconstruction through a UNet architecture with wide skip connections, demonstrating strong generalization capabilities in RSISR. Cha et al. \cite{cha2023meta} proposed a meta-learning-based zero-shot RSI super-resolution (ZSSR) method, which regards multiple LR image sets as a set of ZSSR tasks to learn general super-resolution meta-knowledge, significantly reducing the computational cost of large-scale image processing.

\section{Modal-Specific RSISR Methods}
\label{sec:modal}
Remote sensing modalities such as optical, SAR, infrared, hyperspectral, and LiDAR exhibit distinct sensing mechanisms and data characteristics. These inherent differences demand correspondingly tailored super-resolution strategies. For instance, optical imagery provides rich spectral and spatial textures but is susceptible to atmospheric conditions. Conversely, SAR data offers robustness against weather and lighting variations yet is characterized by inherent speckle noise and nonlinear distortions. Consequently, each modality presents unique RSISR challenges, necessitating specialized architectures and loss functions designed to leverage their respective data priors.

In this section, we systematically review RSISR methods for different remote sensing modalities, including optical, SAR, infrared, and lidar data, analyzing fusion strategies, performance gains, and open challenges to guide future advancements.

\subsection{Optical RSISR Methods}
Typically, optical RSIs exhibit rich spatial details, strong inter-band correlations, and complex degradations caused by atmospheric scattering and optical system aberrations. In recent years, super-resolution for optical RSIs has emerged as a modality-specific field, with innovations driven by the unique imaging mechanisms and statistical characteristics inherent to optical data. In this section, we perform a detailed survey on recent progress in optical RSISR methods.

Some studies leverage the intrinsic inter-band correlations in optical or multispectral data as a powerful prior to enable joint reconstruction. For example, Zhang et al. \cite{chang2023pixel} introduced a pixel-wise attention residual network that adaptively fuses features from multiple bands, significantly enhancing spectral fidelity across Sentinel-2 and Landsat datasets. Building upon this, Wang et al. \cite{10764782} proposed a multi-scale CNN-Transformer hybrid (MSCT), which extracts local spectral features via adaptive residual dense blocks and models inter-band dependencies through Transformer attention, showing improved generalization on both RGB and multispectral inputs. To address the frequent loss of high-frequency texture and edge details in optical super-resolution, several works introduced frequency-aware generative architectures. Zhao et al. \cite{han2023enhancing} developed EHC-DMSR, a diffusion-based model that enforces high-frequency consistency using Fourier-based constraints, improving the reconstruction of fine structures. In addition, Zhang et al. \cite{10781453} proposed a detailed recovery based on GANs (DRGAN) for RSISR, which leverages dynamic convolution and self-attention to enhance high-frequency textures and mitigate artifacts and over-smoothing. Meanwhile, Liu et al. \cite{10453218} a multiscale
residual dense network (MRDN) for RSISR, which reuses multi-scale residuals across layers, preserving subtle details from the original low-resolution inputs.

Considering the specific degradation issues of optical systems, such as point spread functions (PSF), sensor limitations, and atmospheric effects, researchers have introduced physically informed architectures for RSISR. For instance, Liu et al. \cite{9959886} built DLANet based on a dual-resolution design aligned with Gaussian blur priors, simulating optical sampling effects through low-resolution spatial branches. Chen et al. \cite{11078386} proposed RDAF-GAN for RSISR, which incorporated a compound degradation model (blur, noise, compression) to generate realistic LR samples, improving generalization in real-world conditions. Sun et al. \cite{10549773} introduced a distillation-based Transformer-CNN hybrid network, where a teacher model captures long-range degradation cues (e.g., haze) to guide physically plausible reconstruction by a lightweight student network.

Given the deployment constraints of optical satellites and edge devices, computational efficiency has become a critical issue. Wang et al. \cite{9632567} designed the Context Transformation Layer (CTL) as a lightweight substitute for conventional 3×3 convolutions, achieving competitive performance on UC Merced with reduced parameter counts. Chen et al. \cite{zhao2024mamba} adopted a residual Mamba module based on a state-space model to capture long-range dependencies with linear complexity, addressing the scalability bottleneck for large-scale optical imagery. In addition, Li et al. \cite{10767430} proposed an aggregation connection transformer (ACT-SR), which introduces row-column decoupled attention to suppress visual artifacts in Transformer-based reconstruction, improving perceptual consistency in urban and agricultural regions. To enhance interpretability and structural fidelity, Shi et al. \cite{9955482} proposed DMUNet, an ADMM-based dual-branch unfolding network with dedicated edge reconstruction branches and a lightweight multiscale fusion module. 

In summary, modality-driven RSISR research demonstrates compelling advantages by integrating physics-aware architectures with the semantic foundations of optical data. Crucially, advances in harnessing cross-band correlations, embedding degradation physics, and optimizing edge deployment efficiency underscore how deep domain knowledge of optical sensing mechanisms enables designing more effective, interpretable, and hardware-compatible super-resolution systems.

\subsection{SAR RSISR Methods}
Synthetic Aperture Radar (SAR) images possess unique physical properties, such as scattering matrix diversity, elevation sparsity, and speckle-resolution coupling, which limit conventional imaging resolution. These properties manifest differently across modalities: polarimetric systems require multi-channel optimization, tomographic setups are constrained by narrow elevation apertures, and single-channel systems face an inherent tradeoff between speckle suppression and spatial resolution. Recent RSISR techniques exploit these physical structures as reconstruction priors, overcoming traditional radar limits and mitigating acquisition-induced constraints. This section presents a detailed review of the RSISR method based on SAR data modalities.

Early work by Pastina et al. \cite{pastina2003super} pioneered Polarimetric SAR (PolSAR) super-resolution by applying parametric spectral estimation methods to jointly process multi-polarization channels (HH, VV, HV, VH). More recently, some studies employed deep learning techniques to perform detailed reconstruction. For example, Lin et al.\cite{lin2019polarimetric} introduced a CNN with residual compensation for full-polarization SAR SR. In addition, Dai et al. \cite{10843849} proposed a Polarimetric Contexture Convolutional Network (PCCN) for PolSAR SR, which encodes the polarimetric-spatial cube into a matrix representation and employs dual-branch feature extraction with hierarchical fusion to enhance reconstruction fidelity. In Tomographic SAR (TomoSAR), the limited elevation resolution resulting from sparse baseline sampling significantly constrains accurate 3D reconstruction. To address this challenge, recent studies have explored inversion methods that enhance elevation resolution through super-resolution techniques. For example, Zhu et al. \cite{6112799} framed TomoSAR inversion as a spectral analysis problem and proposed the compressive sensing (CS)-based SL1MMER algorithm, exploiting the sparsity of elevation signals for SR. Further, Zhu and Bamler \cite{5966335} rigorously quantified the theoretical limits of SL1MMER in terms of localization accuracy, minimum separable scatterer distance, and required acquisition numbers for robust SR. In addition, Wu et al. \cite{wu2020super} proposed an RSISR approach for MIMO array TomoSAR under low SNR and limited antenna elements, which combined CS-based spatial filtering with deep neural network regression for improved scatterer localization in the elevation direction.

On the other hand, He et al. \cite{he2012learning} proposed a multi-dictionary compressive sensing method for SAR SR using sparse coding spatial pyramids. Building on CS foundations, Wei et al. \cite{9340592} subsequently introduced a hybrid scattering model incorporating line-segment scatterers (LSS) and rectangular-plate scatterers (RPS), solved via Multi-component ADMM. Concurrently, generative approaches gained traction. Specifically, Wang et al. \cite{8634345} developed an SRGAN-based network for reconstruction and noise suppression. For target-level applications, Lee and Lee \cite{lee2023efficient} created a robust method combining CS-based scattering center extraction with super-resolved system impulse response (IRF-S). Meanwhile, solutions addressing speckle noise emerged, including Importance Sampling UKF \cite{8239604} and Deep GANs \cite{8899202}. Finally, Dong et al. \cite{10966888} established a sub-band decomposition framework with cross-domain learning and a novel amplitude-phase evaluation system. Recently, utilizing co-registered optical imagery to guide SAR SR reconstruction has emerged as a powerful strategy. Li et al. \cite{yanshan2022ogsrn} proposed the Optical-Guided Super-Resolution Network (OGSRN), which incorporates an Enhanced Residual Attention Module (ERAM) and optical-space feedback mechanism to guide large-scale SAR SR. Zhao et al. \cite{11045185} further advanced this area by proposing the Hierarchical Selective Fusion Mamba Network (HSFMamba), which uses state-space models and cross-modal feature selection for joint optical-guided SAR SR and denoising, supported by a large-scale aligned dataset.

Increasing attention focuses on resolving coupled degradations (e.g., resolution loss, speckle noise, defocusing) and end-to-end systems. For example, Chen et al. \cite{11084879} proposed a degradation-aware diffusion super-resolution model (DADSR) that leverages contrastive learning for degradation representation and adaptively handles coupled degradations. Meanwhile, Zhang et al. \cite{11071996} introduced an integrated deep learning network (DLIN) achieving end-to-end super-resolution imaging and target recognition through a 2D deep unfolding network and Transformer architecture.

SAR RSISR methods have evolved from modality-specific spectral estimation to deep learning frameworks that leverage intrinsic physical structures as reconstruction priors. While modality-tailored approaches have significantly advanced single-task SR performance, current research mainly focuses on three critical frontiers: multimodal guidance, coupled degradation handling and end-to-end task integration. Future efforts should prioritize the physics-informed architectures, large-scale aligned multi-sensor datasets for complex environments, and unified metrics for coupled SR-denoising evaluation. Real-time onboard processing remains a key challenge, demanding lightweight yet physics-compatible models.

\subsection{Infrared RSISR Methods}
Infrared (IR) imaging often suffers from low spatial resolution and poor texture detail due to the inherent limitations of detectors, optical diffraction, and challenging environmental conditions. These challenges hinder essential tasks such as military reconnaissance and night vision surveillance. To this end, infrared RSISR has emerged as a key solution to overcome hardware-imposed limitations.

Many studies have used multimodal fusion to solve the texture scarcity problem of infrared images, which mainly integrates the complementary visible information of different modal data. For example, Yang et al. \cite{yang2016fast} proposed a Soft-assignment based Multiple Regression (SMR) method that utilizes visible-IR joint training and cluster-adaptive dictionaries to enable real-time high-fidelity reconstruction. In addition, Jiang et al. \cite{jiang2023improved} proposed a sensor-fusion network that utilizes cross-attention transformers and hierarchical distillation to recover thermal details while preserving radiometric accuracy. Chen et al. \cite{chen2024infrared} developed a physics-guided fusion method for infrared super-resolution images, which integrates spectral inference from visible imagery, radiometric modeling, and correction using measured infrared data. Huang et al. \cite{huang2025texture} proposed DASRGAN, a dual-adaptation framework for IR super-resolution, which integrates Texture-Oriented Adaptation (TOA) to enhance texture and Noise-Oriented Adaptation (NOA) to minimize noise transfer, achieving state-of-the-art performance across multiple benchmarks.

Some studies focus on sparsity-driven methods, using IR's inherent sparsity for efficient recovery and deeper exploration. Mao et al. \cite{mao2016infrared} proposed a CS reconstruction method, which utilizes analytical measurement matrices and differential sparsification to achieve hardware-friendly stability. Zhang et al. \cite{zhang2018infrared} proposed a hybrid CS-CNN framework that utilizes sparse-domain initialization and deep denoising to enhance high-frequency components. In addition, advanced deep architectures leverage specialized representation learning for formidable infrared super-resolution, enabling higher-resolution imaging. For instance, He et al. \cite{he2018cascaded} proposed a cascaded multi-receptive field network that leverages stage-wise upscaling with macro-to-micro refinement, achieving 8× super-resolution using a compact architecture. Cao et al. \cite{cao2023cfmb} developed the CFMB-T model, leveraging wavelet frequency separation and MTF-embedded degradation priors to focus reconstruction on critical high-frequency bands. Zhu et al. \cite{zhu2022super} introduced a pseudo-texture transfer algorithm that employs cross-modal feature matching and replacement for semantically aligned visible texture injection. Dan et al. \cite{dan2024pirn} designed the PIRN architecture with phase-invariant augmentation and deformable convolution kernels to enhance spatial feature robustness. Zhang et al. \cite{zhang2023closed} proposed a closed-loop learning system using coupled downsampling-SR generators to exploit real-image distributions across supervision paradigms. Recently, an IRSRMamba framework proposed by Huang et al. \cite{huang2025irsrmamba} addressed Mamba's spatial fragmentation through wavelet-modulated state-space blocks and semantic consistency constraints. In addition, Lu and Su \cite{lu2025super} proposed a multi-source domain adaptation method that utilizes visible-guided enhancement with thermal flux conservation to quadruple Mars IR resolution.                                                                                         
\subsection{LiDAR RSISR Methods}
LiDAR (Light Detection and Ranging) is a pivotal 3D sensing technology employed in autonomous driving, robotics, and environmental monitoring. Due to constraints in sensor cost, size, and energy efficiency, the acquired point clouds are typically sparse and low-resolution, undermining the accuracy of downstream tasks. To address this, LiDAR Point Cloud Super-Resolution (RSISR) has become a critical research focus, aiming to enhance point density and spatial fidelity through algorithmic post-processing.

Shan et al. \cite{shan2020simulation} leveraged simulation-trained CNNs to enhance LiDAR resolution by projecting sparse point clouds into 2D range images, employing Monte Carlo dropout to filter uncertain predictions. Tian et al. \cite{tian2022lidar} introduced a training-free geometric approach that segments and reorganizes ground/non-ground points to generate high-resolution outputs, eliminating training dependencies. Chen et al. \cite{chen2023sgsr} developed SGSR-Net, which integrates structural-semantic attention mechanisms with guided Monte Carlo filtering to boost geometric fidelity and SLAM compatibility. In addition, Ramirez et al. \cite{ramirez2024super} proposed a GAN-based "Hyperheight Data Cube" framework for satellite LiDAR, demonstrating significant spatial resolution enhancement in real-world validation. Collectively, these works demonstrate the field's evolution from image-based enhancement toward geometric-aware modeling, semantic guidance, and generative frameworks in LiDAR RSISR research.

\section{Datasets and evaluation metrics}
\label{sec6}
\subsection{Available remote sensing image datasets}
In the field of RS, due to the high cost of acquiring real HR images and the difficulty of registering corresponding LR images, there are few publicly available datasets for RSISR. To this end, most studies perform bicubic downscaling on RS datasets to generate RSISR datasets. In this section, we summarize the widely used datasets for RSISR in the literature, as shown in Table \ref{tab2}. Specifically, the details of these datasets are as follows:
\begin{table}[t]
  \centering
  \caption{Details of Existing Remote Sensing Image Datasets.}
   \resizebox{0.48\textwidth}{!}{
\begin{tabular}{lcllr}
\toprule
Name & Year & Data Source & Numbers & Resolution \\
\hline
UCMD & 2010 & Air-Spaceborne & 2,100 & $256\times256$  \\

WHU-RS19 & 2012 & Google Earth & 1,005 & $600\times600$  \\

RSSCN7 & 2015 & Google Earth & 2,800 & $400\times400$ \\
AID & 2016 & Aerial & 10,000 & $400\times400$  \\

KOSD & 2016 & Draper Satellite & 1,620 & $3099\times2329$ \\

SIRI-WHU & 2016 & Google Earth & 2,400 & $200\times200$ \\

NWPU-RESISC45 & 2017 & Google Earth & 31,500 & $256\times256$  \\
PatternNet & 2018 & Google Earth & 30,400 & $256\times256$  \\
DOTA & 2018 & JL-1, GF-2 & 2,806 & $4000\times4000$  \\
SpaceNet2 & 2018 & WV-2 & 10,595 & $162\times162$ \\
DFC2019 & 2019 & WV-3 & 2,783 & $1024\times1024$  \\
OPTIMAL-31 & 2019 & Google Earth & 1,860 & $256\times256$ \\
\hdashline
RSI-CB & 2020 & Crowdsource & 60,000 & \makecell[r]{$128\times128$\\$256\times256$} \\
\bottomrule
\end{tabular}
 \label{tab2}}
\end{table}

\begin{itemize}
    \item UCMD (UC Merced Land-Use Dataset)\footnote{\url{https://reurl.cc/Y3XOQX}} \cite{yang2010bag}. This dataset is an RSI dataset for land use research, which is obtained from USGS National Map Urban Area Imagery. Besides, it contains images of 21 different land classes, with 100 images in each class. The pixel resolution of the public domain images is 1 foot (0.3 meters), and the image pixel size is $256\times256$.
    
    \item WHU-RS19\footnote{\url{https://captain-whu.github.io/BED4RS/}} \cite{Dai2011WHURS19}. This dataset consists of HR satellite images exported from Google Earth, with a spatial resolution of up to 0.5 meters. It includes 19 classes of meaningful scenes in high-resolution satellite imagery. Each class contains approximately 50 image samples, providing a diverse and representative collection for research and analysis.
    
    \item RSSCN7\footnote{\url{https://pan.baidu.com/s/1slSn6Vz}} \cite{zou2015deep}. This dataset contains 2800 RS images, which are distributed across seven categories, each with 400 images sampled at four different scales. Each image is $400\times400$. pixels and obtained from Google Earth.
    
    \item AID (Aerial Image Dataset)\footnote{\url{https://captain-whu.github.io/BED4RS/}} \cite{xia2017aid}. AID is a large-scale aerial image dataset, including 10,000 images of 30 different scene types collected from Google Earth, and all images are labeled by experts in the field of RSI interpretation.
    
    \item KOSD (Kaggle Open Source Dataset)\footnote{\url{https://reurl.cc/9nprxa}} \cite{troylau2016draper}. The dataset contains more than 1,000 HR aerial photographs from Southern California, consisting of 350 training images and 1,370 test images.
    
    \item NWPU-RESISC45\footnote{\url{https://1drv.ms/u/s!AmgKYzARBl5ca3HNaHIlzp_IXjs}} \cite{cheng2017remote}. The dataset is a public RSI scene classification dataset created by Northwestern Polytechnical University (NWPU). It contains 31,500 images of 45 scene classes, involving 700 images with the size of $256\times256$ pixels in each class.
    
    \item PatternNet\footnote{\url{https://sites.google.com/view/zhouwx/dataset}} \cite{zhou2018patternnet}. The dataset is a large-scale HR remote sensing dataset for RSI retrieval. It contains 30,400 images of 38 scene classes, involving 800 images with the size of $256\times256$ pixels in each class. The images in PatternNet are collected from Google Earth images or through Google Map API for some cities in the United States.
    
    \item DOTA\footnote{\url{https://captain-whu.github.io/BED4RS/}} \cite{xia2018dota}. DOTA is a large-scale public dataset collected by Google Earth, GF-2, and JL-1 satellites for aerial image object detection. It contains a total of 2,806 HR images with resolutions ranging from $800\times800$ to $4000\times4000$ pixels.
    
    \item SpaceNet\footnote{\url{https://spacenet.ai/datasets/}} \cite{spacenet2018catalog}. SpaceNet is another large-scale satellite imagery dataset, obtained from the VHR WorldView-3(WV-3) satellite.
    
    \item DFC2019 (IEEE Data Fusion Contest 2019)\footnote{\url{https://reurl.cc/6q0a9y}} \cite{ieeedataport2019datafusion}. The dataset contains 2,783 multi-temporal HR satellite images taken by the WV-3 satellite as training sets and 50 test images. All images are provided in the form of regular tiles of $1024\times1024$ pixels with a spatial resolution of sub-meter level.

    \item OPTIMAL-31\footnote{\url{https://reurl.cc/lYeRn6}} \cite{wang2018scene}. The dataset contains 1,860 images of 31 categories collected from Google Maps, each category consists of 60 images with a size of $256\times256$ pixels.

    \item SIRI-WHU\footnote{\url{https://reurl.cc/ekOEnM}} \cite{zhao2015dirichlet}. The dataset includes 2,400 images covering 12 different categories of scenes, collected and produced by the RS-IDEA Research Group of Wuhan University (SIRI-WHU) from Google Earth, mainly covering urban areas in China. Each category contains 200 images, each with a size of $200\times200$ pixels and a spatial resolution of 2m.

    \item RSI-CB \footnote{\url{https://github.com/lehaifeng/RSI-CB}} \cite{li2020RSI-CB} The dataset constructs two sub-datasets of $256\times256$ and $128\times128$ pixel sizes (RSI-CB256 and RSI-CB128, respectively) with 0.3–3-m spatial resolutions. The former contains more than 24,000 images involving 35 categories, and the latter contains more than 36,000 images involving 45 categories. 
\end{itemize}

\subsection{Evaluation metrics}
In the field of RSISR, evaluation metrics are crucial for measuring the performance of algorithms such as reconstruction accuracy and fidelity, which provide an objective reference for the performance comparison of algorithms through quantitative calculation. In this section, we summarize the commonly used quantitative evaluation indicators for RSISR, and the details are as follows:

(1) Structural Similarity Index Measure (SSIM): SSIM \cite{wang2004image} is an important objective evaluation metric that measures the structural similarity between the reconstructed image and the real HR image. Specifically, SSIM simulates the perception of brightness, contrast, and structural information by the human visual system (HVS), which is more in line with subjective visual quality evaluation. Given a super-resolved image $I_{SR}$ and a target image $I_{HR}$, SSIM can be calculated as:
\begin{equation}
    \textit{SSIM} = l(I_{\textit{SR}}, I_{\textit{HR}}) \cdot C(I_{\textit{SR}}, I_{\textit{HR}}) \cdot S(I_{\textit{SR}}, I_{\textit{HR}})
\end{equation}
where $l(I_{\textit{SR}}, I_{\textit{HR}})$, $C(I_{\textit{SR}}, I_{\textit{HR}})$ and $S(I_{\textit{SR}}, I_{\textit{HR}})$ represent the similarity of luminance, contrast and structure respectively. In practice, SSIM is not sensitive enough to the high dynamic range (such as clouds, shadow areas) or complex textures (such as forests, farmlands) of RSIs. Therefore, some studies have proposed the mean structural similarity (MSSIM) and multi-scale structural similarity (MS-SSIM) to evaluate the super-resolution performance. A higher SSIM value implies that the super-resolution image has a higher similarity and better fidelity with the HR image.

(2) Peak-Signal-Noise-Ratio (PSNR): PSNR is another one of the most popular objective evaluation indicators, which is mainly used to quantify the pixel-level fidelity between super-resolved images and true HR images. Specifically, assuming that the size of the super-resolved image $I_{SR}$ and the target image $I_{HR}$ are $M \times N$, the mean square error (MSE) can be defined as: 
\begin{equation}
    \text{MSE} = \frac{1}{H \times W} \sum_{i=1}^{H} \sum_{j=1}^{W} \left( I_{HR}(i,j) - I_{SR}(i,j) \right)^2
\end{equation}
Based on the MSE and the maximum value $L$ of the dynamic range of image pixels, PSNR can be calculated as:
\begin{equation}
    \text{PSNR} = 10 \cdot \log_{10} \left( \frac{L^2}{\text{MSE}} \right)
\end{equation}
Generally speaking, a higher PSNR value indicates that the image has better visual quality.

(3) Learned Perceptual Image Patch Similarity (LPIPS): LPIPS \cite{zhang2018unreasonable} is a perceptual similarity metric based on deep learning, which is closer to HVS evaluation of image quality. Specifically, LPIPS quantifies the perceptual difference between the reconstructed image and the real image. Lower LPIPS means better image quality.

(4) Natural Image Quality Evaluator (NIQE): NIQE \cite{mittal2012making} is a no-reference image quality evaluation metric. This metric is based on natural image statistical modeling and measures the visual quality by comparing the distance between the local statistical features of the image and the statistical distribution. The lower the NIQE value, the better the image visual quality.

(5) Spectral Angle Mapper (SAM): SAM \cite{yuhas1992discrimination} evaluates spectral fidelity by calculating the angle between the test spectrum vector and the reference spectrum vector. Given a test spectrum $\nu$ and a reference spectrum $r$, both with $L$ layer, SAM is calculated as:
\begin{equation}
\operatorname{SAM}(v, w)=\arccos\left(\frac{\sum_{i=1}^{L} v_{i} w_{i}}{\sqrt{\sum_{i=1}^{L} v_{i}^{2}} \sqrt{\sum_{i=1}^{L} w_{i}^{2}}}\right)
\end{equation}
Smaller SAM values indicate a higher spectral fidelity. In particular, angles approaching 0° indicate a high degree of spectral consistency.

(6) Average Gradient (AG): AG \cite{chen2018novel} is used to measure the edge sharpness of image. It is evaluated by calculating the gradient magnitude of all pixels in the image and is defined as:
\begin{equation}
AG = \frac{1}{(h-1)(w-1)} \sum_{x=1}^{w-1} \sum_{y=1}^{h-1} \frac{|G(x, y)|}{\sqrt{2}}
\end{equation}

where $G(x,y)$ represents the gradient at the pixel $(x,y)$, and the denominator $(h-1)(w-1)$ is used for normalization. The larger the AG value, the more details and sharper edges the image contains, and generally has higher clarity.

(7) Perceptual Index (PI): PI \cite{Blau_2018_ECCV_Workshops} was proposed by the 2018 PIRM Super-Resolution Challenge, which combines two no-reference image quality measures: Ma et al. \cite{ma2017learning} and NIQE \cite{mittal2012making}. Specifically, PI can be expressed as follows:
\begin{equation}
PI = \frac{1}{2} \left( (10 - \text{Ma}) + \text{NIQE} \right)
\end{equation}
The smaller the PI value, the better the visual quality of the image.

In addition to the above evaluation metrics, there are some other metrics for quantitative evaluation of RSISR results, such as: ERGAS (Erreur Relative Globale Adimensionnelle de Synthese) \cite{wald2000quality}, Q-index \cite{wang2002universal}, QNR \cite{alparone2008multispectral}, VIF \cite{sheikh2006image}, FSIM \cite{zhang2011fsim}, spectral information divergence (SID) \cite{chang1999spectral}, correlation coefﬁcient (CC) \cite{1525860} and UIQI \cite{wang2002universal}, etc. Here, we do not introduce them in detail.

\section{Performance Evaluation Methods}
\label{sec7}
This section summarizes the performance evaluation methods of RSISR. Generally speaking, most studies combine visual perception and objective metrics to conduct qualitative and quantitative evaluations of RSISR methods.

\subsection{Qualitative Evaluation}
Different from natural images, downstream tasks of RSISR often involve land cover classification, object detection, and change detection, etc. Therefore, the restoration of texture details and the preservation of spatial structure are particularly important, which is mainly evaluated manually and visually. The most common qualitative evaluation methods include the following aspects:
\begin{itemize}
    \item \textit{Visual Comparison}: Super-resolution images are displayed side by side with original images, HR reference images, or other algorithm results to observe the reconstruction quality from the aspects of texture, edge sharpness, structural fidelity, etc.

    \item \textit{Expert visual evaluation}: Experts with remote sensing interpretation experience make comprehensive judgments from the perspectives of object recognition, structure preservation, and edge integrity. This method is particularly suitable for the identification analysis of objects such as roads, buildings, and farmland boundaries in HR images.

    \item \textit{Application-guided evaluation}: Super-resolution results participate in downstream tasks such as land cover classification, object detection, and change detection, etc. Furthermore, the visual quality and semantic preservation are indirectly evaluated through application performance. In addition, the analysis of some typical problems, such as color consistency, artifacts, and over-smoothing may also be involved. This method is highly subjective, and different people may have different standards when checking super-resolution images. To the best of our knowledge, there is no good solution for these issues currently.
\end{itemize}

\subsection{Quantitative Evaluation}
Quantitative evaluation methods objectively check the quality of reconstructed images through metric values, which provide a unified standard for performance comparison among different algorithms. Many evaluation metrics, such as PSNR, SSIM, and information entropy (EN), have been proposed. Specifically, PSNR measures the error between the reconstructed image and the reference image at the pixel level, while SSIM focuses more on the preservation of structural information, which is particularly suitable for the evaluation of highly structural targets such as building outlines and road networks in RSIs. In addition, EN is used to evaluate the richness of spatial information in super-resolution images.

On the other hand, some non-reference metrics, such as NIQE and PI, have also been proposed. They quantify the visual naturalness and quality degradation based on the statistical characteristics of the image, and are particularly suitable for real remote sensing scenes without HR ground truth. In current RSISR research, although quantitative evaluation provides a standardized and repeatable approach, it still faces many challenges. Specifically, different evaluation metrics and test data sets are often used in different studies, which leads to a lack of consistency in experimental settings and result interpretation among various methods, making fair comparisons across methods difficult. In addition, each metric can usually only evaluate one aspect of image quality. Therefore, building a comprehensive and unified evaluation framework has become one of the most critical issues in the field of RSISR.

\section{Recent Development Characteristics}
This section summarizes some new characteristics of the RSISR
ﬁeld shown in recent years.
\label{sec5}
\subsection{More and more types of deep learning models were being applied to RSISR}
In 2015, CNN was first applied to SR tasks, and the milestone work SRCNN \cite{7115171,10.1007/978} was proposed. Further, msiSRCNN \cite{Liebel2016Single-image} was proposed in 2016 for RSISR. Since then, various CNN-based RSISR methods \cite{lei2017super,gu2019deep,dong2019transferred} have emerged. In addition, GANs \cite{ma2018super} and Transformer \cite{ye2021super} were introduced into the RSISR field in 2018 and 2021, respectively. These architectural models have been widely studied and used in the field of RSISR. In 2022, diffusion models \cite{liu2022diffusion} were also introduced into RSISR, and many diffusion model-based RSISR methods have been proposed. Another new architecture is Mamba, which was introduced into RSISR in 2024. We anticipate that an increasing variety of deep learning models will be adopted for RSISR tasks in order to further enhance performance.

\subsection{Most RSISR methods are supervised}
Most RSISR methods involving deep learning are supervised, because RSI usually has the characteristics of high geometric alignment accuracy, which means that paired training samples can be easily constructed by downsampling. In addition, mainstream evaluation metrics such as PSNR and SSIM require ground truth (i.e., real HR image) as a reference, which further demonstrates the advantages of supervised learning. Although unsupervised and self-supervised methods have emerged in recent years, they still face great limitations in the field of remote sensing, such as poor sample diversity and a lack of transferable pre-training. Therefore, most RSISR methods still rely on the supervised paradigm to obtain stable reconstruction performance.

\subsection{RSISR methods for different types of images}
Most existing RSISR methods perform HR reconstruction on optical RSI, which is mainly collected from satellite images such as WorldView, Gaofen, and Sentinel-2. Recently, some researchers started developing methods to conduct SR on different types of RSI. However, it is not uncommon to perform SR on different types of RSI. For example, in 2012, He et al. \cite{he2012learning} proposed an SR method for SAR images. Specifically, they achieved SR reconstruction on a very small test set by combining multi-dictionary and sparse coding spatial pyramid machine training. Similar work includes RSMAP \cite{zhang2019sea} and Guo et al. \cite{guo2025structured}. 

Recently, many SR methods for HSI and MSI have been proposed. These images contain rich spectral information in a large number of bands and have broad research prospects. For example, from 2021 to 2024, many HSISR methods were proposed, covering Transformer-based \cite{hu2022fusformer,zhang2023essaformer}, Diffusion-based \cite{wu2023hsr,chen2024spectral}, and self-supervised learning methods \cite{qian2022selfs2,chen2021hyperspectral,xiao2023degrade}. This diversity of RSISR tasks reflects the characteristics of different types of images, such as noise in SAR, spectral fidelity in HSI, and temporal consistency in satellite images, which require customized solutions. However, it is desirable to develop a general SR framework for different types of RSI, which may face great challenges in model generalization, cross-modal knowledge transfer, and computational efficiency.

\subsection{Research on benchmarks}
Although RSISR has made significant progress in recent years, benchmark research for RSISR still faces challenges. Unlike many basic tasks in computer vision, there are few studies on RSISR benchmarks, and there is a lack of general and high-quality RSISR benchmark datasets, which makes model performance comparisons unstable. In addition, most existing benchmarks \cite{aybar2024comprehensive,wang2022remote} rely on idealized downsampling degradation models, which limit the real-world adaptability of the methods.

Until recently, some researchers have started to create benchmarks in the field of RSISR. For example, Kowaleczko et al. \cite{kowaleczko2023real} proposed the first benchmark dataset MuS2 based on real multi-temporal LR images and HR reference images, which breaks through the limitations of traditional simulated data and enhances the practicality of RSISR. In addition, Wang et al. \cite{10798467} constructed a real degradation benchmark dataset, RRSISR, based on spectral camera imaging to address the mismatch between simulated degradation and real scenes in RSISR, and proposed a reference table-based block exit (RPE) method to improve efficiency through dynamic calculation.

\subsection{Application-oriented RSISR methods}
Most existing RSISR methods do not consider downstream applications in the process of image reconstruction. Therefore, these RSISR methods generate visually pleasing reconstructed images, but this may be suboptimal for downstream tasks. This leads to a growing gap between metric-based performance (e.g., PSNR, SSIM) and actual utility in downstream applications. Bridging this requires RSISR models aligned with task-specific objectives, not just perceptual quality. 

In the past three years, some studies have started deeply coupling the SR process with specific remote sensing application tasks to achieve a paradigm shift from visual quality-oriented to application-oriented. For example, Zhang et al. \cite{zhang2023super} proposed an RSI reconstruction method based on SRGAN to construct a high-quality landslide training set from LR images. In addition, Kong et al. \cite{kong2023super} proposed a dual RSS-GAN method to integrate high-resolution Planet Fusion data to spatially enhance NDVI and NIRv, significantly improving the temporal expression and estimation accuracy of the surface index. Recently, Sun et al.  \cite{sun2024transformer} proposed the RSISR framework based on Transformer and self-supervision mechanism to effectively improve the performance of image interpretation.
Fan et al. \cite{fan2024rmsrgan} proposed a GAN-based RSISR method, which significantly improved the accuracy of vegetation index inversion and yield prediction by improving the spatial resolution of ginkgo tree RSIs. Compared with most RSISR methods, these methods directly consider the performance of downstream applications during the image reconstruction process, which provides reconstructed images that are more suitable for specific applications.

\subsection{Programming frameworks}
We conducted a systematic survey of existing RSISR methods and checked the programming frameworks used. The frequency of use of \textbf{PyTorch} has increased rapidly since 2017, becoming the mainstream framework, thanks to the flexibility of its dynamic graph mechanism, more convenient debugging, and an active open source community. In contrast, while TensorFlow's static computation graph demonstrated notable performance optimization in its early iterations, it has a high threshold for use in scientific research scenarios. MATLAB occasionally appears in some traditional image processing tasks, but is almost never used in deep learning RSISR methods.

\section{Future Prospects}
\label{sec8}
Although the RSISR field has made remarkable progress, significant challenges remain, especially in terms of real-world applications and performance evaluation. In response to the evolving needs of remote sensing technologies, this section expands on emerging directions and the underlying research gaps. Key areas of future development include real-world degradation modeling, the use of foundation models, multi-modal super-resolution (SR), and improvements in model efficiency and evaluation metrics. These emerging trends will define the trajectory of RSISR over the next decade.

\subsection{Real-World Degradation Modeling in RSISR}
Despite growing recognition of its importance, real-world degradation modeling in RSISR remains at an early stage. Traditional RSISR methods predominantly rely on synthetic degradation models, such as bicubic downsampling or noise addition. However, these simplistic approaches fail to capture the full spectrum of complex and compound distortions encountered in real-world remote sensing scenarios. Degradation processes in this domain typically stem from multi-source factors, including atmospheric scattering effects (e.g., haze, turbulence, simulated by atmospheric point spread functions - PSFs), sensor-specific imperfections (e.g., CCD/CMOS noise patterns, lens blur described by modulation transfer functions - MTFs), geometric distortions from platform instability, radiometric inconsistencies, temporal variations across acquisition cycles (e.g., seasonal illumination changes and phenological variations), and preprocessing artifacts. These distortions interact in non-trivial, often sensor-specific ways.

Consequently, most state-of-the-art RSISR models are trained on idealized datasets synthesized using these inadequate techniques, limiting their real-world applicability. A related work \cite{zhang2022single} in 2022 introduced the Residual Balanced Attention Network (RBAN), representing initial steps towards addressing this gap. Nevertheless, these efforts remain exploratory. A critical need persists to systematically develop comprehensive degradation models that explicitly incorporate: \textbf{(1) Physical Priors}: Integrating domain knowledge of atmospheric PSFs and sensor MTFs to ground the modeling in physics. \textbf{(2) Multi-Temporal Dynamics}: Accounting for seasonal illumination changes and phenological variations that alter scene characteristics over time. \textbf{(3) Cross-Sensor Adaptability:} Creating unified frameworks capable of handling the distinct characteristics and degradation profiles of optical, SAR, and hyperspectral sensors, which are the critical requirements for operational multi-source Earth observation.

Achieving this will likely involve leveraging multi-source satellite data with adversarial training to generate realistic degradation patterns, alongside unsupervised domain adaptation techniques to enhance robustness against sensor heterogeneity. Future research should prioritize both data-driven approaches, such as learning degradation priors from unpaired HR/LR image distributions, and physics-informed strategies, like modeling atmospheric scattering via radiative transfer simulations. By systematically bridging the simulation-to-reality gap through such advanced modeling, RSISR can achieve significantly improved generalization capabilities. This advancement is crucial for critical applications such as disaster response and precision agriculture.

\subsection{Foundation Models and Transfer Learning in RSISR}
The remarkable success of foundation models pre-trained on large-scale datasets, like natural language processing and computer vision, prompts a natural question: Can RSISR benefit from similar paradigms? The answer appears promising, albeit with significant challenges. Unlike general vision tasks, RSISR must contend with highly heterogeneous data characterized by variations across sensors, spectral ranges, acquisition conditions, and spatio-temporal resolutions. While foundation models pre-trained on massive natural image datasets (e.g., ImageNet) provide strong general features, they often lack sensitivity to the specific spectral and radiometric nuances inherent in remote sensing imagery.

A more effective approach may involve developing domain-adapted foundation models, pre-trained on large-scale satellite imagery corpora—potentially spanning multiple modalities (optical, SAR, hyperspectral). Subsequently, transfer learning becomes a powerful tool: fine-tuning such a domain-specific foundation model on downstream RSISR tasks (e.g., crop monitoring, disaster assessment) with limited supervision could yield robust, low-data solutions, thereby reducing training data requirements and enabling faster deployment across different domains. This approach allows RSISR models to leverage vast amounts of general computer vision knowledge while efficiently adapting to new sensor types and degradation conditions.

Furthermore, integrating explicit remote sensing priors into foundation architectures, such as atmospheric correction models, sensor-specific calibration, and radiation characteristics, may greatly improve both the accuracy and interpretability of the resulting super-resolved images, enhancing their applicability in real-world scenarios. The future lies in achieving high cross-task, cross-domain transferability, where a unified model trained for RSISR could also effectively support related tasks like change detection, land cover classification, and object recognition with minimal additional training.

\subsection{Multi-Modal RSISR Methods}
Traditional RSISR methods focus primarily on single-modality inputs, such as typically HR optical (RGB or panchromatic) and SAR imagery. However, remote sensing data is inherently multi-modal, with satellite platforms increasingly offering complementary observations including multispectral, hyperspectral, SAR, LiDAR, and thermal infrared data. These modalities are complementary: optical data provide rich spatial and spectral detail under clear skies, SAR enables all-weather, day-night imaging, and hyperspectral data reveal material-level spectral patterns.

Multi-modal SR techniques that combine information from these various sensors offer significant promise for enhancing image resolution and quality, particularly in scenarios where HR reference data for a single modality are unavailable or incomplete. For instance, LR hyperspectral data could be enhanced using HR panchromatic or multispectral guidance, combining fine spatial structure with detailed spectral signatures. Similarly, HR optical images could be fused with LR radar or thermal data to fill in missing details.

This avenue holds great potential for providing a more comprehensive view of the environment, crucial for applications like precision agriculture, infrastructure monitoring, urban planning, and climate science. However, significant challenges remain, such as the alignment and fusion of multi-modal data complicated by sensor misalignments, different geometries, spectral responses, spatial footprints, acquisition times, and temporal asynchrony.

Future research must therefore focus on addressing these technical hurdles, including cross-modal registration, domain alignment, and developing adaptive fusion strategies. Success may lie in techniques like transformer-based attention mechanisms that learn modality-specific contributions, ultimately enabling the full potential of multi-modal SR for unprecedented accuracy in remote sensing applications.

\subsection{Better Evaluation Metrics and Benchmarks}
Most RSISR methods are evaluated qualitatively by using visual performance and quantitatively by using evaluation metrics in existing literature. Current mainstream metrics, such as PSNR, SSIM, and LPIPS, remain overly simplistic and problematic. While PSNR and SSIM primarily assess pixel-level accuracy and are useful for basic comparisons, they fail to capture critical aspects essential for remote sensing, such as perceptual realism, application utility, structural and contextual fidelity, spectral consistency, and multi-modal alignment. LPIPS introduces a valuable perceptual assessment but still lacks interpretability in domain-specific tasks. Crucially, these metrics usually evaluate only one aspect of the super-resolved image, and the inconsistent use of different indicators across papers and diverse test datasets makes fair algorithm comparison difficult. Therefore, developing more comprehensive, robust, and standardized evaluation metrics is imperative. Ideal metrics should be consistent with visual performance, strongly correlated with downstream task performance, and capable of assessing multiple dimensions of quality, including spatial resolution, spectral fidelity, structural fidelity, and cross-modal consistency involving semantic alignment across modalities like SAR-optical fusion. Task-aware and perception-aware metrics, such as evaluating structural similarity in object space using object detection/classification accuracy, represent a promising future direction.

In parallel, the existing benchmarks for RSISR, though evolving, are still in their early stages and face significant limitations. Recent contributions like MuS2 \cite{kowaleczko2023real}, OpenSR-Test \cite{aybar2024comprehensive}, and RRSISR \cite{10798467} represent valuable progress in incorporating real data, addressing multi-temporal pairing, and refining evaluation mechanisms. However, existing RSISR benchmarks face significant limitations. The inherent heterogeneity of remote sensing data leads to inconsistencies in geometric and spectral properties, ground feature variations across time, and mismatches between quantitative metrics and human perception. For example, MuS2 suffers from diverse sensor characteristics, while OpenSR-Test, despite its focus on same-day imaging and manual cleaning, is constrained by the high cost of acquiring high-resolution references and precise registration. These factors reduce evaluation consistency and generalization, making it difficult to fairly compare algorithms. Additionally, current benchmarks struggle with data acquisition costs and lack scalability, limiting their applicability to large-scale, multi-task, and multi-source remote sensing scenarios.

Consequently, there is an urgent need for next-generation benchmarks that emphasize authenticity, universality, and task relevance. Such benchmarks should reflect the full complexity of operational environments by integrating multi-temporal, multi-spectral, and multi-sensor data, while explicitly modeling real-world challenges such as cloud cover, atmospheric interference, and sensor noise. Specifically, an ideal benchmarking suite should be modular and support: \textbf{(1) Multi-source and multi-temporal data integration}. \textbf{(2) Task-embedded evaluations} (e.g., linking SR quality to crop classification or damage detection).\textbf{(3) Realistic degradation conditions}.
Only by assessing models within such diverse, authentic, and task-aligned contexts using robust, multidimensional metrics can we meaningfully evaluate their practical value and push RSISR toward operational maturity.

\subsection{Diffusion-Mamba Based RSISR Methods}
Diffusion models and Mamba have demonstrated excellent performance in many computer vision tasks. However, the application of such models in RSISR is at a very early stage. The diffusion model generates high-fidelity details through iterative denoising, but its computational efficiency is low. In contrast, the Mamba model has high computational efficiency due to its linear complexity and long-range modeling ability, but it still needs to be optimized in recovering high-frequency details. In the future, we expect many RSISR methods based on the diffusion model and Mamba to emerge. Specifically, it is interesting to develop a hybrid architecture of diffusion and Mamba to synergize the advantages of both to improve application performance.

\subsection{Application Scenarios-Oriented RSISR Methods}
A primary motivation for RSISR is to enhance the performance of downstream practical applications. However, a critical limitation persists in current research: most studies overlook the deep-seated quality requirements imposed by specific downstream tasks. This is evident in the prevalent reliance on generic image quality metrics like PSNR and SSIM. These metrics often fail to correlate strongly with the ultimate performance goals of applications such as object detection, land cover classification, or change detection, potentially leading to suboptimal task performance. Consequently, future RSISR research must pivot towards an application-oriented paradigm, establishing it as the mainstream approach in the field.

The necessity for this application scenario-oriented research stems from the vast diversity encountered in real-world scenarios:

(1) \textit{Data Heterogeneity}: Current research is largely driven by benchmark datasets covering a limited range of scenarios. In contrast, practical RSISR applications involve diverse data from various sensors (optical, SAR, multispectral, hyperspectral), acquisition conditions (atmosphere, illumination, season), and temporal resolutions (daily, weekly, monthly).

(2) \textit{Scenario-Specific Requirements}: Different application scenarios impose fundamentally distinct quality requirements on super-resolved imagery. Understanding these requirements is critical for selecting or designing appropriate RSISR methodologies:
\begin{itemize}
    \item \textbf{Agricultural Monitoring}: Super-resolved images must not only achieve high spatial resolution but, critically, preserve spectral fidelity to accurately differentiate crops, weeds, soil conditions, and stress indicators.
    
    \textit{Method Implication: Incorporating spectral constraints (e.g., spectral loss functions, dedicated network branches) is recommended. Evaluation should emphasize spectral fidelity metrics such as the Spectral Angle Mapper (SAM) and downstream classification accuracy.} Typical RSISR methods include ViT-ISRGAN \cite{10836746}, JF-CNNSSR \cite{11045064}, NGSTGAN \cite{zhan2025ngstgan}, CSSF \cite{yin2025cssf} and SSU-Net \cite{zhang2025ssu}, etc.
    
    \item \textbf{Urban Planning}: Accurate reconstruction of fine structural details in buildings, roads, vehicles, and infrastructure is a primary objective. 
    
    \textit{Methodological Implication}: Techniques that enhance spatial detail fidelity and structural integrity, such as those employing perceptual or adversarial losses, or edge-aware mechanisms, are prioritized. Evaluation should incorporate perceptual quality metrics (e.g., LPIPS) and assess object detection and localization performance. Typical RSISR methods include EGSRN \cite{11006733}, EEGAN \cite{jiang2019edge}, CEEGAN \cite{ren2023context}, SeaNet \cite{fang2020soft} and SMSR \cite{9194276}, etc.
    
    \item \textbf{Disaster Management}: Temporal consistency across image sequences acquired at different times becomes paramount for reliably tracking event progression, such as flood extent or fire spread. 
    
    \textit{Method Implication}: Temporal SR or fusion techniques, potentially employing recurrent structures, optical flow, or temporal consistency losses, are crucial. Evaluation must assess temporal stability and change detection accuracy. Typical RSISR methods include ESRGAN \cite{Wang_2018_ECCV_Workshops}, MCW-ESRGAN \cite{karwowska2023mcwesrgan}, GRNN \cite{li2016improved}, FireSRnet \cite{ballard2020firesrnet} and SSR-GAN \cite{dubey2024ssr}, etc.
    
    \item \textbf{Military/Security Surveillance}: This domain necessitates exceptionally high spatial detail for accurate target identification and characterization, demanding precise edge definition and fine texture reproduction. Real-time processing capability may also be required.
    
    \textit{Methodological Implication}: Prioritize SR methods that enhance spatial sharpness (e.g., GAN-based approaches). If real-time operation is mandated, incorporate efficiency optimizations (e.g., lightweight models, acceleration techniques). Evaluation should emphasize target detection/recognition metrics and image interpretability. Typical RSISR methods include LRSD-ADMM-NET \cite{li2024lrsd}, Joint-SRVDNet \cite{mostofa2020joint}, MADNet \cite{lan2020madnet}, YOLOSR-IST \cite{li2023yolosr} and Liu et al. \cite{10663237}, etc.
    
    \item \textbf{Environmental Monitoring}: This application primarily involves tracking land cover, water quality, and ecosystem dynamics over time. It demands high spectral fidelity for accurate material discrimination alongside moderate spatial detail, with critical emphasis on temporal consistency for robust change analysis.
    
    \textit{Methodological Implication}: Techniques that maintain spectral-spatial balance while incorporating multi-temporal data and temporal constraints are optimal. Evaluation must integrate spectral metrics (e.g., SAM, RMSE), change detection accuracy (e.g., F1-score), and temporal consistency measures (e.g., temporal correlation index). Typical RSISR methods include Zhang et al. \cite{zhang2015improvement}, GSSR \cite{wang2023group}, CSRNet \cite{wu2024unsupervised} and STHNN \cite{li2014spatial}, etc.
\end{itemize}

\begin{table*}[t]
\centering
\caption{Recommended RSISR Method Considerations for Key Application Scenarios}
\label{tab:app_scenario_recs}
\renewcommand{\arraystretch}{1.2}
\setlength{\tabcolsep}{6pt}
\resizebox{0.96\textwidth}{!}{
\large
\begin{tabular}{|m{0.25\linewidth} 
|>{\centering\arraybackslash}m{0.3\linewidth} 
|>{\centering\arraybackslash}m{0.48\linewidth} 
|>{\centering\arraybackslash}m{0.48\linewidth}|}
\hline
Application Scenario & Core Quality Requirements & Recommended RSISR Methods & Key Evaluation Metrics \\ 
\hline

Agricultural Monitoring &  \begin{compactitemize}
    \item Spectral Fidelity
    \item Moderate Spatial Detail
\end{compactitemize} & \begin{compactitemize} 
\item \textbf{Explicit spectral constraints}: Spectral loss (SAM, Spectral MSE), physics-informed models, spectral attention.
  \item Prioritize spectral accuracy over sharpness. 
  \item Recommended methods: ViT-ISRGAN \cite{10836746}, JF-CNNSSR \cite{11045064}, NGSTGAN \cite{zhan2025ngstgan}, CSSF \cite{yin2025cssf} and SSU-Net \cite{zhang2025ssu}, etc.
  \end{compactitemize} 
  & \begin{compactitemize}
      \item \textit{Image Quality:} SAM, Spectral RMSE.
      \item \textit{Task:} Crop/weed/soil classification accuracy, stress detection precision.
  \end{compactitemize} \\
  \hline
Urban Planning & \begin{compactitemize}
  \item High Spatial Detail
  \item Structural Fidelity (edges, lines)
  \item Texture Clarity
\end{compactitemize} &
\begin{compactitemize}
  \item Enhancing spatial detail: Perceptual loss, adversarial training, edge-aware losses, attention mechanisms.
  \item Reconstructing sharp man-made structures.
  \item Recommended methods: EGSRN \cite{11006733}, EEGAN \cite{jiang2019edge}, CEEGAN \cite{ren2023context}, SeaNet \cite{fang2020soft} and SMSR \cite{9194276}, etc.
\end{compactitemize} &
\begin{compactitemize}
  \item \textit{Image Quality:} LPIPS, FID (optional), edge sharpness metrics.
  \item \textit{Task:} Building footprint accuracy, road network completeness, vehicle detection \& localization precision.
\end{compactitemize} \\ \hline
Disaster Management & \begin{compactitemize}
  \item Temporal Consistency
  \item Moderate Spatial Detail
  \item Fast Processing 
\end{compactitemize} &
\begin{compactitemize}
  \item \textbf{Temporal SR/Fusion}: Utilizing recurrent networks, optical flow, 3D convolutions, explicit temporal consistency losses.
  \item \textbf{Lightweight} architectures for rapid response.
  \item Recommended methods: ESRGAN \cite{Wang_2018_ECCV_Workshops}, MCW-ESRGAN \cite{karwowska2023mcwesrgan}, GRNN \cite{li2016improved}, FireSRnet \cite{ballard2020firesrnet} and SSR-GAN \cite{dubey2024ssr}, etc.
\end{compactitemize} &
\begin{compactitemize}
  \item \textit{Image Quality:} Temporal PSNR/SSIM, temporal stability index.
  \item \textit{Task:} Change detection accuracy (IoU, F1-score), flood/fire spread mapping precision, response time.
\end{compactitemize} \\ \hline
Military/Security Surveillance & \begin{compactitemize}
  \item Very High Spatial Detail (small targets)
  \item Edge Sharpness
  \item Texture Fidelity 
  \item Real-time (potential)
\end{compactitemize} &
\begin{compactitemize}
  \item High-performance spatial SR: GAN-based methods for sharpness, methods with strong edge/texture enhancement.
  \item \textbf{Lightweight} or accelerated models if real-time is needed. 
  \item Recommended methods: LRSD-ADMM-NET \cite{li2024lrsd}, Joint-SRVDNet \cite{mostofa2020joint}, MADNet \cite{lan2020madnet}, YOLOSR-IST \cite{li2023yolosr} and Liu et al. \cite{10663237}, etc.
\end{compactitemize} &
\begin{compactitemize}
  \item \textit{Image Quality:} High-resolution target visibility (subjective/interpretability scales), LPIPS. 
  \item \textit{Task:} Small target detection rate, target recognition/identification accuracy, inference speed.
\end{compactitemize} \\ \hline
Environmental Monitoring  & \begin{compactitemize}
  \item Spectral Fidelity 
  \item Temporal Consistency
  \item Moderate Spatial Detail
\end{compactitemize} &
\begin{compactitemize}
  \item Methods balancing \textbf{spectral accuracy} and \textbf{temporal stability}: Spectral constraints combined with temporal fusion/consistency mechanisms. 
  \item Robust to seasonal/illumination variations.
  \item Recommended methods: Zhang et al. \cite{zhang2015improvement}, GSSR \cite{wang2023group}, CSRNet \cite{wu2024unsupervised} and STHNN \cite{li2014spatial}, etc.
\end{compactitemize} &
\begin{compactitemize}
  \item \textit{Image Quality:} SAM, Temporal stability metrics.
  \item \textit{Task:} Land cover classification accuracy (long-term), change detection accuracy (slow changes like deforestation, erosion), thematic map quality. 
\end{compactitemize} \\ 
\Xhline{1.25pt}
\end{tabular}}
\end{table*}

Crucially, the ultimate evaluation must be rooted in the downstream application's performance metrics (e.g., detection rate, classification accuracy, change map quality), moving decisively beyond reliance on generic PSNR/SSIM scores. This application-driven evaluation should also guide the design of loss functions during model optimization. This inherent diversity in data types, application scenarios, and core quality requirements dictates that future RSISR methods must be fundamentally tailored: \textbf{(1) Application-Driven Degradation Modeling \& Evaluation}: Degradation models should be developed specifically for target application environments and sensor characteristics. Furthermore, evaluation must move beyond generic metrics (PSNR/SSIM) and incorporate task-embedded performance measures directly linked to the downstream application's objectives, such as object detectability, change detection accuracy, or classification precision. Optimization and loss functions should align with these end-goal metrics. \textbf{(2) Adaptive \& Lightweight Architectures}: Model architectures need to be lightweight for potential deployment on edge devices and adaptive to prioritize the specific quality attributes critical for a given application (e.g., spectral fidelity for agriculture, structural integrity for urban analysis, temporal stability for disaster monitoring). This ensures core requirements are met efficiently.

As a preliminary guidance, Table~\ref{tab:app_scenario_recs} provides an indicative mapping between representative remote sensing application scenarios and corresponding RSISR method characteristics. While not exhaustive, this mapping summarizes emerging consensus from existing literature and offers a conceptual bridge from application-driven demands to method-level decisions.

It is important to emphasize that these recommendations are indicative rather than prescriptive. The actual method choice should consider additional factors such as data availability, model complexity, hardware constraints, and system latency. Ultimately, the path forward for RSISR lies not merely in matching algorithms to applications post hoc, but in embedding downstream task objectives directly into the SR model training so that the super-resolution process becomes an integral component of the application pipeline rather than a preprocessing step.

In summary, the future of RSISR research lies in deeply understanding and responding to the unique demands of specific application scenarios. By advancing application-driven degradation modeling, adopting task-embedded evaluation frameworks, and designing lightweight adaptive architectures, robust RSISR methods can be developed. These methods will genuinely meet the diverse needs of real-world applications, delivering substantial improvements in the performance of these ultimate tasks.
\subsection{Computational Resources and Model Efficiency}
The efficiency of RSISR algorithms is increasingly critical as remote sensing datasets continue to grow in size, resolution, and complexity. Unlike general computer vision tasks, RSISR must handle large-scale multi-source data with high spatial resolution, multiple spectral bands, and dense temporal sequences. A single scene may span gigabytes of data, and tasks such as multi-modal fusion or long-term monitoring significantly increase computational demands. This poses clear challenges for both training and inference, especially in real-time or edge-based applications like UAV monitoring and disaster response.

To address these challenges, future RSISR methods should prioritize efficient model designs that reduce resource consumption without sacrificing performance. One promising direction involves lightweight architectures that integrate convolutional layers for local feature extraction with attention mechanisms (e.g., transformers or Mamba models) for global context modeling. These hybrid approaches enable better accuracy–efficiency trade-offs, especially when applied to large-scale or low-latency tasks. Another important direction is adaptive computation, where models dynamically adjust their inference complexity based on scene content or application requirements. Techniques like early-exit mechanisms, spatial pruning, or confidence-based sampling can significantly reduce redundant computation while maintaining output quality. For instance, simpler processing may suffice for homogeneous regions, while more complex areas receive deeper inference.

At the system level, distributed computing and cloud-based infrastructures offer scalable solutions for processing high-volume satellite data. Efficient workload partitioning across temporal or spatial segments enables parallelism during both training and deployment. Leveraging heterogeneous hardware platforms, such as GPU clusters or specialized accelerators, can further reduce latency and improve throughput, especially in large-scale or urgent-response scenarios.

Looking ahead, RSISR systems must be designed with end-to-end efficiency in mind, from sensor-level data handling to inference and downstream integration. By combining compact architectures, adaptive computation strategies, and scalable infrastructure, future models can achieve robust performance across diverse operational contexts, balancing precision with real-world constraints.

\section{Conclusion}
\label{sec9}
In this paper, a detailed and comprehensive review of RSISR methods is presented, covering methodology, datasets, and evaluation metrics. Specifically, we carefully group the existing RSISR methods into supervised and unsupervised. Furthermore, representative methods in each category are analyzed and discussed in detail. We also discuss the recent developments in this field based on a survey of more than 400 papers. Through these reviews and analyses, we provide future perspectives on some issues we consider critical. We expect this survey to provide a suitable reference for researchers in the field of RSISR.

\section*{CRediT authorship contribution statement}
\textbf{Yunliang Qi}: Writing - original draft, Visualization, Validation, Methodology, Investigation. \textbf{Meng Lou}: Writing – review \& editing, Visualization, Methodology. \textbf{Yimin Liu}: Supervision, Investigation.
\textbf{Lu Li}: Writing – review \& editing, Supervision, Investigation.
\textbf{Zhen Yang}: Supervision, Investigation, \textbf{Wen Nie}: Supervision, Investigation. 

\section*{Declaration of competing interest}
The authors declare that they have no known competing financial interests or personal relationships that could have appeared to influence the work reported in this paper.

\section*{Acknowledgements}
This work was jointly supported by the Natural Science Foundation of Gansu Province, China (No.22JR5RA492) and the Fundamental Research Funds for the Central Universities of China (No. lzujbky-2017-it72 and lzujbky-2022-pd12).

\bibliographystyle{unsrt}
\bibliography{cas-refs}

\begin{thebibliography}{100}

\bibitem{farsiu2004fast}
Sina Farsiu, M~Dirk Robinson, Michael Elad, and Peyman Milanfar.
\newblock Fast and robust multiframe super resolution.
\newblock {\em IEEE transactions on image processing}, 13(10):1327--1344, 2004.

\bibitem{wang2022comprehensive}
Peijuan Wang, Bulent Bayram, and Elif Sertel.
\newblock A comprehensive review on deep learning based remote sensing image super-resolution methods.
\newblock {\em Earth-Science Reviews}, 232:104110, 2022.

\bibitem{shi2013cardiac}
Wenzhe Shi, Jose Caballero, Christian Ledig, Xiahai Zhuang, Wenjia Bai, Kanwal Bhatia, Antonio M Simoes~Monteiro de~Marvao, Tim Dawes, Declan O’Regan, and Daniel Rueckert.
\newblock Cardiac image super-resolution with global correspondence using multi-atlas patchmatch.
\newblock In {\em Medical Image Computing and Computer-Assisted Intervention--MICCAI 2013: 16th International Conference, Nagoya, Japan, September 22-26, 2013, Proceedings, Part III 16}, pages 9--16. Springer, 2013.

\bibitem{zou2011very}
Wilman~WW Zou and Pong~C Yuen.
\newblock Very low resolution face recognition problem.
\newblock {\em IEEE Transactions on image processing}, 21(1):327--340, 2011.

\bibitem{xie2021super}
Jie Xie, Leyuan Fang, Bob Zhang, Jocelyn Chanussot, and Shutao Li.
\newblock Super resolution guided deep network for land cover classification from remote sensing images.
\newblock {\em IEEE Transactions on Geoscience and Remote Sensing}, 60:1--12, 2021.

\bibitem{jia2024enhancing}
Xiaofeng Jia, Xinyan Li, Zirui Wang, Zhen Hao, Dong Ren, Hui Liu, Yun Du, and Feng Ling.
\newblock Enhancing cropland mapping with spatial super-resolution reconstruction by optimizing training samples for image super-resolution models.
\newblock {\em Remote Sensing}, 16(24):4678, 2024.

\bibitem{fu2022toward}
Xuanchao Fu, Toru Kouyama, Hang Yang, Ryosuke Nakamura, and Ichiro Yoshikawa.
\newblock Toward faster and accurate post-disaster damage assessment: Development of end-to-end building damage detection framework with super-resolution architecture.
\newblock In {\em IGARSS 2022-2022 IEEE International Geoscience and Remote Sensing Symposium}, pages 1588--1591. IEEE, 2022.

\bibitem{li2016improved}
Linyi Li, Tingbao Xu, and Yun Chen.
\newblock Improved urban flooding mapping from remote sensing images using generalized regression neural network-based super-resolution algorithm.
\newblock {\em Remote Sensing}, 8(8):625, 2016.

\bibitem{Fernandez2017single}
Ruben Fernandez-Beltran, Pedro Latorre-Carmona, and Filiberto Pla.
\newblock Single-frame super-resolution in remote sensing: a practical overview.
\newblock {\em International Journal of Remote Sensing.}, 38:314–354, 2017.

\bibitem{7284770}
Laetitia Loncan, Luis~B. de~Almeida, Jose~M. Bioucas-Dias, Xavier Briottet, Jocelyn Chanussot, Nicolas Dobigeon, Sophie Fabre, Wenzhi Liao, Giorgio~A. Licciardi, Miguel Simões, Jean-Yves Tourneret, Miguel~Angel Veganzones, Gemine Vivone, Qi~Wei, and Naoto Yokoya.
\newblock Hyperspectral pansharpening: A review.
\newblock {\em IEEE Geoscience and Remote Sensing Magazine}, 3(3):27--46, 2015.

\bibitem{chavez2014super}
Herminio Chavez-Roman and Volodymyr Ponomaryov.
\newblock Super resolution image generation using wavelet domain interpolation with edge extraction via a sparse representation.
\newblock {\em IEEE Geoscience and remote sensing Letters}, 11(10):1777--1781, 2014.

\bibitem{demirel2011discrete}
Hasan Demirel and Gholamreza Anbarjafari.
\newblock Discrete wavelet transform-based satellite image resolution enhancement.
\newblock {\em IEEE transactions on geoscience and remote sensing}, 49(6):1997--2004, 2011.

\bibitem{wang2021channel}
Peijuan Wang and Elif Sertel.
\newblock Channel--spatial attention-based pan-sharpening of very high-resolution satellite images.
\newblock {\em Knowledge-Based Systems}, 229:107324, 2021.

\bibitem{yang2015remote}
Daiqin Yang, Zimeng Li, Yatong Xia, and Zhenzhong Chen.
\newblock Remote sensing image super-resolution: Challenges and approaches.
\newblock In {\em 2015 IEEE international conference on digital signal processing (DSP)}, pages 196--200. IEEE, 2015.

\bibitem{ling2013interpolation}
Feng Ling, Yun Du, Xiaodong Li, Wenbo Li, Fei Xiao, and Yihang Zhang.
\newblock Interpolation-based super-resolution land cover mapping.
\newblock {\em Remote sensing letters}, 4(7):629--638, 2013.

\bibitem{ma2019achieving}
Wen Ma, Zongxu Pan, Jiayi Guo, and Bin Lei.
\newblock Achieving super-resolution remote sensing images via the wavelet transform combined with the recursive res-net.
\newblock {\em IEEE Transactions on Geoscience and Remote Sensing}, 57(6):3512--3527, 2019.

\bibitem{wang2023multi}
Zheng Wang, Yanwei Zhao, and Jiacheng Chen.
\newblock Multi-scale fast fourier transform based attention network for remote-sensing image super-resolution.
\newblock {\em IEEE Journal of Selected Topics in Applied Earth Observations and Remote Sensing}, 16:2728--2740, 2023.

\bibitem{shao2019remote}
Zhenfeng Shao, Lei Wang, Zhongyuan Wang, and Juan Deng.
\newblock Remote sensing image super-resolution using sparse representation and coupled sparse autoencoder.
\newblock {\em IEEE Journal of Selected Topics in Applied Earth Observations and Remote Sensing}, 12(8):2663--2674, 2019.

\bibitem{yang2019multi}
Jingxiang Yang, Yong-Qiang Zhao, Jonathan Cheung-Wai Chan, and Liang Xiao.
\newblock A multi-scale wavelet 3d-cnn for hyperspectral image super-resolution.
\newblock {\em Remote sensing}, 11(13):1557, 2019.

\bibitem{he2012learning}
Chu He, Longzhu Liu, Lianyu Xu, Ming Liu, and Mingsheng Liao.
\newblock Learning based compressed sensing for sar image super-resolution.
\newblock {\em IEEE Journal of Selected Topics in Applied Earth Observations and Remote Sensing}, 5(4):1272--1281, 2012.

\bibitem{dong2016hyperspectral}
Weisheng Dong, Fazuo Fu, Guangming Shi, Xun Cao, Jinjian Wu, Guangyu Li, and Xin Li.
\newblock Hyperspectral image super-resolution via non-negative structured sparse representation.
\newblock {\em IEEE Transactions on Image Processing}, 25(5):2337--2352, 2016.

\bibitem{wu2016new}
Wei Wu, Xiaomin Yang, Kai Liu, Yiguang Liu, Binyu Yan, et~al.
\newblock A new framework for remote sensing image super-resolution: sparse representation-based method by processing dictionaries with multi-type features.
\newblock {\em Journal of Systems Architecture}, 64:63--75, 2016.

\bibitem{tuna2018single}
Caglayan Tuna, Gozde Unal, and Elif Sertel.
\newblock Single-frame super resolution of remote-sensing images by convolutional neural networks.
\newblock {\em International journal of remote sensing}, 39(8):2463--2479, 2018.

\bibitem{gargiulo2019advances}
Massimiliano Gargiulo.
\newblock Advances on cnn-based super-resolution of sentinel-2 images.
\newblock In {\em IGARSS 2019-2019 IEEE International Geoscience and Remote Sensing Symposium}, pages 3165--3168. IEEE, 2019.

\bibitem{wang2023review}
Xuan Wang, Lijun Sun, Abdellah Chehri, and Yongchao Song.
\newblock A review of gan-based super-resolution reconstruction for optical remote sensing images.
\newblock {\em Remote Sensing}, 15(20):5062, 2023.

\bibitem{jia2022multiattention}
Sen Jia, Zhihao Wang, Qingquan Li, Xiuping Jia, and Meng Xu.
\newblock Multiattention generative adversarial network for remote sensing image super-resolution.
\newblock {\em IEEE Transactions on Geoscience and Remote Sensing}, 60:1--15, 2022.

\bibitem{tu2024rgtgan}
Ziming Tu, Xiubin Yang, Xi~He, Jiapu Yan, and Tingting Xu.
\newblock Rgtgan: Reference-based gradient-assisted texture-enhancement gan for remote sensing super-resolution.
\newblock {\em IEEE Transactions on Geoscience and Remote Sensing}, 2024.

\bibitem{kang2024efficient}
Xudong Kang, Puhong Duan, Jier Li, and Shutao Li.
\newblock Efficient swin transformer for remote sensing image super-resolution.
\newblock {\em IEEE Transactions on Image Processing}, 2024.

\bibitem{xiao2024ttst}
Yi~Xiao, Qiangqiang Yuan, Kui Jiang, Jiang He, Chia-Wen Lin, and Liangpei Zhang.
\newblock Ttst: A top-k token selective transformer for remote sensing image super-resolution.
\newblock {\em IEEE Transactions on Image Processing}, 2024.

\bibitem{xiao2024remote}
Yi~Xiao and Qiangqiang Yuan.
\newblock Remote sensing image super-resolution with top-k token selective transformer.
\newblock In {\em IGARSS 2024-2024 IEEE International Geoscience and Remote Sensing Symposium}, pages 3159--3162. IEEE, 2024.

\bibitem{zhi2024mambaformersr}
Ruicong Zhi, Xiaopei Fan, and Jingye Shi.
\newblock Mambaformersr: A lightweight model for remote-sensing image super-resolution.
\newblock {\em IEEE Geoscience and Remote Sensing Letters}, 2024.

\bibitem{7115171}
Chao Dong, Chen~Change Loy, Kaiming He, and Xiaoou Tang.
\newblock Image super-resolution using deep convolutional networks.
\newblock {\em IEEE Transactions on Pattern Analysis and Machine Intelligence}, 38(2):295--307, 2016.

\bibitem{10.1007/978}
Chao Dong, Chen~Change Loy, Kaiming He, and Xiaoou Tang.
\newblock Learning a deep convolutional network for image super-resolution.
\newblock In {\em Computer Vision -- ECCV 2014}, pages 184--199, Cham, 2014. Springer International Publishing.

\bibitem{liu2021research}
Hui Liu, Yurong Qian, Xiwu Zhong, Long Chen, and Guangqi Yang.
\newblock Research on super-resolution reconstruction of remote sensing images: A comprehensive review.
\newblock {\em Optical Engineering}, 60(10):100901--100901, 2021.

\bibitem{wang2022review}
Xuan Wang, Jinglei Yi, Jian Guo, Yongchao Song, Jun Lyu, Jindong Xu, Weiqing Yan, Jindong Zhao, Qing Cai, and Haigen Min.
\newblock A review of image super-resolution approaches based on deep learning and applications in remote sensing.
\newblock {\em Remote Sensing}, 14(21):5423, 2022.

\bibitem{karwowska2022using}
Kinga Karwowska and Damian Wierzbicki.
\newblock Using super-resolution algorithms for small satellite imagery: A systematic review.
\newblock {\em IEEE Journal of Selected Topics in Applied Earth Observations and Remote Sensing}, 15:3292--3312, 2022.

\bibitem{lepcha2023image}
Dawa~Chyophel Lepcha, Bhawna Goyal, Ayush Dogra, and Vishal Goyal.
\newblock Image super-resolution: A comprehensive review, recent trends, challenges and applications.
\newblock {\em Information Fusion}, 91:230--260, 2023.

\bibitem{chen2023review}
Chi Chen, Yongcheng Wang, Ning Zhang, Yuxi Zhang, and Zhikang Zhao.
\newblock A review of hyperspectral image super-resolution based on deep learning.
\newblock {\em Remote Sensing}, 15(11):2853, 2023.

\bibitem{al2024single}
Hanadi Al-Mekhlafi and Shiguang Liu.
\newblock Single image super-resolution: a comprehensive review and recent insight.
\newblock {\em Frontiers of Computer Science}, 18(1):181702, 2024.

\bibitem{qian2022selfs2}
Xiao Qian, Tai-Xiang Jiang, and Xi-Le Zhao.
\newblock Selfs2: Self-supervised transfer learning for sentinel-2 multispectral image super-resolution.
\newblock {\em IEEE Journal of Selected Topics in Applied Earth Observations and Remote Sensing}, 16:215--227, 2022.

\bibitem{liu2024spectral}
Jianjun Liu, Zebin Wu, and Liang Xiao.
\newblock A spectral diffusion prior for unsupervised hyperspectral image super-resolution.
\newblock {\em IEEE Transactions on Geoscience and Remote Sensing}, 2024.

\bibitem{he2025pan}
Xuanhua He, Ke~Cao, Jie Zhang, Keyu Yan, Yingying Wang, Rui Li, Chengjun Xie, Danfeng Hong, and Man Zhou.
\newblock Pan-mamba: Effective pan-sharpening with state space model.
\newblock {\em Information Fusion}, 115:102779, 2025.

\bibitem{chan2009neighbor}
Tak-Ming Chan, Junping Zhang, Jian Pu, and Hua Huang.
\newblock Neighbor embedding based super-resolution algorithm through edge detection and feature selection.
\newblock {\em Pattern Recognition Letters}, 30(5):494--502, 2009.

\bibitem{zhang2011super}
Zhaocai Zhang, Xiangjun Wang, Jinju Ma, and Guimin Jia.
\newblock Super resolution reconstruction of three view remote sensing images based on global weighted pocs algorithm.
\newblock In {\em 2011 International Conference on Remote Sensing, Environment and Transportation Engineering}, pages 3615--3618. IEEE, 2011.

\bibitem{zhou2012interpolation}
Fei Zhou, Wenming Yang, and Qingmin Liao.
\newblock Interpolation-based image super-resolution using multisurface fitting.
\newblock {\em IEEE Transactions on Image Processing}, 21(7):3312--3318, 2012.

\bibitem{timofte2013anchored}
Radu Timofte, Vincent De~Smet, and Luc Van~Gool.
\newblock Anchored neighborhood regression for fast example-based super-resolution.
\newblock In {\em Proceedings of the IEEE international conference on computer vision}, pages 1920--1927, 2013.

\bibitem{6805627}
Shuyuan Yang, Zhiyi Wang, Liao Zhang, and Min Wang.
\newblock Dual-geometric neighbor embedding for image super resolution with sparse tensor.
\newblock {\em IEEE Transactions on Image Processing}, 23(7):2793--2803, 2014.

\bibitem{ma2014robust}
Jianglin Ma, Jonathan Cheung-Wai Chan, and Frank Canters.
\newblock Robust locally weighted regression for superresolution enhancement of multi-angle remote sensing imagery.
\newblock {\em IEEE journal of selected topics in applied earth observations and remote sensing}, 7(4):1357--1371, 2014.

\bibitem{zhang2014example}
Yihang Zhang, Yun Du, Feng Ling, Shiming Fang, and Xiaodong Li.
\newblock Example-based super-resolution land cover mapping using support vector regression.
\newblock {\em IEEE Journal of Selected Topics in Applied Earth Observations and Remote Sensing}, 7(4):1271--1283, 2014.

\bibitem{dai2015jointly}
Dengxin Dai, Radu Timofte, and Luc Van~Gool.
\newblock Jointly optimized regressors for image super-resolution.
\newblock In {\em Computer Graphics Forum}, volume~34, pages 95--104. Wiley Online Library, 2015.

\bibitem{xinlei2016super}
Wang Xinlei and Liu Naifeng.
\newblock Super-resolution of remote sensing images via sparse structural manifold embedding.
\newblock {\em Neurocomputing}, 173:1402--1411, 2016.

\bibitem{rs8080625}
Linyi Li, Tingbao Xu, and Yun Chen.
\newblock Improved urban flooding mapping from remote sensing images using generalized regression neural network-based super-resolution algorithm.
\newblock {\em Remote Sensing}, 8(8), 2016.

\bibitem{liu2016improved}
Jinsong Liu, Shaosheng Dai, Zhongyuan Guo, and Dezhou Zhang.
\newblock An improved pocs super-resolution infrared image reconstruction algorithm based on visual mechanism.
\newblock {\em Infrared Physics \& Technology}, 78:92--98, 2016.

\bibitem{lei2017super}
Sen Lei, Zhenwei Shi, and Zhengxia Zou.
\newblock Super-resolution for remote sensing images via local--global combined network.
\newblock {\em IEEE Geoscience and Remote Sensing Letters}, 14(8):1243--1247, 2017.

\bibitem{haut2018new}
Juan~Mario Haut, Ruben Fernandez-Beltran, Mercedes~E Paoletti, Javier Plaza, Antonio Plaza, and Filiberto Pla.
\newblock A new deep generative network for unsupervised remote sensing single-image super-resolution.
\newblock {\em IEEE Transactions on Geoscience and Remote sensing}, 56(11):6792--6810, 2018.

\bibitem{solanki2018efficient}
Pooja Solanki, Dippal Israni, and Arpita Shah.
\newblock An efficient satellite image super resolution technique for shift-variant images using improved new edge directed interpolation.
\newblock {\em Statistics, Optimization \& Information Computing}, 6(4):619--632, 2018.

\bibitem{irmak2018map}
Hasan Irmak, Gozde~Bozdagi Akar, and Seniha~Esen Yuksel.
\newblock A map-based approach for hyperspectral imagery super-resolution.
\newblock {\em IEEE Transactions on Image Processing}, 27(6):2942--2951, 2018.

\bibitem{nayak2018enhanced}
Rajashree Nayak and Dipti Patra.
\newblock Enhanced iterative back-projection based super-resolution reconstruction of digital images.
\newblock {\em Arabian Journal for Science and Engineering}, 43(12):7521--7547, 2018.

\bibitem{jiang2019edge}
Kui Jiang, Zhongyuan Wang, Peng Yi, Guangcheng Wang, Tao Lu, and Junjun Jiang.
\newblock Edge-enhanced gan for remote sensing image superresolution.
\newblock {\em IEEE Transactions on Geoscience and Remote Sensing}, 57(8):5799--5812, 2019.

\bibitem{li2019enhanced}
Linyi Li, Yun Chen, Tingbao Xu, Kaifang Shi, Chang Huang, Rui Liu, Binbin Lu, and Lingkui Meng.
\newblock Enhanced super-resolution mapping of urban floods based on the fusion of support vector machine and general regression neural network.
\newblock {\em IEEE Geoscience and Remote Sensing Letters}, 16(8):1269--1273, 2019.

\bibitem{gu2019deep}
Jun Gu, Xian Sun, Yue Zhang, Kun Fu, and Lei Wang.
\newblock Deep residual squeeze and excitation network for remote sensing image super-resolution.
\newblock {\em Remote Sensing}, 11(15):1817, 2019.

\bibitem{haut2019remote}
Juan~Mario Haut, Ruben Fernandez-Beltran, Mercedes~E Paoletti, Javier Plaza, and Antonio Plaza.
\newblock Remote sensing image superresolution using deep residual channel attention.
\newblock {\em IEEE Transactions on Geoscience and Remote Sensing}, 57(11):9277--9289, 2019.

\bibitem{dong2019transferred}
Xiaoyu Dong, Zhihong Xi, Xu~Sun, and Lianru Gao.
\newblock Transferred multi-perception attention networks for remote sensing image super-resolution.
\newblock {\em Remote Sensing}, 11(23):2857, 2019.

\bibitem{zhang2019sea}
Yin Zhang, Qiping Zhang, Changlin Li, Yongchao Zhang, Yulin Huang, and Jianyu Yang.
\newblock Sea-surface target angular superresolution in forward-looking radar imaging based on maximum a posteriori algorithm.
\newblock {\em IEEE Journal of Selected Topics in Applied Earth Observations and Remote Sensing}, 12(8):2822--2834, 2019.

\bibitem{li2020fused}
Xinyao Li, Dongyang Zhang, Zhenwen Liang, Deqiang Ouyang, and Jie Shao.
\newblock Fused recurrent network via channel attention for remote sensing satellite image super-resolution.
\newblock In {\em 2020 IEEE international conference on multimedia and expo (ICME)}, pages 1--6. IEEE, 2020.

\bibitem{zhang2020scene}
Shu Zhang, Qiangqiang Yuan, Jie Li, Jing Sun, and Xuguo Zhang.
\newblock Scene-adaptive remote sensing image super-resolution using a multiscale attention network.
\newblock {\em IEEE Transactions on Geoscience and Remote Sensing}, 58(7):4764--4779, 2020.

\bibitem{wang2020non}
Huan Wang, Qian Hu, Chengdong Wu, Jianning Chi, and Xiaosheng Yu.
\newblock Non-locally up-down convolutional attention network for remote sensing image super-resolution.
\newblock {\em IEEE Access}, 8:166304--166319, 2020.

\bibitem{lei2019coupled}
Sen Lei, Zhenwei Shi, and Zhengxia Zou.
\newblock Coupled adversarial training for remote sensing image super-resolution.
\newblock {\em IEEE Transactions on Geoscience and Remote Sensing}, 58(5):3633--3643, 2019.

\bibitem{sheikholeslami2020efficient}
Mohammad~Moein Sheikholeslami, Saeed Nadi, Amin~Alizadeh Naeini, and Pedram Ghamisi.
\newblock An efficient deep unsupervised superresolution model for remote sensing images.
\newblock {\em IEEE Journal of Selected Topics in Applied Earth Observations and Remote Sensing}, 13:1937--1945, 2020.

\bibitem{dong2020remote}
Xiaoyu Dong, Xu~Sun, Xiuping Jia, Zhihong Xi, Lianru Gao, and Bing Zhang.
\newblock Remote sensing image super-resolution using novel dense-sampling networks.
\newblock {\em IEEE Transactions on Geoscience and Remote Sensing}, 59(2):1618--1633, 2020.

\bibitem{guo2021remote}
Dongen Guo, Ying Xia, Liming Xu, Weisheng Li, and Xiaobo Luo.
\newblock Remote sensing image super-resolution using cascade generative adversarial nets.
\newblock {\em Neurocomputing}, 443:117--130, 2021.

\bibitem{wang2021unsupervised}
Jiaming Wang, Zhenfeng Shao, Tao Lu, Xiao Huang, Ruiqian Zhang, and Yu~Wang.
\newblock Unsupervised remoting sensing super-resolution via migration image prior.
\newblock In {\em 2021 IEEE International Conference on Multimedia and Expo (ICME)}, pages 1--6. IEEE, 2021.

\bibitem{zhang2020nonpairwise}
Haopeng Zhang, Pengrui Wang, and Zhiguo Jiang.
\newblock Nonpairwise-trained cycle convolutional neural network for single remote sensing image super-resolution.
\newblock {\em IEEE Transactions on Geoscience and Remote Sensing}, 59(5):4250--4261, 2020.

\bibitem{wang2021enhanced}
Jiaming Wang, Zhenfeng Shao, Xiao Huang, Tao Lu, Ruiqian Zhang, and Jiayi Ma.
\newblock Enhanced image prior for unsupervised remoting sensing super-resolution.
\newblock {\em Neural Networks}, 143:400--412, 2021.

\bibitem{chen2021hyperspectral}
Wenjing Chen, Xiangtao Zheng, and Xiaoqiang Lu.
\newblock Hyperspectral image super-resolution with self-supervised spectral-spatial residual network.
\newblock {\em Remote Sensing}, 13(7):1260, 2021.

\bibitem{wang2021improved}
Yulei Wang, Xinxin He, Yao Shi, Qingyu Zhu, and Haoyang Yu.
\newblock An improved hyperspectral image super resolution restoration algorithm based on pocs.
\newblock In {\em 2021 IEEE International Geoscience and Remote Sensing Symposium IGARSS}, pages 2460--2463. IEEE, 2021.

\bibitem{wang2022fenet}
Zheyuan Wang, Liangliang Li, Yuan Xue, Chenchen Jiang, Jiawen Wang, Kaipeng Sun, and Hongbing Ma.
\newblock Fenet: Feature enhancement network for lightweight remote-sensing image super-resolution.
\newblock {\em IEEE Transactions on Geoscience and Remote Sensing}, 60:1--12, 2022.

\bibitem{zhang2022single}
Jizhou Zhang, Tingfa Xu, Jianan Li, Shenwang Jiang, and Yuhan Zhang.
\newblock Single-image super resolution of remote sensing images with real-world degradation modeling.
\newblock {\em Remote Sensing}, 14(12):2895, 2022.

\bibitem{hu2022fusformer}
Jin-Fan Hu, Ting-Zhu Huang, Liang-Jian Deng, Hong-Xia Dou, Danfeng Hong, and Gemine Vivone.
\newblock Fusformer: A transformer-based fusion network for hyperspectral image super-resolution.
\newblock {\em IEEE Geoscience and Remote Sensing Letters}, 19:1--5, 2022.

\bibitem{lei2021transformer}
Sen Lei, Zhenwei Shi, and Wenjing Mo.
\newblock Transformer-based multistage enhancement for remote sensing image super-resolution.
\newblock {\em IEEE Transactions on Geoscience and Remote Sensing}, 60:1--11, 2021.

\bibitem{mutai2022cubic}
Victor~Kipkoech Mutai, Elijah Mwangi, and Wa~Maina Ciira.
\newblock A cubic b-splines approximation method combined with dwt and ibp for single image super-resolution.
\newblock 2022.

\bibitem{ye2022bayesian}
Fei Ye, Zebin Wu, Yang Xu, Hongyi Liu, and Zhihui Wei.
\newblock Bayesian hyperspectral image super-resolution in the presence of spectral variability.
\newblock {\em IEEE Transactions on Geoscience and Remote Sensing}, 60:1--13, 2022.

\bibitem{deka2023joint}
Bhabesh Deka, Helal~Uddin Mullah, Trishna Barman, and Sumit Datta.
\newblock Joint sparse representation-based single image super-resolution for remote sensing applications.
\newblock {\em IEEE Journal of Selected Topics in Applied Earth Observations and Remote Sensing}, 16:2352--2365, 2023.

\bibitem{wang2023hybrid}
Jiarui Wang, Binglu Wang, Xiaoxu Wang, Yongqiang Zhao, and Teng Long.
\newblock Hybrid attention-based u-shaped network for remote sensing image super-resolution.
\newblock {\em IEEE Transactions on Geoscience and Remote Sensing}, 61:1--15, 2023.

\bibitem{wu2023hsr}
Chanyue Wu, Dong Wang, Yunpeng Bai, Hanyu Mao, Ying Li, and Qiang Shen.
\newblock Hsr-diff: Hyperspectral image super-resolution via conditional diffusion models.
\newblock In {\em Proceedings of the IEEE/CVF International Conference on Computer Vision}, pages 7083--7093, 2023.

\bibitem{zhang2023essaformer}
Mingjin Zhang, Chi Zhang, Qiming Zhang, Jie Guo, Xinbo Gao, and Jing Zhang.
\newblock Essaformer: Efficient transformer for hyperspectral image super-resolution.
\newblock In {\em Proceedings of the IEEE/CVF International Conference on Computer Vision}, pages 23073--23084, 2023.

\bibitem{tao2023fssbp}
Jingzhe Tao, Weihan Ni, Chuanming Song, and Xianghai Wang.
\newblock Fssbp: Fast spatial--spectral back projection based on pan-sharpening iterative optimization.
\newblock {\em Remote Sensing}, 15(18):4543, 2023.

\bibitem{liu2023ran}
Baodi Liu, Lifei Zhao, Shuai Shao, Weifeng Liu, Dapeng Tao, Weijia Cao, and Yicong Zhou.
\newblock Ran: Region-aware network for remote sensing image super-resolution.
\newblock {\em IEEE Transactions on Geoscience and Remote Sensing}, 61:1--13, 2023.

\bibitem{mishra2023clsr}
Divya Mishra and Ofer Hadar.
\newblock Clsr: Contrastive learning for semi-supervised remote sensing image super-resolution.
\newblock {\em IEEE Geoscience and Remote Sensing Letters}, 20:1--5, 2023.

\bibitem{xiao2023degrade}
Yi~Xiao, Qiangqiang Yuan, Kui Jiang, Jiang He, Yuan Wang, and Liangpei Zhang.
\newblock From degrade to upgrade: Learning a self-supervised degradation guided adaptive network for blind remote sensing image super-resolution.
\newblock {\em Information Fusion}, 96:297--311, 2023.

\bibitem{mishra2023self}
Divya Mishra and Ofer Hadar.
\newblock Self-fusenet: Data free unsupervised remote sensing image super-resolution.
\newblock {\em IEEE Journal of Selected Topics in Applied Earth Observations and Remote Sensing}, 16:1710--1727, 2023.

\bibitem{cha2023meta}
Zhangzhao Cha, Dongmei Xu, Yi~Tang, and Zuo Jiang.
\newblock Meta-learning for zero-shot remote sensing image super-resolution.
\newblock {\em Mathematics}, 11(7):1653, 2023.

\bibitem{wang2024two}
Jiarui Wang, Yuting Lu, Shunzhou Wang, Binglu Wang, Xiaoxu Wang, and Teng Long.
\newblock Two-stage spatial-frequency joint learning for large-factor remote sensing image super-resolution.
\newblock {\em IEEE Transactions on Geoscience and Remote Sensing}, 62:1--13, 2024.

\bibitem{chen2024spectral}
Bowen Chen, Liqin Liu, Chenyang Liu, Zhengxia Zou, and Zhenwei Shi.
\newblock Spectral-cascaded diffusion model for remote sensing image spectral super-resolution.
\newblock {\em IEEE Transactions on Geoscience and Remote Sensing}, 2024.

\bibitem{meng2024conditional}
Fanen Meng, Yijun Chen, Haoyu Jing, Laifu Zhang, Yiming Yan, Yingchao Ren, Sensen Wu, Tian Feng, Renyi Liu, and Zhenhong Du.
\newblock A conditional diffusion model with fast sampling strategy for remote sensing image super-resolution.
\newblock {\em IEEE Transactions on Geoscience and Remote Sensing}, 2024.

\bibitem{an2023efficient}
Tai An, Bin Xue, Chunlei Huo, Shiming Xiang, and Chunhong Pan.
\newblock Efficient remote sensing image super-resolution via lightweight diffusion models.
\newblock {\em IEEE Geoscience and Remote Sensing Letters}, 21:1--5, 2023.

\bibitem{zhong2024ssdiff}
Yu~Zhong, Xiao Wu, Zihan Cao, Hong-Xia Dou, and Liang-Jian Deng.
\newblock Ssdiff: Spatial-spectral integrated diffusion model for remote sensing pansharpening.
\newblock {\em Advances in Neural Information Processing Systems}, 37:77962--77986, 2024.

\bibitem{lu2024effective}
Xiangyu Lu, Jianlin Zhang, Rui Yang, Qina Yang, Mengyuan Chen, Hongxing Xu, Pinjun Wan, Jiawen Guo, and Fei Liu.
\newblock Effective variance attention-enhanced diffusion model for crop field aerial image super resolution.
\newblock {\em ISPRS Journal of Photogrammetry and Remote Sensing}, 218:50--68, 2024.

\bibitem{10353979}
Yi~Xiao, Qiangqiang Yuan, Kui Jiang, Jiang He, Xianyu Jin, and Liangpei Zhang.
\newblock Ediffsr: An efficient diffusion probabilistic model for remote sensing image super-resolution.
\newblock {\em IEEE Transactions on Geoscience and Remote Sensing}, 62:1--14, 2024.

\bibitem{10375518}
Fanen Meng, Sensen Wu, Yadong Li, Zhe Zhang, Tian Feng, Renyi Liu, and Zhenhong Du.
\newblock Single remote sensing image super-resolution via a generative adversarial network with stratified dense sampling and chain training.
\newblock {\em IEEE Transactions on Geoscience and Remote Sensing}, 62:1--22, 2024.

\bibitem{10746331}
Xudong Kang, Puhong Duan, Jier Li, and Shutao Li.
\newblock Efficient swin transformer for remote sensing image super-resolution.
\newblock {\em IEEE Transactions on Image Processing}, 33:6367--6379, 2024.

\bibitem{jiao2024symswin}
Dian Jiao, Nan Su, Yiming Yan, Ying Liang, Shou Feng, Chunhui Zhao, and Guangjun He.
\newblock Symswin: Multi-scale-aware super-resolution of remote sensing images based on swin transformers.
\newblock {\em Remote Sensing}, 16(24):4734, 2024.

\bibitem{wang2024ttsr}
Yi~Wang, Shichao Jin, Zekun Yang, Hongcan Guan, Yu~Ren, Kai Cheng, Xiaoqian Zhao, Xiaoqiang Liu, Mengxi Chen, Yu~Liu, et~al.
\newblock Ttsr: A transformer-based topography neural network for digital elevation model super-resolution.
\newblock {\em IEEE Transactions on Geoscience and Remote Sensing}, 62:1--19, 2024.

\bibitem{10683775}
Rui Li and Xiaowei Zhao.
\newblock Lswinsr: Uav imagery super-resolution based on linear swin transformer.
\newblock {\em IEEE Transactions on Geoscience and Remote Sensing}, 62:1--13, 2024.

\bibitem{zhu2024convmambasr}
Qiwei Zhu, Guojing Zhang, Xuechao Zou, Xiaoying Wang, Jianqiang Huang, and Xilai Li.
\newblock Convmambasr: Leveraging state-space models and cnns in a dual-branch architecture for remote sensing imagery super-resolution.
\newblock {\em Remote Sensing}, 16(17):3254, 2024.

\bibitem{10817590}
Yi~Xiao, Qiangqiang Yuan, Kui Jiang, Yuzeng Chen, Qiang Zhang, and Chia-Wen Lin.
\newblock Frequency-assisted mamba for remote sensing image super-resolution.
\newblock {\em IEEE Transactions on Multimedia}, pages 1--14, 2024.

\bibitem{wang2024efficient}
Xuan Wang, Lijun Sun, Jinglei Yi, Yongchao Song, Qiang Zheng, and Abdellah Chehri.
\newblock Efficient degradation representation learning network for remote sensing image super-resolution.
\newblock {\em Computer Vision and Image Understanding}, 249:104182, 2024.

\bibitem{zhai2024transcyclegan}
Lujun Zhai, Yonghui Wang, Suxia Cui, and Yu~Zhou.
\newblock Transcyclegan: An approach for remote sensing image super-resolution.
\newblock In {\em 2024 IEEE Southwest Symposium on Image Analysis and Interpretation (SSIAI)}, pages 61--64. IEEE, 2024.

\bibitem{10509697}
Jiang Zhu, Van Kwan~Zhi Koh, Zhiping Lin, and Bihan Wen.
\newblock Tm-gan: A transformer-based multi-modal generative adversarial network for guided depth image super-resolution.
\newblock {\em IEEE Journal on Emerging and Selected Topics in Circuits and Systems}, 14(2):261--274, 2024.

\bibitem{10829708}
Alireza Sharifi and Mohammad~Mahdi Safari.
\newblock Enhancing the spatial resolution of sentinel-2 images through super-resolution using transformer-based deep-learning models.
\newblock {\em IEEE Journal of Selected Topics in Applied Earth Observations and Remote Sensing}, 18:4805--4820, 2025.

\bibitem{10947187}
Wu-Ding Weng, Chao-Wei Zheng, Jian-Nan Su, Guang-Yong Chen, and Min Gan.
\newblock Efficient high-frequency texture recovery diffusion model for remote sensing image super-resolution.
\newblock {\em IEEE Transactions on Instrumentation and Measurement}, 74:1--14, 2025.

\bibitem{wang2025semantic}
Ce~Wang and Wanjie Sun.
\newblock Semantic guided large scale factor remote sensing image super-resolution with generative diffusion prior.
\newblock {\em ISPRS Journal of Photogrammetry and Remote Sensing}, 220:125--138, 2025.

\bibitem{10781453}
Yongchao Song, Lijun Sun, Jiping Bi, Siwen Quan, and Xuan Wang.
\newblock Drgan: A detail recovery-based model for optical remote sensing images super-resolution.
\newblock {\em IEEE Transactions on Geoscience and Remote Sensing}, 63:1--13, 2025.

\bibitem{guo2025structured}
Yiheng Guo, Yujie Liang, Yi~Liang, and Xiangwei Sun.
\newblock Structured bayesian super-resolution forward-looking imaging for maneuvering platforms based on enhanced sparsity model.
\newblock {\em Remote Sensing}, 17(5):775, 2025.

\bibitem{10836746}
Yifeng Yang, Hengqian Zhao, Xiadan Huangfu, Zihan Li, and Pan Wang.
\newblock Vit-isrgan: A high-quality super-resolution reconstruction method for multispectral remote sensing images.
\newblock {\em IEEE Journal of Selected Topics in Applied Earth Observations and Remote Sensing}, 18:3973--3988, 2025.

\bibitem{wang2020deep}
Zhihao Wang, Jian Chen, and Steven~CH Hoi.
\newblock Deep learning for image super-resolution: A survey.
\newblock {\em IEEE transactions on pattern analysis and machine intelligence}, 43(10):3365--3387, 2020.

\bibitem{yang2008image}
Jianchao Yang, John Wright, Thomas Huang, and Yi~Ma.
\newblock Image super-resolution as sparse representation of raw image patches.
\newblock In {\em 2008 IEEE conference on computer vision and pattern recognition}, pages 1--8. IEEE, 2008.

\bibitem{yang2010image}
Jianchao Yang, John Wright, Thomas~S Huang, and Yi~Ma.
\newblock Image super-resolution via sparse representation.
\newblock {\em IEEE transactions on image processing}, 19(11):2861--2873, 2010.

\bibitem{zhihui2011single}
Zheng Zhihui, Wang Bo, and Sun Kang.
\newblock Single remote sensing image super-resolution and denoising via sparse representation.
\newblock In {\em 2011 international workshop on multi-platform/multi-sensor remote sensing and mapping}, pages 1--5. IEEE, 2011.

\bibitem{zhang2013remote}
Yingying Zhang, Wei Wu, Yong Dai, Xiaomin Yang, Binyu Yan, and Wei Lu.
\newblock Remote sensing images super-resolution based on sparse dictionaries and residual dictionaries.
\newblock In {\em 2013 IEEE 11th International Conference on Dependable, Autonomic and Secure Computing}, pages 318--323. IEEE, 2013.

\bibitem{gou2014remote}
Shuiping Gou, Shuzhen Liu, Shuyuan Yang, and Licheng Jiao.
\newblock Remote sensing image super-resolution reconstruction based on nonlocal pairwise dictionaries and double regularization.
\newblock {\em IEEE Journal of Selected Topics in Applied Earth Observations and Remote Sensing}, 7(12):4784--4792, 2014.

\bibitem{roweis2000nonlinear}
Sam~T Roweis and Lawrence~K Saul.
\newblock Nonlinear dimensionality reduction by locally linear embedding.
\newblock {\em science}, 290(5500):2323--2326, 2000.

\bibitem{chang2004super}
Hong Chang, Dit-Yan Yeung, and Yimin Xiong.
\newblock Super-resolution through neighbor embedding.
\newblock In {\em Proceedings of the 2004 IEEE Computer Society Conference on Computer Vision and Pattern Recognition, 2004. CVPR 2004.}, volume~1, pages I--I. IEEE, 2004.

\bibitem{XU2021108033}
Jian Xu, Yan Gao, Jun Xing, Jiulun Fan, Qiannan Gao, and Shaojie Tang.
\newblock Two-direction self-learning super-resolution propagation based on neighbor embedding.
\newblock {\em Signal Processing}, 183:108033, 2021.

\bibitem{bevilacqua2012low}
Marco Bevilacqua, Aline Roumy, Christine Guillemot, and Marie~Line Alberi-Morel.
\newblock Low-complexity single-image super-resolution based on nonnegative neighbor embedding.
\newblock 2012.

\bibitem{zhang2011scale}
Hankui Zhang and Bo~Huang.
\newblock Scale conversion of multi sensor remote sensing image using single frame super resolution technology.
\newblock In {\em 2011 19th International Conference on Geoinformatics}, pages 1--5. IEEE, 2011.

\bibitem{zhou2014single}
Fei Zhou, Tingrong Yuan, Wenming Yang, and Qingmin Liao.
\newblock Single-image super-resolution based on compact kpca coding and kernel regression.
\newblock {\em IEEE Signal Processing Letters}, 22(3):336--340, 2014.

\bibitem{kanakaraj2020adaptive}
Sithara Kanakaraj, Madhu~S Nair, and Saidalavi Kalady.
\newblock Adaptive importance sampling unscented kalman filter with kernel regression for sar image super-resolution.
\newblock {\em IEEE Geoscience and Remote Sensing Letters}, 19:1--5, 2020.

\bibitem{babacan2010variational}
S~Derin Babacan, Rafael Molina, and Aggelos~K Katsaggelos.
\newblock Variational bayesian super resolution.
\newblock {\em IEEE Transactions on Image Processing}, 20(4):984--999, 2010.

\bibitem{akhtar2015bayesian}
Naveed Akhtar, Faisal Shafait, and Ajmal Mian.
\newblock Bayesian sparse representation for hyperspectral image super resolution.
\newblock In {\em Proceedings of the IEEE conference on computer vision and pattern recognition}, pages 3631--3640, 2015.

\bibitem{liebel2016single}
Lukas Liebel and Marco K{\"o}rner.
\newblock Single-image super resolution for multispectral remote sensing data using convolutional neural networks.
\newblock {\em The International Archives of the Photogrammetry, Remote Sensing and Spatial Information Sciences}, 41:883--890, 2016.

\bibitem{dong2014learning}
Chao Dong, Chen~Change Loy, Kaiming He, and Xiaoou Tang.
\newblock Learning a deep convolutional network for image super-resolution.
\newblock In {\em Computer Vision--ECCV 2014: 13th European Conference, Zurich, Switzerland, September 6-12, 2014, Proceedings, Part IV 13}, pages 184--199. Springer, 2014.

\bibitem{kim2010single}
Kwang~In Kim and Younghee Kwon.
\newblock Single-image super-resolution using sparse regression and natural image prior.
\newblock {\em IEEE transactions on pattern analysis and machine intelligence}, 32(6):1127--1133, 2010.

\bibitem{he2011single}
He~He and Wan-Chi Siu.
\newblock Single image super-resolution using gaussian process regression.
\newblock In {\em CVPR 2011}, pages 449--456. IEEE, 2011.

\bibitem{zhang2016single}
Kaibing Zhang, Xinbo Gao, Jie Li, and Hongxing Xia.
\newblock Single image super-resolution using regularization of non-local steering kernel regression.
\newblock {\em Signal Processing}, 123:53--63, 2016.

\bibitem{zhang2012single}
Kaibing Zhang, Xinbo Gao, Dacheng Tao, and Xuelong Li.
\newblock Single image super-resolution with non-local means and steering kernel regression.
\newblock {\em IEEE Transactions on Image Processing}, 21(11):4544--4556, 2012.

\bibitem{9242259}
Sithara Kanakaraj, Madhu~S. Nair, and Saidalavi Kalady.
\newblock Adaptive importance sampling unscented kalman filter with kernel regression for sar image super-resolution.
\newblock {\em IEEE Geoscience and Remote Sensing Letters}, 19:1--5, 2022.

\bibitem{zhang2015improvement}
Yihang Zhang, Yun Du, Feng Ling, and Xiaodong Li.
\newblock Improvement of the example-regression-based super-resolution land cover mapping algorithm.
\newblock {\em IEEE Geoscience and Remote Sensing Letters}, 12(8):1740--1744, 2015.

\bibitem{specht1991general}
Donald~F Specht et~al.
\newblock A general regression neural network.
\newblock {\em IEEE transactions on neural networks}, 2(6):568--576, 1991.

\bibitem{yang2013fast}
Chih-Yuan Yang and Ming-Hsuan Yang.
\newblock Fast direct super-resolution by simple functions.
\newblock In {\em Proceedings of the IEEE international conference on computer vision}, pages 561--568, 2013.

\bibitem{yang2016fast}
Xiaomin Yang, Wei Wu, Kai Liu, Kai Zhou, and Binyu Yan.
\newblock Fast multisensor infrared image super-resolution scheme with multiple regression models.
\newblock {\em Journal of Systems Architecture}, 64:11--25, 2016.

\bibitem{gao2018self}
Dongsheng Gao, Zhentao Hu, and Renzhen Ye.
\newblock Self-dictionary regression for hyperspectral image super-resolution.
\newblock {\em Remote Sensing}, 10(10):1574, 2018.

\bibitem{726791}
Y.~Lecun, L.~Bottou, Y.~Bengio, and P.~Haffner.
\newblock Gradient-based learning applied to document recognition.
\newblock {\em Proceedings of the IEEE}, 86(11):2278--2324, 1998.

\bibitem{2012ImageNet}
Alex Krizhevsky, I.~Sutskever, and G.~Hinton.
\newblock Imagenet classification with deep convolutional neural networks.
\newblock In {\em NIPS}, 2012.

\bibitem{Simonyan15}
Karen Simonyan and Andrew Zisserman.
\newblock Very deep convolutional networks for large-scale image recognition.
\newblock In {\em International Conference on Learning Representations}, 2015.

\bibitem{He_2016_CVPR}
Kaiming He, Xiangyu Zhang, Shaoqing Ren, and Jian Sun.
\newblock Deep residual learning for image recognition.
\newblock In {\em Proceedings of the IEEE Conference on Computer Vision and Pattern Recognition (CVPR)}, June 2016.

\bibitem{lou2025overlock}
Meng Lou and Yizhou Yu.
\newblock Overlock: An overview-first-look-closely-next convnet with context-mixing dynamic kernels.
\newblock In {\em Proceedings of the Computer Vision and Pattern Recognition Conference}, pages 128--138, 2025.

\bibitem{9356353}
Shervin Minaee, Yuri Boykov, Fatih Porikli, Antonio Plaza, Nasser Kehtarnavaz, and Demetri Terzopoulos.
\newblock Image segmentation using deep learning: A survey.
\newblock {\em IEEE Transactions on Pattern Analysis and Machine Intelligence}, 44(7):3523--3542, 2022.

\bibitem{7485869}
Shaoqing Ren, Kaiming He, Ross Girshick, and Jian Sun.
\newblock Faster r-cnn: Towards real-time object detection with region proposal networks.
\newblock {\em IEEE Transactions on Pattern Analysis and Machine Intelligence}, 39(6):1137--1149, 2017.

\bibitem{Qi2022ImageEnhancement}
Yunliang Qi, Zhen Yang, Wenhao Sun, Meng Lou, Jing Lian, Wenwei Zhao, Xiangyu Deng, and Yide Ma.
\newblock A comprehensive overview of image enhancement techniques.
\newblock {\em Archives of Computational Methods in Engineering}, 29:583--607, 2022.

\bibitem{Liebel2016Single-image}
Lukas Liebel and Marco Korner.
\newblock Single-image super resolution for multispectral remote sensing data using convolutional neural networks.
\newblock {\em The International Archives of the Photogrammetry, Remote Sensing and Spatial Information Sciences}, 41:883--890, 2016.

\bibitem{8518584}
Jie Fu, Yuhong Liu, and Feng Li.
\newblock In {\em IGARSS 2018 - 2018 IEEE International Geoscience and Remote Sensing Symposium}, pages 8014--8017, 2018.

\bibitem{keshk2021obtaining}
Hatem~Magdy Keshk and Xu-Cheng Yin.
\newblock Obtaining super-resolution satellites images based on enhancement deep convolutional neural network.
\newblock {\em International Journal of Aeronautical and Space Sciences}, 22(1):195--202, 2021.

\bibitem{ran2020remote}
Qiong Ran, Xiaodong Xu, Shizhi Zhao, Wei Li, and Qian Du.
\newblock Remote sensing images super-resolution with deep convolution networks.
\newblock {\em Multimedia Tools and Applications}, 79(13):8985--9001, 2020.

\bibitem{lu2019satellite}
Tao Lu, Jiaming Wang, Yanduo Zhang, Zhongyuan Wang, and Junjun Jiang.
\newblock Satellite image super-resolution via multi-scale residual deep neural network.
\newblock {\em Remote Sensing}, 11(13):1588, 2019.

\bibitem{9194276}
Xiaoyu Dong, Longguang Wang, Xu~Sun, Xiuping Jia, Lianru Gao, and Bing Zhang.
\newblock Remote sensing image super-resolution using second-order multi-scale networks.
\newblock {\em IEEE Transactions on Geoscience and Remote Sensing}, 59(4):3473--3485, 2021.

\bibitem{wang2020remote}
Xinying Wang, Yingdan Wu, Yang Ming, and Hui Lv.
\newblock Remote sensing imagery super resolution based on adaptive multi-scale feature fusion network.
\newblock {\em Sensors}, 20(4):1142, 2020.

\bibitem{Huan2022Remote}
Hai Huan, Nan Zou, Yi~Zhang, Yaqin Xie, and Chao Wang.
\newblock Remote sensing image reconstruction using an asymmetric multi-scale super-resolution network.
\newblock {\em The Journal of Supercomputing}, 78(17):18524--18550, 2022.

\bibitem{jiang2018deep}
Kui Jiang, Zhongyuan Wang, Peng Yi, Junjun Jiang, Jing Xiao, and Yuan Yao.
\newblock Deep distillation recursive network for remote sensing imagery super-resolution.
\newblock {\em Remote Sensing}, 10(11):1700, 2018.

\bibitem{yin2023super}
Zhixiang Yin, Yanlan Wu, Penghai Wu, Zhen Hao, and Feng Ling.
\newblock Super-resolution mapping with a fraction error eliminating cnn model.
\newblock {\em IEEE Transactions on Geoscience and Remote Sensing}, 61:1--18, 2023.

\bibitem{huan2021end}
Hai Huan, Pengcheng Li, Nan Zou, Chao Wang, Yaqin Xie, Yong Xie, and Dongdong Xu.
\newblock End-to-end super-resolution for remote-sensing images using an improved multi-scale residual network.
\newblock {\em Remote Sensing}, 13(4):666, 2021.

\bibitem{deeba2021multi}
Farah Deeba, Yuanchun Zhou, Fayaz~Ali Dharejo, Yi~Du, Xuezhi Wang, and She Kun.
\newblock Multi-scale single image super-resolution with remote-sensing application using transferred wide residual network.
\newblock {\em Wireless Personal Communications}, 120(1):323--342, 2021.

\bibitem{10480122}
Wujian Ye, Bili Lin, Junming Lao, Yijun Liu, and Zhenyi Lin.
\newblock Mra-idn: A lightweight super-resolution framework of remote sensing images based on multiscale residual attention fusion mechanism.
\newblock {\em IEEE Journal of Selected Topics in Applied Earth Observations and Remote Sensing}, 17:7781--7800, 2024.

\bibitem{10720033}
Tianren Wu, Rundong Zhao, Ming Lv, Zhenhong Jia, Liangliang Li, Zheyuan Wang, and Hongbing Ma.
\newblock Lightweight remote sensing image super-resolution via background-based multiscale feature enhancement network.
\newblock {\em IEEE Geoscience and Remote Sensing Letters}, 21:1--5, 2024.

\bibitem{10453218}
Dezhi Kong, Lingjia Gu, Xiaofeng Li, and Fang Gao.
\newblock Multiscale residual dense network for the super-resolution of remote sensing images.
\newblock {\em IEEE Transactions on Geoscience and Remote Sensing}, 62:1--12, 2024.

\bibitem{xiao2025dual}
Huanling Xiao, Xintong Chen, Liuhui Luo, and Cong Lin.
\newblock A dual-path feature reuse multi-scale network for remote sensing image super-resolution.
\newblock {\em The Journal of Supercomputing}, 81(1):1--28, 2025.

\bibitem{chen2023remote}
Xitong Chen, Yuntao Wu, Tao Lu, Quan Kong, Jiaming Wang, and Yu~Wang.
\newblock Remote sensing image super-resolution with residual split attention mechanism.
\newblock {\em IEEE Journal of Selected Topics in Applied Earth Observations and Remote Sensing}, 16:1--13, 2023.

\bibitem{10585320}
Zhanlong Chen, Xiaoyi Han, and Xiaochuan Ma.
\newblock Combining contextual information by integrated attention mechanism in convolutional neural networks for digital elevation model super-resolution.
\newblock {\em IEEE Transactions on Geoscience and Remote Sensing}, 62:1--16, 2024.

\bibitem{peng2021pre}
Yali Peng, Xuning Wang, Junwei Zhang, and Shigang Liu.
\newblock Pre-training of gated convolution neural network for remote sensing image super-resolution.
\newblock {\em IET Image Processing}, 15(5):1179--1188, 2021.

\bibitem{zhao2024structure}
Kanghui Zhao, Tao Lu, Yanduo Zhang, Junjun Jiang, Zhongyuan Wang, and Zixiang Xiong.
\newblock Structure-texture dual preserving for remote sensing image super resolution.
\newblock {\em IEEE Journal of Selected Topics in Applied Earth Observations and Remote Sensing}, 17:5527--5540, 2024.

\bibitem{10543045}
Wenjing Wang, Tingkui Mu, Qiuxia Li, Haoyang Li, and Qiujie Yang.
\newblock A hybrid spectral attention-enabled multiscale spatial–spectral learning network for hyperspectral image superresolution.
\newblock {\em IEEE Journal of Selected Topics in Applied Earth Observations and Remote Sensing}, 17:11016--11033, 2024.

\bibitem{chen2021remote}
Long Chen, Hui Liu, Minhang Yang, Yurong Qian, Zhengqing Xiao, and Xiwu Zhong.
\newblock Remote sensing image super-resolution via residual aggregation and split attentional fusion network.
\newblock {\em IEEE Journal of Selected Topics in Applied Earth Observations and Remote Sensing}, 14:9546--9556, 2021.

\bibitem{ma2021remote}
Yunchuan Ma, Pengyuan Lv, Hao Liu, Xuehong Sun, and Yanfei Zhong.
\newblock Remote sensing image super-resolution based on dense channel attention network.
\newblock {\em Remote Sensing}, 13(15):2966, 2021.

\bibitem{10453952}
Allen Patnaik, M.~K. Bhuyan, and Karl~F. MacDorman.
\newblock A two-branch multiscale residual attention network for single image super-resolution in remote sensing imagery.
\newblock {\em IEEE Journal of Selected Topics in Applied Earth Observations and Remote Sensing}, 17:6003--6013, 2024.

\bibitem{zhang2020remote}
Dongyang Zhang, Jie Shao, Xinyao Li, and Heng~Tao Shen.
\newblock Remote sensing image super-resolution via mixed high-order attention network.
\newblock {\em IEEE Transactions on Geoscience and Remote Sensing}, 59(6):5183--5196, 2020.

\bibitem{huang2021deep}
Bo~Huang, Boyong He, Liaoni Wu, and Zhiming Guo.
\newblock Deep residual dual-attention network for super-resolution reconstruction of remote sensing images.
\newblock {\em Remote Sensing}, 13(14):2784, 2021.

\bibitem{wang2021lightweight}
Jin Wang, Yiming Wu, Liu Wang, Lei Wang, Osama Alfarraj, and Amr Tolba.
\newblock Lightweight feedback convolution neural network for remote sensing images super-resolution.
\newblock {\em IEEE Access}, 9:15992--16003, 2021.

\bibitem{ren2021remote}
Chao Ren, Xiaohai He, Linbo Qing, Yuanyuan Wu, and Yifei Pu.
\newblock Remote sensing image recovery via enhanced residual learning and dual-luminance scheme.
\newblock {\em Knowledge-Based Systems}, 222:107013, 2021.

\bibitem{wang2018aerial}
Tingwei Wang, Wenjian Sun, Hairong Qi, and Peng Ren.
\newblock Aerial image super resolution via wavelet multiscale convolutional neural networks.
\newblock {\em IEEE Geoscience and Remote Sensing Letters}, 15(5):769--773, 2018.

\bibitem{yang2019hyperspectral}
Jingxiang Yang, Yong-Qiang Zhao, and Jonathan Cheung-Wai Chan.
\newblock Hyperspectral image super-resolution based on multi-scale wavelet 3d convolutional neural network.
\newblock In {\em IGARSS 2019-2019 IEEE International Geoscience and Remote Sensing Symposium}, pages 2770--2773. IEEE, 2019.

\bibitem{aburaed2020super}
Nour Aburaed, Alavikunhu Panthakkan, Mina Al-Saad, Marwa~Chendeb El~Rai, Saeed Al~Mansoori, Hussain Al-Ahmad, and Stephen Marshall.
\newblock Super-resolution of satellite imagery using a wavelet multiscale-based deep convolutional neural network model.
\newblock In {\em Image and Signal Processing for Remote Sensing XXVI}, volume 11533, pages 305--311. SPIE, 2020.

\bibitem{deeba2021plexus}
Farah Deeba, Yuanchun Zhou, Fayaz~Ali Dharejo, Muhammad~Ashfaq Khan, Bhagwan Das, Xuezhi Wang, and Yi~Du.
\newblock A plexus-convolutional neural network framework for fast remote sensing image super-resolution in wavelet domain.
\newblock {\em IET Image Processing}, 15(8):1679--1687, 2021.

\bibitem{thool2023combined}
KV~Thool, S~Deivalakshmi, et~al.
\newblock Combined wavelet and multiwavelet-based convolutional neural network (cwm-cnn) for remote sensing image super resolution.
\newblock In {\em International Conference on Computer Vision and Internet of Things 2023 (ICCVIoT'23)}, volume 2023, pages 102--107. IET, 2023.

\bibitem{lanaras2018super}
Charis Lanaras, Jos{\'e} Bioucas-Dias, Silvano Galliani, Emmanuel Baltsavias, and Konrad Schindler.
\newblock Super-resolution of sentinel-2 images: Learning a globally applicable deep neural network.
\newblock {\em ISPRS Journal of Photogrammetry and Remote Sensing}, 146:305--319, 2018.

\bibitem{yin2021cascaded}
Zhixiang Yin, Feng Ling, Xinyan Li, Xiaobin Cai, Hong Chi, Xiaodong Li, Lihui Wang, Yihang Zhang, and Yun Du.
\newblock A cascaded spectral--spatial cnn model for super-resolution river mapping with modis imagery.
\newblock {\em IEEE Transactions on Geoscience and Remote Sensing}, 60:1--13, 2021.

\bibitem{arun2020cnn}
Pattathal~V Arun, Krishna~Mohan Buddhiraju, Alok Porwal, and Jocelyn Chanussot.
\newblock Cnn based spectral super-resolution of remote sensing images.
\newblock {\em Signal Processing}, 169:107394, 2020.

\bibitem{wagner2019deep}
Lena Wagner, Lukas Liebel, and Marco K{\"o}rner.
\newblock Deep residual learning for single-image super-resolution of multi-spectral satellite imagery.
\newblock {\em ISPRS Annals of the Photogrammetry, Remote Sensing and Spatial Information Sciences}, 4:189--196, 2019.

\bibitem{vasilescu2023cnn}
Vlad Vasilescu, Mihai Datcu, and Daniela Faur.
\newblock A cnn-based sentinel-2 image super-resolution method using multiobjective training.
\newblock {\em IEEE Transactions on Geoscience and Remote Sensing}, 61:1--14, 2023.

\bibitem{10499252}
Chi Chen, Yongcheng Wang, Yuxi Zhang, Zhikang Zhao, and Hao Feng.
\newblock Remote sensing hyperspectral image super-resolution via multidomain spatial information and multiscale spectral information fusion.
\newblock {\em IEEE Transactions on Geoscience and Remote Sensing}, 62:1--16, 2024.

\bibitem{mei2020spatial}
Shaohui Mei, Ruituo Jiang, Xu~Li, and Qian Du.
\newblock Spatial and spectral joint super-resolution using convolutional neural network.
\newblock {\em IEEE Transactions on Geoscience and Remote Sensing}, 58(7):4590--4603, 2020.

\bibitem{muller2020super}
MU~M{\"u}ller, N~Ekhtiari, RM~Almeida, and C~Rieke.
\newblock Super-resolution of multispectral satellite images using convolutional neural networks.
\newblock {\em ISPRS Annals of the Photogrammetry, Remote Sensing and Spatial Information Sciences}, 1:33--40, 2020.

\bibitem{xu2018high}
Wenjia Xu, XU~Guangluan, Yang Wang, Xian Sun, Daoyu Lin, and WU~Yirong.
\newblock High quality remote sensing image super-resolution using deep memory connected network.
\newblock In {\em IGARSS 2018-2018 IEEE International Geoscience and Remote Sensing Symposium}, pages 8889--8892. IEEE, 2018.

\bibitem{chang2019bidirectional}
Yunpeng Chang and Bin Luo.
\newblock Bidirectional convolutional lstm neural network for remote sensing image super-resolution.
\newblock {\em Remote Sensing}, 11(20):2333, 2019.

\bibitem{zhu2020super}
Xiang Zhu, Hossein Talebi, Xinwei Shi, Feng Yang, and Peyman Milanfar.
\newblock Super-resolving commercial satellite imagery using realistic training data.
\newblock In {\em 2020 IEEE International Conference on Image Processing (ICIP)}, pages 498--502. IEEE, 2020.

\bibitem{deeba2020single}
Farah Deeba, Fayaz~Ali Dharejo, Yuanchun Zhou, Abdul Ghaffar, Mujahid~Hussain Memon, and She Kun.
\newblock Single image super-resolution with application to remote-sensing image.
\newblock In {\em 2020 Global Conference on Wireless and Optical Technologies (GCWOT)}, pages 1--6. IEEE, 2020.

\bibitem{wei2021construction}
Zikang Wei and Yunqing Liu.
\newblock Construction of super-resolution model of remote sensing image based on deep convolutional neural network.
\newblock {\em Computer Communications}, 178:191--200, 2021.

\bibitem{10613839}
Kanghui Zhao, Tao Lu, Jiaming Wang, Yanduo Zhang, Junjun Jiang, and Zixiang Xiong.
\newblock Hyper-laplacian prior for remote sensing image super-resolution.
\newblock {\em IEEE Transactions on Geoscience and Remote Sensing}, 62:1--14, 2024.

\bibitem{yin2024super}
Zhixiang Yin, Penghai Wu, Xinyan Li, Zhen Hao, Xiaoshuang Ma, Ruirui Fan, Chun Liu, and Feng Ling.
\newblock Super-resolution water body mapping with a feature collaborative cnn model by fusing sentinel-1 and sentinel-2 images.
\newblock {\em International Journal of Applied Earth Observation and Geoinformation}, 134:104176, 2024.

\bibitem{10843849}
Lin-Yu Dai, Ming-Dian Li, and Si-Wei Chen.
\newblock Pccn: Polarimetric contexture convolutional network for polsar image super-resolution.
\newblock {\em IEEE Journal of Selected Topics in Applied Earth Observations and Remote Sensing}, 18:4664--4679, 2025.

\bibitem{ledig2017photo}
Christian Ledig, Lucas Theis, Ferenc Husz{\'a}r, Jose Caballero, Andrew Cunningham, Alejandro Acosta, Andrew Aitken, Alykhan Tejani, Johannes Totz, Zehan Wang, et~al.
\newblock Photo-realistic single image super-resolution using a generative adversarial network.
\newblock In {\em Proceedings of the IEEE conference on computer vision and pattern recognition}, pages 4681--4690, 2017.

\bibitem{ma2018super}
Wen Ma, Zongxu Pan, Jiayi Guo, and Bin Lei.
\newblock Super-resolution of remote sensing images based on transferred generative adversarial network.
\newblock In {\em IGARSS 2018-2018 IEEE International Geoscience and Remote Sensing Symposium}, pages 1148--1151. IEEE, 2018.

\bibitem{ma2019super}
Wen Ma, Zongxu Pan, Feng Yuan, and Bin Lei.
\newblock Super-resolution of remote sensing images via a dense residual generative adversarial network.
\newblock {\em Remote Sensing}, 11(21):2578, 2019.

\bibitem{wang2020ultra}
Zhongyuan Wang, Kui Jiang, Peng Yi, Zhen Han, and Zheng He.
\newblock Ultra-dense gan for satellite imagery super-resolution.
\newblock {\em Neurocomputing}, 398:328--337, 2020.

\bibitem{chen2022super}
Yu-Zhang Chen, Tsung-Jung Liu, and Kuan-Hsien Liu.
\newblock Super-resolution of satellite images based on two-dimensional rrdb and edge-enhanced generative adversarial network.
\newblock In {\em 2022 IEEE International Conference on Consumer Electronics (ICCE)}, pages 1--4. IEEE, 2022.

\bibitem{pang2023use}
Boyu Pang, Siwei Zhao, and Yinnian Liu.
\newblock The use of a stable super-resolution generative adversarial network (ssrgan) on remote sensing images.
\newblock {\em Remote Sensing}, 15(20):5064, 2023.

\bibitem{liu2020super}
Bo~Liu, Heng Li, Yutao Zhou, Yuqing Peng, Ahmed Elazab, and Changmiao Wang.
\newblock A super resolution method for remote sensing images based on cascaded conditional wasserstein gans.
\newblock In {\em 2020 IEEE 3rd International Conference on Information Communication and Signal Processing (ICICSP)}, pages 284--289. IEEE, 2020.

\bibitem{guo2022ndsrgan}
Mingqiang Guo, Zeyuan Zhang, Heng Liu, and Ying Huang.
\newblock Ndsrgan: A novel dense generative adversarial network for real aerial imagery super-resolution reconstruction.
\newblock {\em Remote Sensing}, 14(7):1574, 2022.

\bibitem{sustika2020generative}
Rika Sustika, Andriyan~Bayu Suksmono, Donny Danudirdjo, and Ketut Wikantika.
\newblock Generative adversarial network with residual dense generator for remote sensing image super resolution.
\newblock In {\em 2020 International Conference on Radar, Antenna, Microwave, Electronics, and Telecommunications (ICRAMET)}, pages 34--39. IEEE, 2020.

\bibitem{zhao2023forest}
Yafeng Zhao, Shuai Zhang, and Junfeng Hu.
\newblock Forest single-frame remote sensing image super-resolution using gans.
\newblock {\em Forests}, 14(11):2188, 2023.

\bibitem{moustafa2021satellite}
Marwa~S Moustafa and Sayed~A Sayed.
\newblock Satellite imagery super-resolution using squeeze-and-excitation-based gan.
\newblock {\em International journal of aeronautical and space sciences}, 22(6):1481--1492, 2021.

\bibitem{li2021single}
Yadong Li, Sebastien Mavromatis, Feng Zhang, Zhenhong Du, Jean Sequeira, Zhongyi Wang, Xianwei Zhao, and Renyi Liu.
\newblock Single-image super-resolution for remote sensing images using a deep generative adversarial network with local and global attention mechanisms.
\newblock {\em IEEE Transactions on Geoscience and Remote Sensing}, 60:1--24, 2021.

\bibitem{zhang2023super}
Huajun Zhang, Chengming Ye, Yuzhan Zhou, Rong Tang, and Ruilong Wei.
\newblock A super-resolution network for high-resolution reconstruction of landslide main bodies in remote sensing imagery using coordinated attention mechanisms and deep residual blocks.
\newblock {\em Remote Sensing}, 15(18):4498, 2023.

\bibitem{wang2023msagan}
Chunyang Wang, Xian Zhang, Wei Yang, Xingwang Li, Bibo Lu, and Jianlong Wang.
\newblock Msagan: a new super-resolution algorithm for multispectral remote sensing image based on a multiscale attention gan network.
\newblock {\em IEEE Geoscience and Remote Sensing Letters}, 20:1--5, 2023.

\bibitem{ren2023context}
Zhihan Ren, Lijun He, and Jichuan Lu.
\newblock Context aware edge-enhanced gan for remote sensing image super-resolution.
\newblock {\em IEEE Journal of Selected Topics in Applied Earth Observations and Remote Sensing}, 17:1363--1376, 2023.

\bibitem{10461030}
Yinggan Tang, Tianjiao Wang, and Defeng Liu.
\newblock Mffagan: Generative adversarial network with multilevel feature fusion attention mechanism for remote sensing image super-resolution.
\newblock {\em IEEE Journal of Selected Topics in Applied Earth Observations and Remote Sensing}, 17:6860--6874, 2024.

\bibitem{guo2024tdegan}
Mingqiang Guo, Feng Xiong, Baorui Zhao, Ying Huang, Zhong Xie, Liang Wu, Xueye Chen, and Jiaming Zhang.
\newblock Tdegan: A texture-detail-enhanced dense generative adversarial network for remote sensing image super-resolution.
\newblock {\em Remote Sensing}, 16(13):2312, 2024.

\bibitem{hu2024super}
Wenyi Hu, Lei Ju, Yujia Du, and Yuxia Li.
\newblock A super-resolution reconstruction model for remote sensing image based on generative adversarial networks.
\newblock {\em Remote Sensing}, 16(8):1460, 2024.

\bibitem{fan2024rmsrgan}
Kaixuan Fan, Min Hu, Maocheng Zhao, Liang Qi, Weijun Xie, Hongyan Zou, Bin Wu, Shuaishuai Zhao, and Xiwei Wang.
\newblock Rmsrgan: A real multispectral imagery super-resolution reconstruction for enhancing ginkgo biloba yield prediction.
\newblock {\em Forests}, 15(5):859, 2024.

\bibitem{chung2023enhancing}
Minkyung Chung, Minyoung Jung, and Yongil Kim.
\newblock Enhancing remote sensing image super-resolution guided by bicubic-downsampled low-resolution image.
\newblock {\em Remote Sensing}, 15(13):3309, 2023.

\bibitem{karwowska2023mcwesrgan}
Kinga Karwowska and Damian Wierzbicki.
\newblock Mcwesrgan: Improving enhanced super-resolution generative adversarial network for satellite images.
\newblock {\em IEEE Journal of Selected Topics in Applied Earth Observations and Remote Sensing}, 16:9459--9479, 2023.

\bibitem{zhu2023improved}
Fuzhen Zhu, Chen Wang, Bing Zhu, Ce~Sun, and Chengxiao Qi.
\newblock An improved generative adversarial networks for remote sensing image super-resolution reconstruction via multi-scale residual block.
\newblock {\em The Egyptian Journal of Remote Sensing and Space Science}, 26(1):151--160, 2023.

\bibitem{guo2023improved}
Jifeng Guo, Feicai Lv, Jiayou Shen, Jing Liu, and Mingzhi Wang.
\newblock An improved generative adversarial network for remote sensing image super-resolution.
\newblock {\em IET Image Processing}, 17(6):1852--1863, 2023.

\bibitem{kong2023super}
Juwon Kong, Youngryel Ryu, Sungchan Jeong, Zilong Zhong, Wonseok Choi, Jongmin Kim, Kyungdo Lee, Joongbin Lim, Keunchang Jang, Junghwa Chun, et~al.
\newblock Super resolution of historic landsat imagery using a dual generative adversarial network (gan) model with cubesat constellation imagery for spatially enhanced long-term vegetation monitoring.
\newblock {\em ISPRS Journal of Photogrammetry and Remote Sensing}, 200:1--23, 2023.

\bibitem{10528671}
Guihou Sun, Yuehong Chen, Jiamei Huang, Qiang Ma, and Yong Ge.
\newblock Digital surface model super-resolution by integrating high-resolution remote sensing imagery using generative adversarial networks.
\newblock {\em IEEE Journal of Selected Topics in Applied Earth Observations and Remote Sensing}, 17:10636--10647, 2024.

\bibitem{pineda2020generative}
Ferdinand Pineda, Victor Ayma, and C{\'e}sar Beltran.
\newblock A generative adversarial network approach for super-resolution of sentinel-2 satellite images.
\newblock {\em The International Archives of the Photogrammetry, Remote Sensing and Spatial Information Sciences}, 43:9--14, 2020.

\bibitem{huang2020super}
Zhi-Xing Huang and Chang-Wei Jing.
\newblock Super-resolution reconstruction method of remote sensing image based on multi-feature fusion.
\newblock {\em Ieee Access}, 8:18764--18771, 2020.

\bibitem{wang2023multik}
Peijuan Wang and Elif Sertel.
\newblock Multi-frame super-resolution of remote sensing images using attention-based gan models.
\newblock {\em Knowledge-Based Systems}, 266:110387, 2023.

\bibitem{gong2021enlighten}
Yuanfu Gong, Puyun Liao, Xiaodong Zhang, Lifei Zhang, Guanzhou Chen, Kun Zhu, Xiaoliang Tan, and Zhiyong Lv.
\newblock Enlighten-gan for super resolution reconstruction in mid-resolution remote sensing images.
\newblock {\em Remote Sensing}, 13(6):1104, 2021.

\bibitem{wang2024super}
Xinyu Wang, Zurui Ao, Runhao Li, Yingchun Fu, Yufei Xue, and Yunxin Ge.
\newblock Super-resolution image reconstruction method between sentinel-2 and gaofen-2 based on cascaded generative adversarial networks.
\newblock {\em Applied Sciences}, 14(12):5013, 2024.

\bibitem{vaswani2017attention}
Ashish Vaswani, Noam Shazeer, Niki Parmar, Jakob Uszkoreit, Llion Jones, Aidan~N Gomez, {\L}ukasz Kaiser, and Illia Polosukhin.
\newblock Attention is all you need.
\newblock {\em Advances in neural information processing systems}, 30, 2017.

\bibitem{dosovitskiy2020image}
Alexey Dosovitskiy, Lucas Beyer, Alexander Kolesnikov, Dirk Weissenborn, Xiaohua Zhai, Thomas Unterthiner, Mostafa Dehghani, Matthias Minderer, Georg Heigold, Sylvain Gelly, et~al.
\newblock An image is worth 16x16 words: Transformers for image recognition at scale.
\newblock {\em arXiv preprint arXiv:2010.11929}, 2020.

\bibitem{ye2021super}
Chongjun Ye, Lingyu Yan, Yucheng Zhang, Jun Zhan, Jie Yang, and Junfang Wang.
\newblock A super-resolution method of remote sensing image using transformers.
\newblock In {\em 2021 11th IEEE International Conference on Intelligent Data Acquisition and Advanced Computing Systems: Technology and Applications (IDAACS)}, volume~2, pages 905--910. IEEE, 2021.

\bibitem{shi2023multisource}
Mengyang Shi, Yesheng Gao, Lin Chen, and Xingzhao Liu.
\newblock Multisource information fusion network for optical remote sensing image super-resolution.
\newblock {\em IEEE Journal of Selected Topics in Applied Earth Observations and Remote Sensing}, 16:3805--3818, 2023.

\bibitem{shang2023hybrid}
Jianrun Shang, Mingliang Gao, Qilei Li, Jinfeng Pan, Guofeng Zou, and Gwanggil Jeon.
\newblock Hybrid-scale hierarchical transformer for remote sensing image super-resolution.
\newblock {\em Remote Sensing}, 15(13):3442, 2023.

\bibitem{lu2024enhanced}
Yuting Lu, Shunzhou Wang, Binglu Wang, Xin Zhang, Xiaoxu Wang, and Yongqiang Zhao.
\newblock Enhanced window-based self-attention with global and multi-scale representations for remote sensing image super-resolution.
\newblock {\em Remote Sensing}, 16(15):2837, 2024.

\bibitem{cao2023cfmb}
Yuan Cao, Ligang Li, Bo~Liu, Wenbo Zhou, Zengyi Li, and Wei Ni.
\newblock Cfmb-t: A cross-frequency multi-branch transformer for low-quality infrared remote sensing image super-resolution.
\newblock {\em Infrared Physics \& Technology}, 133:104861, 2023.

\bibitem{sun2024transformer}
Yu~Sun, Xiyang Zhi, Shikai Jiang, Guanghua Fan, Tianjun Shi, and Xu~Yan.
\newblock Transformer-based self-supervised image super-resolution method for rotating synthetic aperture system via multi-temporal fusion.
\newblock {\em Information Fusion}, 108:102372, 2024.

\bibitem{10742930}
Zexin Xie, Jian Wang, Wei Song, Yanling Du, Huifang Xu, and Qinhan Yang.
\newblock Cfformer: Channel fourier transformer for remote sensing super resolution.
\newblock {\em IEEE Journal of Selected Topics in Applied Earth Observations and Remote Sensing}, 18:569--583, 2025.

\bibitem{liu2025wtt}
Jingyi Liu and Xiaomin Yang.
\newblock Wtt: combining wavelet transform with transformer for remote sensing image super-resolution.
\newblock {\em Machine Vision and Applications}, 36(2):1--14, 2025.

\bibitem{li2023spectral}
Zengyi Li, Ligang Li, Bo~Liu, Yuan Cao, Wenbo Zhou, Wei Ni, and Zhen Yang.
\newblock Spectral-learning-based transformer network for the spectral super-resolution of remote-sensing degraded images.
\newblock {\em IEEE Geoscience and Remote Sensing Letters}, 20:1--5, 2023.

\bibitem{cai2022t}
Durong Cai and Peng Zhang.
\newblock T$^3${SR}: Texture transfer transformer for remote sensing image superresolution.
\newblock {\em IEEE Journal of Selected Topics in Applied Earth Observations and Remote Sensing}, 15:7346--7358, 2022.

\bibitem{lu2023cross}
Yuting Lu, Lingtong Min, Binglu Wang, Le~Zheng, Xiaoxu Wang, Yongqiang Zhao, Le~Yang, and Teng Long.
\newblock Cross-spatial pixel integration and cross-stage feature fusion-based transformer network for remote sensing image super-resolution.
\newblock {\em IEEE Transactions on Geoscience and Remote Sensing}, 61:1--16, 2023.

\bibitem{mao2024desat}
Yujie Mao, Guojin He, Guizhou Wang, Ranyu Yin, Yan Peng, and Bin Guan.
\newblock Desat: A distance-enhanced strip attention transformer for remote sensing image super-resolution.
\newblock {\em Remote Sensing}, 16(22):4251, 2024.

\bibitem{10767430}
Yingdong Kang, Xinyu Wang, Xuemin Zhang, Shaoju Wang, and Guang Jin.
\newblock Act-sr: Aggregation connection transformer for remote sensing image super-resolution.
\newblock {\em IEEE Journal of Selected Topics in Applied Earth Observations and Remote Sensing}, pages 1--12, 2024.

\bibitem{10753509}
Jinglei Hao, Wukai Li, Yuting Lu, Yang Jin, Yongqiang Zhao, Shunzhou Wang, and Binglu Wang.
\newblock Scale-aware backprojection transformer for single remote sensing image super-resolution.
\newblock {\em IEEE Transactions on Geoscience and Remote Sensing}, 62:1--13, 2024.

\bibitem{gu2023mamba}
Albert Gu and Tri Dao.
\newblock Mamba: Linear-time sequence modeling with selective state spaces.
\newblock {\em arXiv preprint arXiv:2312.00752}, 2023.

\bibitem{liu2024vmamba}
Yue Liu, Yunjie Tian, Yuzhong Zhao, Hongtian Yu, Lingxi Xie, Yaowei Wang, Qixiang Ye, Jianbin Jiao, and Yunfan Liu.
\newblock Vmamba: Visual state space model.
\newblock {\em Advances in neural information processing systems}, 37:103031--103063, 2024.

\bibitem{dong2024fusion}
Wenhao Dong, Haodong Zhu, Shaohui Lin, Xiaoyan Luo, Yunhang Shen, Xuhui Liu, Juan Zhang, Guodong Guo, and Baochang Zhang.
\newblock Fusion-mamba for cross-modality object detection.
\newblock {\em arXiv preprint arXiv:2404.09146}, 2024.

\bibitem{ma2024u}
Jun Ma, Feifei Li, and Bo~Wang.
\newblock U-mamba: Enhancing long-range dependency for biomedical image segmentation.
\newblock {\em arXiv preprint arXiv:2401.04722}, 2024.

\bibitem{10663411}
Ruicong Zhi, Xiaopei Fan, and Jingye Shi.
\newblock Mambaformersr: A lightweight model for remote-sensing image super-resolution.
\newblock {\em IEEE Geoscience and Remote Sensing Letters}, 21:1--5, 2024.

\bibitem{zhao2024mamba}
Gongbo Zhao and Feng Wang.
\newblock Mamba-based residual network for remote sensing image super-resolution.
\newblock In {\em 2024 7th International Conference on Pattern Recognition and Artificial Intelligence (PRAI)}, pages 316--321. IEEE, 2024.

\bibitem{liu2022diffusion}
Jinzhe Liu, Zhiqiang Yuan, Zhaoying Pan, Yiqun Fu, Li~Liu, and Bin Lu.
\newblock Diffusion model with detail complement for super-resolution of remote sensing.
\newblock {\em Remote Sensing}, 14(19):4834, 2022.

\bibitem{han2023enhancing}
Lintao Han, Yuchen Zhao, Hengyi Lv, Yisa Zhang, Hailong Liu, Guoling Bi, and Qing Han.
\newblock Enhancing remote sensing image super-resolution with efficient hybrid conditional diffusion model.
\newblock {\em Remote Sensing}, 15(13):3452, 2023.

\bibitem{sui2024denoising}
Jialu Sui, Qianqian Wu, and Man-On Pun.
\newblock Denoising diffusion probabilistic model with adversarial learning for remote sensing super-resolution.
\newblock {\em Remote Sensing}, 16(7):1219, 2024.

\bibitem{zhang2024tcdm}
Yan Zhang, Hanqi Liu, Zhenghao Li, Xinbo Gao, Guangyao Shi, and Jianan Jiang.
\newblock Tcdm: Effective large-factor image super-resolution via texture consistency diffusion.
\newblock {\em IEEE Transactions on Geoscience and Remote Sensing}, 62:1--13, 2024.

\bibitem{zhu2025taming}
Chao Zhu, Yong Liu, Shan Huang, and Fei Wang.
\newblock Taming a diffusion model to revitalize remote sensing image super-resolution.
\newblock {\em Remote Sensing}, 17(8):1348, 2025.

\bibitem{he2022dster}
Jiang He, Qiangqiang Yuan, Jie Li, Yi~Xiao, Xinxin Liu, and Yun Zou.
\newblock Dster: A dense spectral transformer for remote sensing spectral super-resolution.
\newblock {\em International Journal of Applied Earth Observation and Geoinformation}, 109:102773, 2022.

\bibitem{tu2022swcgan}
Jingzhi Tu, Gang Mei, Zhengjing Ma, and Francesco Piccialli.
\newblock Swcgan: Generative adversarial network combining swin transformer and cnn for remote sensing image super-resolution.
\newblock {\em IEEE Journal of Selected Topics in Applied Earth Observations and Remote Sensing}, 15:5662--5673, 2022.

\bibitem{wang2023landsat}
Chunyang Wang, Xian Zhang, Wei Yang, Gaige Wang, Zongze Zhao, Xuan Liu, and Bibo Lu.
\newblock Landsat-8 to sentinel-2 satellite imagery super-resolution-based multiscale dilated transformer generative adversarial networks.
\newblock {\em Remote Sensing}, 15(22):5272, 2023.

\bibitem{peng2023context}
Guangwen Peng, Minghong Xie, and Liuyang Fang.
\newblock Context-aware lightweight remote-sensing image super-resolution network.
\newblock {\em Frontiers in Neurorobotics}, 17:1220166, 2023.

\bibitem{10549773}
Cong Lin, Xin Mao, Chenghao Qiu, and Lilan Zou.
\newblock Dtcnet: Transformer-cnn distillation for super-resolution of remote sensing image.
\newblock {\em IEEE Journal of Selected Topics in Applied Earth Observations and Remote Sensing}, 17:11117--11133, 2024.

\bibitem{10562345}
Mingyang Hou, Zhiyong Huang, Zhi Yu, Yan Yan, Yunlan Zhao, and Xiao Han.
\newblock Cswt-sr: Conv-swin transformer for blind remote sensing image super-resolution with amplitude-phase learning and structural detail alternating learning.
\newblock {\em IEEE Transactions on Geoscience and Remote Sensing}, 62:1--14, 2024.

\bibitem{wang2024lightweight}
Jie Wang, Hongwei Li, Yifan Li, and Zilong Qin.
\newblock A lightweight cnn-transformer implemented via structural re-parameterization and hybrid attention for remote sensing image super-resolution.
\newblock {\em ISPRS International Journal of Geo-Information}, 14(1):8, 2024.

\bibitem{huang2024infrared}
Feng Huang, Yunxiang Li, Xiaojing Ye, and Jing Wu.
\newblock Infrared image super-resolution network utilizing the enhanced transformer and u-net.
\newblock {\em Sensors}, 24(14):4686, 2024.

\bibitem{guo2024activated}
Yongde Guo, Chengying Gong, and Jun Yan.
\newblock Activated sparsely sub-pixel transformer for remote sensing image super-resolution.
\newblock {\em Remote Sensing}, 16(11):1895, 2024.

\bibitem{liu2025remote}
Denghui Liu, Lin Zhong, Haiyang Wu, Songyang Li, and Yida Li.
\newblock Remote sensing image super-resolution reconstruction by fusing multi-scale receptive fields and hybrid transformer.
\newblock {\em Scientific Reports}, 15(1):2140, 2025.

\bibitem{10345595}
Junjie Li, Yizhuo Meng, Chongxin Tao, Zhen Zhang, Xining Yang, Zhe Wang, Xi~Wang, Linyi Li, and Wen Zhang.
\newblock Convformersr: Fusing transformers and convolutional neural networks for cross-sensor remote sensing imagery super-resolution.
\newblock {\em IEEE Transactions on Geoscience and Remote Sensing}, 62:1--15, 2024.

\bibitem{10494508}
Chunyang Wang, Xian Zhang, Wei Yang, GaiGe Wang, Xingwang Li, Jianlong Wang, and Bibo Lu.
\newblock Mswagan: Multispectral remote sensing image super-resolution based on multiscale window attention transformer.
\newblock {\em IEEE Transactions on Geoscience and Remote Sensing}, 62:1--15, 2024.

\bibitem{yu2023estugan}
Chunhe Yu, Lingyue Hong, Tianpeng Pan, Yufeng Li, and Tingting Li.
\newblock Estugan: enhanced swin transformer with u-net discriminator for remote sensing image super-resolution.
\newblock {\em Electronics}, 12(20):4235, 2023.

\bibitem{lin2024trans}
Zhenyi Lin, Bili Lin, Wujian Ye, and Yijun Liu.
\newblock Trans-cnn gan: Self-attention generative adversarial networkd for remote sensing image super-resolution.
\newblock In {\em 2024 4th International Conference on Neural Networks, Information and Communication (NNICE)}, pages 456--459. IEEE, 2024.

\bibitem{huo2024stgan}
Wei Huo, Xiaodan Zhang, Shaojie You, Yongkun Zhang, Qiyuan Zhang, and Naihao Hu.
\newblock Stgan: Swin transformer-based gan to achieve remote sensing image super-resolution reconstruction.
\newblock {\em Applied Sciences}, 15(1):305, 2024.

\bibitem{tao2003superresolution}
Hongjiu Tao, Xinjian Tang, Jian Liu, and Jinwen Tian.
\newblock Superresolution remote sensing image processing algorithm based on wavelet transform and interpolation.
\newblock In {\em Image Processing and Pattern Recognition in Remote Sensing}, volume 4898, pages 259--263. SPIE, 2003.

\bibitem{han2015wavelet}
Su~Young Han, Nam~Hun Park, and Kil~Hong Joo.
\newblock Wavelet transform based image interpolation for remote sensing image.
\newblock {\em International Journal of Software Engineering and Its Applications}, 9(2):59--66, 2015.

\bibitem{mareboyana2018super}
Manohar Mareboyana and Jaqueline Le~Moigne.
\newblock Super-resolution of remote sensing images using edge-directed radial basis functions.
\newblock In {\em Signal processing, sensor/information fusion, and target recognition XXVII}, volume 10646, pages 216--221. SPIE, 2018.

\bibitem{patti1997superresolution}
Andrew~J Patti, M~Ibrahim Sezan, and A~Murat Tekalp.
\newblock Superresolution video reconstruction with arbitrary sampling lattices and nonzero aperture time.
\newblock {\em IEEE transactions on image processing}, 6(8):1064--1076, 1997.

\bibitem{patti1997robust}
Andrew~J Patti, M~Ibrahim Sezan, and A~Murat Tekalp.
\newblock Robust methods for high-quality stills from interlaced video in the presence of dominant motion.
\newblock {\em IEEE Transactions on Circuits and Systems for Video Technology}, 7(2):328--342, 1997.

\bibitem{aguena2006multispectral}
Marcia~LS Aguena and Nelson~DA Mascarenhas.
\newblock Multispectral image data fusion using pocs and super-resolution.
\newblock {\em Computer Vision and Image Understanding}, 102(2):178--187, 2006.

\bibitem{xie2009blind}
Wei Xie, Feiyan Zhang, Hao Chen, and Qianqing Qin.
\newblock Blind super-resolution image reconstruction based on pocs model.
\newblock In {\em 2009 international conference on measuring technology and mechatronics automation}, volume~1, pages 437--440. IEEE, 2009.

\bibitem{fan2017projections}
Chong Fan, Chaoyun Wu, Grand Li, and Jun Ma.
\newblock Projections onto convex sets super-resolution reconstruction based on point spread function estimation of low-resolution remote sensing images.
\newblock {\em Sensors}, 17(2):362, 2017.

\bibitem{dai2017study}
Shaosheng Dai, Junjie Cui, Dezhou Zhang, Qin Liu, and Xiaoxiao Zhang.
\newblock Study on infrared image super-resolution reconstruction based on an improved pocs algorithm.
\newblock {\em Journal of Semiconductors}, 38(4):044010, 2017.

\bibitem{irani1991improving}
Michal Irani and Shmuel Peleg.
\newblock Improving resolution by image registration.
\newblock {\em CVGIP: Graphical models and image processing}, 53(3):231--239, 1991.

\bibitem{lu2002pyramid}
Yao Lu and Minoru Imamura.
\newblock Pyramid-based super-resolution of the undersampled and subpixel shifted image sequence.
\newblock {\em International journal of imaging systems and technology}, 12(6):254--263, 2002.

\bibitem{li2006improved}
Feng Li, Donald Fraser, and Xiuping Jia.
\newblock Improved ibp for super-resolving remote sensing images.
\newblock {\em Geographic Information Sciences}, 12(2):106--111, 2006.

\bibitem{li2007efficient}
Feng Li, Donald Fraser, and Xiuping Jia.
\newblock Efficient ibp with super resolution for alos imagery.
\newblock In {\em Geoinformatics 2007: Remotely Sensed Data and Information}, volume 6752, pages 1423--1430. SPIE, 2007.

\bibitem{yan2009super}
Ziye Yan and Yao Lu.
\newblock Super resolution of mri using improved ibp.
\newblock In {\em 2009 International Conference on Computational Intelligence and Security}, volume~1, pages 643--647. IEEE, 2009.

\bibitem{patel2011hybrid}
Vaishali Patel, Chintan~K Modi, Chirag~N Paunwala, and Suprava Patnaik.
\newblock Hybrid approach for single image super resolution using isef and ibp.
\newblock In {\em 2011 International Conference on Communication Systems and Network Technologies}, pages 495--499. IEEE, 2011.

\bibitem{bareja2012effective}
Milan~N Bareja and Chintan~K Modi.
\newblock An effective iterative back projection based single image super resolution approach.
\newblock In {\em 2012 international conference on communication systems and Network technologies}, pages 95--99. IEEE, 2012.

\bibitem{nayak2013spatial}
Rajashree Nayak, S~Monalisa, and Dipti Patra.
\newblock Spatial super resolution based image reconstruction using hibp.
\newblock In {\em 2013 Annual IEEE India Conference (INDICON)}, pages 1--6. IEEE, 2013.

\bibitem{cong2013effective}
Wang Cong, Li~Weifeng, Wang Longbiao, and Liao Qingming.
\newblock An effective framework of ibp for single facial image super resolution.
\newblock In {\em 3rd International Conference on Multimedia Technology (ICMT-13)}, pages 986--993. Atlantis Press, 2013.

\bibitem{nayak2014morphology}
Rajashree Nayak, Saka Harshavardhan, and Dipti Patra.
\newblock Morphology based iterative back-projection for super-resolution reconstruction of image.
\newblock In {\em 2014 2nd international conference on emerging technology trends in electronics, communication and networking}, pages 1--6. IEEE, 2014.

\bibitem{tipping2002bayesian}
Michael Tipping and Christopher Bishop.
\newblock Bayesian image super-resolution.
\newblock {\em Advances in neural information processing systems}, 15, 2002.

\bibitem{molina2005new}
Rafael Molina, Javier Mateos, Aggelos~K Katsaggelos, R~Zurita-Milla, S~Liang, J~Liu, X~Li, R~Liu, and M~Schaepman.
\newblock A new super resolution bayesian method for pansharpening landsat etm+ imagery.
\newblock In {\em 9th International Symposium on Physical Measurements and Signatures in Remote Sensing (ISPMSRS)}, pages 280--283. ISPRS WG VII/1, 2005.

\bibitem{molina2008variational}
Rafael Molina, Miguel Vega, Javier Mateos, and Aggelos~K Katsaggelos.
\newblock Variational posterior distribution approximation in bayesian super resolution reconstruction of multispectral images.
\newblock {\em Applied and Computational Harmonic Analysis}, 24(2):251--267, 2008.

\bibitem{molina2006parameter}
Rafael Molina, Miguel Vega, Javier Mateos, and Aggelos~K Katsaggelos.
\newblock Parameter estimation in bayesian reconstruction of multispectral images using super resolution techniques.
\newblock In {\em 2006 International Conference on Image Processing}, pages 1749--1752. IEEE, 2006.

\bibitem{molina2006hierarchical}
Rafael Molina, Miguel Vega, Javier Mateos, and Aggelos~K Katsaggelos.
\newblock Hierarchical bayesian super resolution reconstruction of multispectral images.
\newblock In {\em 2006 14th European Signal Processing Conference}, pages 1--5. IEEE, 2006.

\bibitem{wang2018high}
Tingting Wang, Faming Fang, Fang Li, and Guixu Zhang.
\newblock High-quality bayesian pansharpening.
\newblock {\em IEEE Transactions on Image Processing}, 28(1):227--239, 2018.

\bibitem{armannsson2021comparison}
Sveinn~E Armannsson, Magnus~O Ulfarsson, Jakob Sigurdsson, Han~V Nguyen, and Johannes~R Sveinsson.
\newblock A comparison of optimized sentinel-2 super-resolution methods using wald’s protocol and bayesian optimization.
\newblock {\em Remote Sensing}, 13(11):2192, 2021.

\bibitem{li2021forward}
Weixin Li, Ming Li, Lei Zuo, Hao Sun, Hongmeng Chen, and Yachao Li.
\newblock Forward-looking super-resolution imaging for sea-surface target with multi-prior bayesian method.
\newblock {\em Remote Sensing}, 14(1):26, 2021.

\bibitem{tan2021novel}
Ke~Tan, Xingyu Lu, Jianchao Yang, Weimin Su, and Hong Gu.
\newblock A novel bayesian super-resolution method for radar forward-looking imaging based on markov random field model.
\newblock {\em Remote Sensing}, 13(20):4115, 2021.

\bibitem{shen2024super}
Jiahao Shen, Deqing Mao, Yin Zhang, Yulin Huang, Haiguang Yang, and Jianyu Yang.
\newblock Super-resolution imaging method for forward-looking scanning radar based on two-layer bayesian model.
\newblock In {\em 2024 IEEE Radar Conference (RadarConf24)}, pages 1--6. IEEE, 2024.

\bibitem{chantas2007super}
Giannis~K Chantas, Nikolaos~P Galatsanos, and Nathan~A Woods.
\newblock Super-resolution based on fast registration and maximum a posteriori reconstruction.
\newblock {\em IEEE Transactions on Image Processing}, 16(7):1821--1830, 2007.

\bibitem{wang2010spectral}
Suyu Wang, Li~Zhuo, and Xiaoguang Li.
\newblock Spectral imagery super resolution by using of a high resolution panchromatic image.
\newblock In {\em 2010 3rd International Conference on Computer Science and Information Technology}, volume~4, pages 220--224. IEEE, 2010.

\bibitem{belekos2010maximum}
Stefanos~P Belekos, Nikolaos~P Galatsanos, and Aggelos~K Katsaggelos.
\newblock Maximum a posteriori video super-resolution using a new multichannel image prior.
\newblock {\em IEEE transactions on image processing}, 19(6):1451--1464, 2010.

\bibitem{li2009super}
Feng Li, Xiuping Jia, Donald Fraser, and Andrew Lambert.
\newblock Super resolution for remote sensing images based on a universal hidden markov tree model.
\newblock {\em IEEE Transactions on Geoscience and Remote Sensing}, 48(3):1270--1278, 2009.

\bibitem{guan2014maximum}
Jinchen Guan, Jianyu Yang, Yulin Huang, and Wenchao Li.
\newblock Maximum a posteriori--based angular superresolution for scanning radar imaging.
\newblock {\em IEEE Transactions on Aerospace and Electronic Systems}, 50(3):2389--2398, 2014.

\bibitem{yuan2014remote}
Qiangqiang Yuan, Li~Yan, Jiancheng Li, and Liangpei Zhang.
\newblock Remote sensing image super-resolution via regional spatially adaptive total variation model.
\newblock In {\em 2014 IEEE Geoscience and Remote Sensing Symposium}, pages 3073--3076. IEEE, 2014.

\bibitem{li2017super}
Feng Li, Lei Xin, Yi~Guo, Dongsheng Gao, Xianghao Kong, and Xiuping Jia.
\newblock Super-resolution for gaofen-4 remote sensing images.
\newblock {\em IEEE Geoscience and Remote Sensing Letters}, 15(1):28--32, 2017.

\bibitem{irmak2016super}
Hasan Irmak, G{\"o}zde~B Akar, Seniha~Esen Yuksel, and Hakan Aytaylan.
\newblock Super-resolution reconstruction of hyperspectral images via an improved map-based approach.
\newblock In {\em 2016 IEEE International Geoscience and Remote Sensing Symposium (IGARSS)}, pages 7244--7247. IEEE, 2016.

\bibitem{tan2018penalized}
Ke~Tan, Wenchao Li, Qian Zhang, Yulin Huang, Junjie Wu, and Jianyu Yang.
\newblock Penalized maximum likelihood angular super-resolution method for scanning radar forward-looking imaging.
\newblock {\em Sensors}, 18(3):912, 2018.

\bibitem{tan2018q}
Ke~Tan, Wenchao Li, Jifang Pei, Yulin Huang, and Jianyu Yang.
\newblock An i/q-channel modeling maximum likelihood super-resolution imaging method for forward-looking scanning radar.
\newblock {\em IEEE Geoscience and Remote Sensing Letters}, 15(6):863--867, 2018.

\bibitem{wu2021super}
Xinyi Wu, Yu~Fang, and Daoqing Wu.
\newblock Super-resolution imaging method based on fast maximum likelihood.
\newblock In {\em 2021 IEEE 5th Advanced Information Technology, Electronic and Automation Control Conference (IAEAC)}, pages 1545--1548. IEEE, 2021.

\bibitem{li2013super}
Xiaodong Li, Yun Du, and Feng Ling.
\newblock Super-resolution mapping of forests with bitemporal different spatial resolution images based on the spatial-temporal markov random field.
\newblock {\em IEEE Journal of Selected Topics in Applied Earth Observations and Remote Sensing}, 7(1):29--39, 2013.

\bibitem{aghighi2015fully}
Hossein Aghighi, John Trinder, Samsung Lim, and Yuliya Tarabalka.
\newblock Fully spatially adaptive smoothing parameter estimation for markov random field super-resolution mapping of remotely sensed images.
\newblock {\em International Journal of Remote Sensing}, 36(11):2851--2879, 2015.

\bibitem{welikanna2024fuzzy}
DR~Welikanna and M~Tamura.
\newblock Fuzzy parameters integrated markov random field (mrf) model for super resolution mapping (srm) over vague land cover regions.
\newblock {\em Journal of Geospatial Surveying}, 4(1), 2024.

\bibitem{choi2020no}
Yeonju Choi and Yongwoo Kim.
\newblock A no-reference super resolution for satellite image quality enhancement for kompsat-3.
\newblock In {\em IGARSS 2020-2020 IEEE International Geoscience and Remote Sensing Symposium}, pages 220--223. IEEE, 2020.

\bibitem{nguyen2021self}
Ngoc~Long Nguyen, J{\'e}r{\'e}my Anger, Axel Davy, Pablo Arias, and Gabriele Facciolo.
\newblock Self-supervised multi-image super-resolution for push-frame satellite images.
\newblock In {\em Proceedings of the IEEE/CVF Conference on Computer Vision and Pattern Recognition}, pages 1121--1131, 2021.

\bibitem{hong2023decoupled}
Danfeng Hong, Jing Yao, Chenyu Li, Deyu Meng, Naoto Yokoya, and Jocelyn Chanussot.
\newblock Decoupled-and-coupled networks: Self-supervised hyperspectral image super-resolution with subpixel fusion.
\newblock {\em IEEE Transactions on Geoscience and Remote Sensing}, 61:1--12, 2023.

\bibitem{mishra2023accelerating}
Divya Mishra and Ofer Hadar.
\newblock Accelerating neural style-transfer using contrastive learning for unsupervised satellite image super-resolution.
\newblock {\em IEEE Transactions on Geoscience and Remote Sensing}, 61:1--14, 2023.

\bibitem{wang2019unsupervised}
Pengrui Wang, Haopeng Zhang, Feng Zhou, and Zhiguo Jiang.
\newblock Unsupervised remote sensing image super-resolution using cycle cnn.
\newblock In {\em IGARSS 2019-2019 IEEE International Geoscience and Remote Sensing Symposium}, pages 3117--3120. IEEE, 2019.

\bibitem{zhang2020unsupervised}
Ning Zhang, Yongcheng Wang, Xin Zhang, Dongdong Xu, and Xiaodong Wang.
\newblock An unsupervised remote sensing single-image super-resolution method based on generative adversarial network.
\newblock {\em IEEE Access}, 8:29027--29039, 2020.

\bibitem{yang2022enhanced}
Yu~Yang and Yajuan Wu.
\newblock Enhanced zero-shot learning algorithm for super-resolution reconstruction of remote sensing images.
\newblock In {\em 2022 IEEE 8th International Conference on Computer and Communications (ICCC)}, pages 1994--1998. IEEE, 2022.

\bibitem{bose2021zero}
Rupak Bose, Vikrant Rangnekar, Biplab Banerjee, and Subhasis Chaudhuri.
\newblock Zero-shot remote sensing image super-resolution based on image continuity and self tessellations.
\newblock In {\em DAGM German Conference on Pattern Recognition}, pages 649--662. Springer, 2021.

\bibitem{chang2023pixel}
Yali Chang, Gang Chen, and Jifa Chen.
\newblock Pixel-wise attention residual network for super-resolution of optical remote sensing images.
\newblock {\em Remote Sensing}, 15(12):3139, 2023.

\bibitem{10764782}
Xudong Yao, Haopeng Zhang, Sizhe Wen, Zhenwei Shi, and Zhiguo Jiang.
\newblock Single-image superresolution for rgb remote sensing imagery via multiscale cnn-transformer feature fusion.
\newblock {\em IEEE Journal of Selected Topics in Applied Earth Observations and Remote Sensing}, 18:1302--1316, 2025.

\bibitem{9959886}
Mengyang Shi, Yesheng Gao, Lin Chen, and Xingzhao Liu.
\newblock Dual-resolution local attention unfolding network for optical remote sensing image super-resolution.
\newblock {\em IEEE Geoscience and Remote Sensing Letters}, 19:1--5, 2022.

\bibitem{11078386}
Lijun Sun, Jiping Bi, Yongchao Song, Zhaowei Liu, and Xuan Wang.
\newblock Improving optical remote sensing image quality through random degradation and adaptive fusion super-resolution networks.
\newblock {\em IEEE Transactions on Geoscience and Remote Sensing}, 63:1--17, 2025.

\bibitem{9632567}
Shunzhou Wang, Tianfei Zhou, Yao Lu, and Huijun Di.
\newblock Contextual transformation network for lightweight remote-sensing image super-resolution.
\newblock {\em IEEE Transactions on Geoscience and Remote Sensing}, 60:1--13, 2022.

\bibitem{9955482}
Mengyang Shi, Yesheng Gao, Lin Chen, and Xingzhao Liu.
\newblock Dual-branch multiscale channel fusion unfolding network for optical remote sensing image super-resolution.
\newblock {\em IEEE Geoscience and Remote Sensing Letters}, 19:1--5, 2022.

\bibitem{pastina2003super}
Debora Pastina, Pierfrancesco Lombardo, Alfonso Farina, and Piero Daddi.
\newblock Super-resolution of polarimetric sar images of ship targets.
\newblock {\em Signal Processing}, 83(8):1737--1748, 2003.

\bibitem{lin2019polarimetric}
Liupeng Lin, Jie Li, Qiangqiang Yuan, and Huanfeng Shen.
\newblock Polarimetric sar image super-resolution via deep convolutional neural network.
\newblock In {\em IGARSS 2019-2019 IEEE International Geoscience and Remote Sensing Symposium}, pages 3205--3208. IEEE, 2019.

\bibitem{6112799}
Xiao~Xiang Zhu and Richard Bamler.
\newblock Demonstration of super-resolution for tomographic sar imaging in urban environment.
\newblock {\em IEEE Transactions on Geoscience and Remote Sensing}, 50(8):3150--3157, 2012.

\bibitem{5966335}
Xiao~Xiang Zhu and Richard Bamler.
\newblock Super-resolution power and robustness of compressive sensing for spectral estimation with application to spaceborne tomographic sar.
\newblock {\em IEEE Transactions on Geoscience and Remote Sensing}, 50(1):247--258, 2012.

\bibitem{wu2020super}
Chunxiao Wu, Zenghui Zhang, Longyong Chen, and Wenxian Yu.
\newblock Super-resolution for mimo array sar 3-d imaging based on compressive sensing and deep neural network.
\newblock {\em IEEE Journal of Selected Topics in Applied Earth Observations and Remote Sensing}, 13:3109--3124, 2020.

\bibitem{9340592}
Yangkai Wei, Yinchuan Li, Zegang Ding, Yan Wang, Tao Zeng, and Teng Long.
\newblock Sar parametric super-resolution image reconstruction methods based on admm and deep neural network.
\newblock {\em IEEE Transactions on Geoscience and Remote Sensing}, 59(12):10197--10212, 2021.

\bibitem{8634345}
Longgang Wang, Mana Zheng, Wenbo Du, Menglin Wei, and Lianlin Li.
\newblock Super-resolution sar image reconstruction via generative adversarial network.
\newblock In {\em 2018 12th International Symposium on Antennas, Propagation and EM Theory (ISAPE)}, pages 1--4, 2018.

\bibitem{lee2023efficient}
Seung-Jae Lee and Sun-Gu Lee.
\newblock Efficient super-resolution method for targets observed by satellite sar.
\newblock {\em Sensors}, 23(13):5893, 2023.

\bibitem{8239604}
Sithara Kanakaraj, Madhu~S. Nair, and Saidalavi Kalady.
\newblock Sar image super resolution using importance sampling unscented kalman filter.
\newblock {\em IEEE Journal of Selected Topics in Applied Earth Observations and Remote Sensing}, 11(2):562--571, 2018.

\bibitem{8899202}
Feng Gu, Hong Zhang, Chao Wang, and Fan Wu.
\newblock Sar image super-resolution based on noise-free generative adversarial network.
\newblock In {\em IGARSS 2019 - 2019 IEEE International Geoscience and Remote Sensing Symposium}, pages 2575--2578, 2019.

\bibitem{10966888}
Ganggang Dong, Yao Wang, Hongwei Liu, and Songlin Liu.
\newblock Complex-valued sar image super-resolution via subaperture learning and fusion.
\newblock {\em IEEE Transactions on Geoscience and Remote Sensing}, 63:1--14, 2025.

\bibitem{yanshan2022ogsrn}
LI~Yanshan, ZHOU Li, XU~Fan, and CHEN Shifu.
\newblock Ogsrn: Optical-guided super-resolution network for sar image.
\newblock {\em Chinese Journal of Aeronautics}, 35(5):204--219, 2022.

\bibitem{11045185}
Zhicheng Zhao, Qing Gao, Jinquan Yan, Chenglong Li, and Jin Tang.
\newblock Hsfmamba: Hierarchical selective fusion mamba network for optics-guided joint super-resolution and denoising of noise-corrupted sar images.
\newblock {\em IEEE Journal of Selected Topics in Applied Earth Observations and Remote Sensing}, 18:16445--16461, 2025.

\bibitem{11084879}
Zilong Chen, Caiguang Zhang, Chenyu Wan, Siqian Zhang, and Boli Xiong.
\newblock Dadsr: Degradation-aware diffusion super-resolution model for object-level sar image.
\newblock {\em IEEE Journal of Selected Topics in Applied Earth Observations and Remote Sensing}, pages 1--14, 2025.

\bibitem{11071996}
Hongwei Zhang, Xiaolin Zhao, Xuan Hao, Weilong Li, Hang Hu, Jiacheng Ni, and Ying Luo.
\newblock Sar super-resolution imaging and recognition integrated network based on deep learning framework.
\newblock {\em IEEE Journal of Selected Topics in Applied Earth Observations and Remote Sensing}, pages 1--19, 2025.

\bibitem{jiang2023improved}
Yichun Jiang, Yunqing Liu, Weida Zhan, and Depeng Zhu.
\newblock Improved thermal infrared image super-resolution reconstruction method base on multimodal sensor fusion.
\newblock {\em Entropy}, 25(6):914, 2023.

\bibitem{chen2024infrared}
Wenbin Chen, Shikai Jiang, Fuhai Wang, Xiyang Zhi, Jianming Hu, Yin Zhang, and Wei Zhang.
\newblock Infrared remote-sensing image super-resolution based on physical characteristic deduction.
\newblock {\em Results in Physics}, 64:107897, 2024.

\bibitem{huang2025texture}
Yongsong Huang, Tomo Miyazaki, Xiaofeng Liu, Yafei Dong, and Shinichiro Omachi.
\newblock Texture and noise dual adaptation for infrared image super-resolution.
\newblock {\em Pattern Recognition}, 163:111449, 2025.

\bibitem{mao2016infrared}
Yuxing Mao, Yan Wang, Jintao Zhou, and Haiwei Jia.
\newblock An infrared image super-resolution reconstruction method based on compressive sensing.
\newblock {\em Infrared Physics \& Technology}, 76:735--739, 2016.

\bibitem{zhang2018infrared}
Xudong Zhang, Chunlai Li, Qingpeng Meng, Shijie Liu, Yue Zhang, and Jianyu Wang.
\newblock Infrared image super resolution by combining compressive sensing and deep learning.
\newblock {\em Sensors}, 18(8):2587, 2018.

\bibitem{he2018cascaded}
Zewei He, Siliang Tang, Jiangxin Yang, Yanlong Cao, Michael~Ying Yang, and Yanpeng Cao.
\newblock Cascaded deep networks with multiple receptive fields for infrared image super-resolution.
\newblock {\em IEEE transactions on circuits and systems for video technology}, 29(8):2310--2322, 2018.

\bibitem{zhu2022super}
Shengyan Zhu, Caiqiu Zhou, and Yongjian Wang.
\newblock Super resolution reconstruction method for infrared images based on pseudo transferred features.
\newblock {\em Displays}, 74:102187, 2022.

\bibitem{dan2024pirn}
Jun Dan, Tao Jin, Hao Chi, Mushui Liu, Jiawang Yu, Keying Cao, Xinjing Yang, Luo Zhao, and Haoran Xie.
\newblock Pirn: Phase invariant reconstruction network for infrared image super-resolution.
\newblock {\em Neurocomputing}, 599:128221, 2024.

\bibitem{zhang2023closed}
Haopeng Zhang, Cong Zhang, Fengying Xie, and Zhiguo Jiang.
\newblock A closed-loop network for single infrared remote sensing image super-resolution in real world.
\newblock {\em Remote Sensing}, 15(4):882, 2023.

\bibitem{huang2025irsrmamba}
Yongsong Huang, Tomo Miyazaki, Xiaofeng Liu, and Shinichiro Omachi.
\newblock Irsrmamba: Infrared image super-resolution via mamba-based wavelet transform feature modulation model.
\newblock {\em IEEE Transactions on Geoscience and Remote Sensing}, 2025.

\bibitem{lu2025super}
Chenyan Lu and Cheng Su.
\newblock Super resolution reconstruction of mars thermal infrared remote sensing images integrating multi-source data.
\newblock {\em Remote Sensing}, 17(13):2115, 2025.

\bibitem{shan2020simulation}
Tixiao Shan, Jinkun Wang, Fanfei Chen, Paul Szenher, and Brendan Englot.
\newblock Simulation-based lidar super-resolution for ground vehicles.
\newblock {\em Robotics and Autonomous Systems}, 134:103647, 2020.

\bibitem{tian2022lidar}
Di~Tian, Dangjun Zhao, Dongyang Cheng, and Junchao Zhang.
\newblock Lidar super-resolution based on segmentation and geometric analysis.
\newblock {\em IEEE Transactions on Instrumentation and Measurement}, 71:1--17, 2022.

\bibitem{chen2023sgsr}
Chi Chen, Ang Jin, Zhiye Wang, Yongwei Zheng, Bisheng Yang, Jian Zhou, Yuhang Xu, and Zhigang Tu.
\newblock Sgsr-net: structure semantics guided lidar super-resolution network for indoor lidar slam.
\newblock {\em IEEE transactions on multimedia}, 26:1842--1854, 2023.

\bibitem{ramirez2024super}
A~Ramirez-Jaime, N~Porras-Diaz, GR~Arce, D~Harding, M~Stephen, and J~MacKinnon.
\newblock Super-resolution of satellite lidars for forest studies via generative adversarial networks.
\newblock In {\em IGARSS 2024-2024 IEEE International Geoscience and Remote Sensing Symposium}, pages 2271--2274. IEEE, 2024.

\bibitem{yang2010bag}
Yi~Yang and Shawn Newsam.
\newblock Bag-of-visual-words and spatial extensions for land-use classification.
\newblock In {\em Proceedings of the 18th SIGSPATIAL international conference on advances in geographic information systems}, pages 270--279, 2010.

\bibitem{Dai2011WHURS19}
Dengxin Dai and Wen Yang.
\newblock Satellite image classification via two-layer sparse coding with biased image representation.
\newblock {\em IEEE Transactions on Geoscience and Remote Sensing}, 8(1):173--176, 2011.

\bibitem{zou2015deep}
Qin Zou, Lihao Ni, Tong Zhang, and Qian Wang.
\newblock Deep learning based feature selection for remote sensing scene classification.
\newblock {\em IEEE Geoscience and remote sensing letters}, 12(11):2321--2325, 2015.

\bibitem{xia2017aid}
Gui-Song Xia, Jingwen Hu, Fan Hu, Baoguang Shi, Xiang Bai, Yanfei Zhong, Liangpei Zhang, and Xiaoqiang Lu.
\newblock Aid: A benchmark data set for performance evaluation of aerial scene classification.
\newblock {\em IEEE Transactions on Geoscience and Remote Sensing}, 55(7):3965--3981, 2017.

\bibitem{troylau2016draper}
troylau and Will Cukierski.
\newblock {Draper Satellite Image Chronology}.
\newblock \url{https://kaggle.com/competitions/draper-satellite-image-chronology}, 2016.
\newblock Accessed: [Insert Access Date].

\bibitem{cheng2017remote}
Gong Cheng, Junwei Han, and Xiaoqiang Lu.
\newblock Remote sensing image scene classification: Benchmark and state of the art.
\newblock {\em Proceedings of the IEEE}, 105(10):1865--1883, 2017.

\bibitem{zhou2018patternnet}
Weixun Zhou, Shawn Newsam, Congmin Li, and Zhenfeng Shao.
\newblock Patternnet: A benchmark dataset for performance evaluation of remote sensing image retrieval.
\newblock {\em ISPRS journal of photogrammetry and remote sensing}, 145:197--209, 2018.

\bibitem{xia2018dota}
Gui-Song Xia, Xiang Bai, Jian Ding, Zhen Zhu, Serge Belongie, Jiebo Luo, Mihai Datcu, Marcello Pelillo, and Liangpei Zhang.
\newblock Dota: A large-scale dataset for object detection in aerial images.
\newblock In {\em Proceedings of the IEEE conference on computer vision and pattern recognition}, pages 3974--3983, 2018.

\bibitem{spacenet2018catalog}
{SpaceNet on AWS}.
\newblock {The SpaceNet Catalog}.
\newblock \url{https://spacenet.ai/datasets/}, 2018.
\newblock Accessed: August 1, 2021.

\bibitem{ieeedataport2019datafusion}
{Data Fusion Contest 2019 (DFC2019)}.
\newblock \url{https://ieee-dataport.org/open-access/data-fusion-contest-2019-dfc2019}, 2019.
\newblock Accessed: [Insert Access Date].

\bibitem{wang2018scene}
Qi~Wang, Shaoteng Liu, Jocelyn Chanussot, and Xuelong Li.
\newblock Scene classification with recurrent attention of vhr remote sensing images.
\newblock {\em IEEE Transactions on Geoscience and Remote Sensing}, 57(2):1155--1167, 2018.

\bibitem{zhao2015dirichlet}
Bei Zhao, Yanfei Zhong, Gui-Song Xia, and Liangpei Zhang.
\newblock Dirichlet-derived multiple topic scene classification model for high spatial resolution remote sensing imagery.
\newblock {\em IEEE Transactions on Geoscience and Remote Sensing}, 54(4):2108--2123, 2015.

\bibitem{li2020RSI-CB}
Haifeng Li, Xin Dou, Chao Tao, Zhixiang Wu, Jie Chen, Jian Peng, Min Deng, and Ling Zhao.
\newblock Rsi-cb: A large-scale remote sensing image classification benchmark using crowdsourced data.
\newblock {\em Sensors}, 20(6):1594, 2020.

\bibitem{wang2004image}
Zhou Wang, Alan~C Bovik, Hamid~R Sheikh, and Eero~P Simoncelli.
\newblock Image quality assessment: from error visibility to structural similarity.
\newblock {\em IEEE transactions on image processing}, 13(4):600--612, 2004.

\bibitem{zhang2018unreasonable}
Richard Zhang, Phillip Isola, Alexei~A Efros, Eli Shechtman, and Oliver Wang.
\newblock The unreasonable effectiveness of deep features as a perceptual metric.
\newblock In {\em Proceedings of the IEEE conference on computer vision and pattern recognition}, pages 586--595, 2018.

\bibitem{mittal2012making}
Anish Mittal, Rajiv Soundararajan, and Alan~C Bovik.
\newblock Making a “completely blind” image quality analyzer.
\newblock {\em IEEE Signal processing letters}, 20(3):209--212, 2012.

\bibitem{yuhas1992discrimination}
Roberta~H Yuhas, Alexander~FH Goetz, and Joe~W Boardman.
\newblock Discrimination among semi-arid landscape endmembers using the spectral angle mapper (sam) algorithm.
\newblock In {\em JPL, Summaries of the Third Annual JPL Airborne Geoscience Workshop. Volume 1: AVIRIS Workshop}, 1992.

\bibitem{chen2018novel}
Aixiang~Andy Chen, Xiaolong Chai, Bingchuan Chen, Rui Bian, and Qingliang Chen.
\newblock A novel stochastic stratified average gradient method: Convergence rate and its complexity.
\newblock In {\em 2018 International Joint Conference on Neural Networks (IJCNN)}, pages 1--8. IEEE, 2018.

\bibitem{Blau_2018_ECCV_Workshops}
Yochai Blau, Roey Mechrez, Radu Timofte, Tomer Michaeli, and Lihi Zelnik-Manor.
\newblock The 2018 pirm challenge on perceptual image super-resolution.
\newblock In {\em Proceedings of the European Conference on Computer Vision (ECCV) Workshops}, September 2018.

\bibitem{ma2017learning}
Chao Ma, Chih-Yuan Yang, Xiaokang Yang, and Ming-Hsuan Yang.
\newblock Learning a no-reference quality metric for single-image super-resolution.
\newblock {\em Computer Vision and Image Understanding}, 158:1--16, 2017.

\bibitem{wald2000quality}
Lucien Wald.
\newblock Quality of high resolution synthesised images: Is there a simple criterion?
\newblock In {\em Third conference" Fusion of Earth data: merging point measurements, raster maps and remotely sensed images"}, pages 99--103. SEE/URISCA, 2000.

\bibitem{wang2002universal}
Zhou Wang and Alan~C Bovik.
\newblock A universal image quality index.
\newblock {\em IEEE signal processing letters}, 9(3):81--84, 2002.

\bibitem{alparone2008multispectral}
Luciano Alparone, Bruno Aiazzi, Stefano Baronti, Andrea Garzelli, Filippo Nencini, and Massimo Selva.
\newblock Multispectral and panchromatic data fusion assessment without reference.
\newblock {\em Photogrammetric Engineering \& Remote Sensing}, 74(2):193--200, 2008.

\bibitem{sheikh2006image}
Hamid~R Sheikh and Alan~C Bovik.
\newblock Image information and visual quality.
\newblock {\em IEEE Transactions on image processing}, 15(2):430--444, 2006.

\bibitem{zhang2011fsim}
Lin Zhang, Lei Zhang, Xuanqin Mou, and David Zhang.
\newblock Fsim: A feature similarity index for image quality assessment.
\newblock {\em IEEE transactions on Image Processing}, 20(8):2378--2386, 2011.

\bibitem{chang1999spectral}
Chein-I Chang.
\newblock Spectral information divergence for hyperspectral image analysis.
\newblock In {\em IEEE 1999 International Geoscience and Remote Sensing Symposium. IGARSS'99 (Cat. No. 99CH36293)}, volume~1, pages 509--511. IEEE, 1999.

\bibitem{1525860}
Qian Du, O.~Gungor, and Jie Shan.
\newblock Performance evaluation for pan-sharpening techniques.
\newblock In {\em Proceedings. 2005 IEEE International Geoscience and Remote Sensing Symposium, 2005. IGARSS '05.}, volume~6, pages 4264--4266, 2005.

\bibitem{aybar2024comprehensive}
Cesar Aybar, David Montero, Simon Donike, Freddie Kalaitzis, and Luis G{\'o}mez-Chova.
\newblock A comprehensive benchmark for optical remote sensing image super-resolution.
\newblock {\em IEEE Geoscience and Remote Sensing Letters}, 2024.

\bibitem{wang2022remote}
Yi~Wang, Syed Muhammad~Arsalan Bashir, Mahrukh Khan, Qudrat Ullah, Rui Wang, Yilin Song, Zhe Guo, and Yilong Niu.
\newblock Remote sensing image super-resolution and object detection: Benchmark and state of the art.
\newblock {\em Expert Systems with Applications}, 197:116793, 2022.

\bibitem{kowaleczko2023real}
Pawel Kowaleczko, Tomasz Tarasiewicz, Maciej Ziaja, Daniel Kostrzewa, Jakub Nalepa, Przemyslaw Rokita, and Michal Kawulok.
\newblock A real-world benchmark for sentinel-2 multi-image super-resolution.
\newblock {\em Scientific Data}, 10(1):644, 2023.

\bibitem{10798467}
Jia Wang, Liuyu Xiang, Lei Liu, Jiaochong Xu, Peipei Li, Qizhi Xu, and Zhaofeng He.
\newblock Toward real-world remote sensing image super-resolution: A new benchmark and an efficient model.
\newblock {\em IEEE Transactions on Geoscience and Remote Sensing}, 63:1--13, 2025.

\bibitem{11045064}
Wang Benlin, Yu~Qinglin, Wang Zuo, Li~Weitao, Wang Yong, Liu Huan, Gu~Shuangxi, Zhang Lingling, and Lv~Dong.
\newblock Spectral super-resolution reconstruction of multispectral remote sensing images via clustering-based spectral feature.
\newblock {\em IEEE Journal of Selected Topics in Applied Earth Observations and Remote Sensing}, 18:16227--16245, 2025.

\bibitem{zhan2025ngstgan}
Chao Zhan, Chunyang Wang, Bibo Lu, Wei Yang, Xian Zhang, and Gaige Wang.
\newblock Ngstgan: N-gram swin transformer and multi-attention u-net discriminator for efficient multi-spectral remote sensing image super-resolution.
\newblock {\em Remote Sensing}, 17(12):2079, 2025.

\bibitem{yin2025cssf}
Zhixiang Yin, Xinyan Li, Penghai Wu, Jie Lu, and Feng Ling.
\newblock Cssf: Collaborative spatial-spectral fusion for generating fine-resolution land cover maps from coarse-resolution multi-spectral remote sensing images.
\newblock {\em ISPRS Journal of Photogrammetry and Remote Sensing}, 226:33--53, 2025.

\bibitem{zhang2025ssu}
Wenjuan Zhang, Mengnan Jin, Bing Zhang, Zhen Li, Wentao Song, and Jie Pan.
\newblock Ssu-net: A novel spectral-spatial dual-branch u-net for spectral super-resolution in wide-area multispectral remote sensing imagery.
\newblock {\em IEEE Journal of Selected Topics in Applied Earth Observations and Remote Sensing}, 2025.

\bibitem{11006733}
Yining Wang, Zhixiong Huang, Xinying Wang, Shaodong Zhang, Shenglan Liu, and Lin Feng.
\newblock Lightweight edge-guided super-resolution network for remote sensing images.
\newblock {\em IEEE Transactions on Geoscience and Remote Sensing}, 63:1--14, 2025.

\bibitem{fang2020soft}
Faming Fang, Juncheng Li, and Tieyong Zeng.
\newblock Soft-edge assisted network for single image super-resolution.
\newblock {\em IEEE Transactions on Image Processing}, 29:4656--4668, 2020.

\bibitem{Wang_2018_ECCV_Workshops}
Xintao Wang, Ke~Yu, Shixiang Wu, Jinjin Gu, Yihao Liu, Chao Dong, Yu~Qiao, and Chen Change~Loy.
\newblock Esrgan: Enhanced super-resolution generative adversarial networks.
\newblock In {\em Proceedings of the European Conference on Computer Vision (ECCV) Workshops}, September 2018.

\bibitem{ballard2020firesrnet}
Tristan Ballard and Gopal Erinjippurath.
\newblock Firesrnet: Geoscience-driven super-resolution of future fire risk from climate change.
\newblock {\em arXiv preprint arXiv:2011.12353}, 2020.

\bibitem{dubey2024ssr}
Vinay Dubey and Rahul Katarya.
\newblock Ssr-gan: super resolution-based generative adversarial networks model for flood image enhancement.
\newblock {\em Signal, Image and Video Processing}, 18(8):5763--5773, 2024.

\bibitem{li2024lrsd}
Wenchao Li, Boyang Zhang, Kefeng Li, Jianyu Yang, Junjie Wu, Yin Zhang, and Yulin Huang.
\newblock Lrsd-admm-net: Simultaneous super-resolution imaging and target detection for forward-looking scanning radar.
\newblock {\em IEEE Journal of Selected Topics in Applied Earth Observations and Remote Sensing}, 17:4052--4061, 2024.

\bibitem{mostofa2020joint}
Moktari Mostofa, Syeda~Nyma Ferdous, Benjamin~S Riggan, and Nasser~M Nasrabadi.
\newblock Joint-srvdnet: Joint super resolution and vehicle detection network.
\newblock {\em IEEE Access}, 8:82306--82319, 2020.

\bibitem{lan2020madnet}
Rushi Lan, Long Sun, Zhenbing Liu, Huimin Lu, Cheng Pang, and Xiaonan Luo.
\newblock Madnet: A fast and lightweight network for single-image super resolution.
\newblock {\em IEEE transactions on cybernetics}, 51(3):1443--1453, 2020.

\bibitem{li2023yolosr}
Ronghao Li and Ying Shen.
\newblock Yolosr-ist: A deep learning method for small target detection in infrared remote sensing images based on super-resolution and yolo.
\newblock {\em Signal Processing}, 208:108962, 2023.

\bibitem{10663237}
Jiahang Liu, Jinlong Zhang, Yue Ni, Weijian Chi, and Zitong Qi.
\newblock Small-object detection in remote sensing images with super-resolution perception.
\newblock {\em IEEE Journal of Selected Topics in Applied Earth Observations and Remote Sensing}, 17:15721--15734, 2024.

\bibitem{wang2023group}
Xinya Wang, Yingsong Cheng, Xiaoguang Mei, Junjun Jiang, and Jiayi Ma.
\newblock Group shuffle and spectral-spatial fusion for hyperspectral image super-resolution.
\newblock {\em IEEE Transactions on Computational Imaging}, 8:1223--1236, 2023.

\bibitem{wu2024unsupervised}
Suqin Wu, Kefei Zhang, Xuexi Liu, Shuangshuang Shi, Chaofa Bian, et~al.
\newblock Unsupervised blind spectral--spatial cross-super-resolution network for hsi and msi fusion.
\newblock {\em IEEE Transactions on Geoscience and Remote Sensing}, 62:1--14, 2024.

\bibitem{li2014spatial}
Xiaodong Li, Feng Ling, Yun Du, Qi~Feng, and Yihang Zhang.
\newblock A spatial--temporal hopfield neural network approach for super-resolution land cover mapping with multi-temporal different resolution remotely sensed images.
\newblock {\em ISPRS journal of photogrammetry and remote sensing}, 93:76--87, 2014.

\end{thebibliography}





\end{document}